\documentstyle[preprint,epsfig,eqsecnum,aps]{revtex}
\tightenlines
\newcommand{\bea}{\begin{eqnarray}}
\newcommand{\eea}{\end{eqnarray}}
\newcommand{\beq}{\begin{equation}}
\newcommand{\eeq}{\end{equation}}
\newcommand{\bay}{\begin{array}}
\newcommand{\eay}{\end{array}}
\begin{document}
\preprint{\parbox{6cm}{\flushright CLNS 97/1500\\TECHNION-PH 97-06}}
\title{$1/N_c$ Expansion for Excited Baryons}
\author{Dan Pirjol}
\address{Department of Physics, Technion - Israel Institute of Technology,
32000 Haifa, Israel}
\author{Tung-Mow Yan}
\address{Floyd R. Newman Laboratory of Nuclear Studies, Cornell University,
Ithaca, New York 14853}
\date{\today}
\maketitle
\begin{abstract}
We derive consistency conditions which constrain the possible form of the
strong couplings of the excited baryons to the pions. The consistency conditions 
follow from
requiring the pion-excited baryon scattering amplitudes to satisfy the
large-$N_c$ Witten counting rules and are analogous to consistency conditions
used by Dashen, Jenkins and Manohar and others for s-wave baryons.
The consistency conditions are explicitly solved, giving the most general
allowed form of the strong vertices for excited baryons in the large-$N_c$ limit.
We show that the solutions to the large-$N_c$ consistency conditions coincide 
with the predictions of the nonrelativistic quark model for these states, extending
the results previously obtained  for the s-wave baryons.
The $1/N_c$ corrections to these predictions are studied in the
quark model with arbitrary number of colors $N_c$.
\end{abstract}

\narrowtext
\section{Introduction}

The successes of the nonrelativistic quark model (NRQM) in describing
the baryon spectroscopy and couplings \cite{Close} have remained for a 
long time something of a mystery. Recent work by several groups 
\cite{DM,Jen,JenMan,DJM1,DJM2,MR1,MR2,CarGeOso}, 
most notably by Dashen, Jenkins and Manohar (see also earlier related work in 
\cite{GervSak}) helped to shed light on this
problem and clarify the relation of the NRQM to QCD.
These works showed that the predictions of the NRQM for low-lying s-wave baryons 
follow from QCD in the large-$N_c$ limit \cite{LargeNc} as a consequence of the counting
rules of Witten \cite{Wi,Coleman,Lam} for pion-baryon scattering amplitudes.
In this way they have been able to derive consistency conditions 
which constrain the mass splittings, pion couplings and magnetic
moments of ground-state baryons up to order $O(1/N_c)$ in the $1/N_c$ expansion.

The nonrelativistic quark model has been used to describe also the properties of
the orbitally excited baryons. 
The realization of the fact that these states can be accounted for in the quark
model has been one of the first significant arguments in its favor \cite{Green}.
Later works applied the quark model to explaining the phenomenology of the
strong decays of the $L=1$ baryons to the ground state baryons. The measured
decay widths have been found to be well described by a fit to the
quark model predictions \cite{FaiPla,Hey,CGKM}. When supplemented with dynamic
assumptions, the quark model can be also used to make more detailed predictions
about the mass spectrum and decay properties of these states 
\cite{IsgKa1,IsgKa2,IsgKon}.

In addition to the quark model, various other approaches have been
employed to describe the orbital excitations of baryons. 
Among them the Skyrme model, which is closely related to
the large-$N_c$ approximation, has been used to construct these
states as bound states of a soliton and a meson
\cite{Dann,SchSubb,OhPa,ChWi,CK}. 
A bag model description of these states has been given in \cite{MyhWro}.
The properties of the negative parity baryons have been investigated
also with the help of the QCD sum rules in \cite{QCDSR0,QCDSR1,QCDSR2}.
More recently, in \cite{Goity}
the structure of the mass spectrum of the excited baryons has been studied
using an effective Hamiltonian motivated by large-$N_c$ arguments.

Following the recent progress in understanding the predictions of the
quark model for ground state baryons, some effort has been also directed
into  explaining the analogous predictions for the excited baryons
sector. Thus, in \cite{CGKM} the data on the strong decays of these states
have been used to test the idea that the large-$N_c$ limit might provide an
explanation for the validity of the quark model description. 
The authors of \cite{CGKM} adopted a Hartree description with the number
of quarks in the baryon fixed to its physical value $N_c=3$. The large-$N_c$ 
expansion has been
implemented at the level of operators mediating the strong decays, which can
be classified according to their order in $1/N_c$.
A fit to the experimental data on strong decays of the $L=1$ baryons
in the {\bf 70} of SU(6) gave the result that the naive
quark model, containing only one-body operators, reproduces the experimental 
data to a good precision. On the other hand, two-body
operators which could contribute to same order in $1/N_c$ as those 
kept in the quark model, appear to be suppressed in Nature for reasons
seemingly unrelated to the large-$N_c$ expansion.
From this, the authors of \cite{CGKM} concluded that there might be
more than large-$N_c$ to the success of the quark model relations.

In this paper we study the strong pion couplings of the orbitally excited
baryons, both light and heavy, in the large-$N_c$ limit using as input 
constraints on pion-baryon
scattering amplitudes following from the counting rules of Witten.
This approach is closer in spirit to the one used in \cite{DM,Jen} by
Dashen, Jenkins and Manohar. We derive in this way consistency conditions
which constrain the possible form of the strong coupling vertices,
which are then solved explicitly. Our final conclusion is that the
model-independent results obtained from solving the consistency conditions
are the same as those following from the quark model in the large-$N_c$ 
limit, thus extending the statements of \cite{DM,Jen,JenMan,DJM1,DJM2} 
to the excited baryons' sector. 
We stress that our results do not conflict with the conclusions of \cite{CGKM}.
A detailed discussion of our results in the language of \cite{CGKM} ca be found 
in Appendix B. Rather, the findings of \cite{CGKM} can be formulated in the light 
of our results as a new puzzle: why does the quark model work better than it 
should?

Our paper is structured as follows. We begin by introducing in
Section II the spectrum of the orbital excitations and constructing its generalization
to the large-$N_c$ limit. The structure of these states is more complex
than for the case of the s-wave baryons. We introduce the concept of $P$-spin
to deal with the mixed symmetry spin-flavor states 
and point out an additional
problem  connected with the appearance of spurious
unphysical states in the $N_c > 3$ case. Section III contains the
derivation of the consistency conditions for strong coupling vertices.
These arise from a mismatch between the scaling power with $N_c$ of the 
meson-baryon vertices and the Witten scaling law for the meson-baryon 
scattering amplitudes. The consistency conditions are explicitly
solved in Section III giving the most general solution for $S$-, $P$- and $D$-wave
pion couplings in the large-$N_c$ limit.
We show in Section IV that the solutions to the consistency conditions
actually coincide with the predictions of the constituent quark model in the
large-$N_c$ limit. The orbital excitations are first explicitly constructed
in the quark model with $u$ and $d$ quarks only and
arbitrary number of colors $N_c$. Armed with these wavefunctions,
we develop the machinery necessary to compute the strong coupling
vertices
of these states. A by-product of this quark model calculation is a determination
of the large-$N_c$ scaling law of the decay vertices, which exhibits a 
surprising dependence on the symmetry type of the excited state involved. 
Section V contains an exact calculation in the quark model of the strong 
coupling vertices for an arbitrary value of $N_c$. These results are used
to examine the structure of the $1/N_c$ corrections to the large-$N_c$ 
relations obtained in Section III.
We conclude in Section VI with a summary and outlook on our results.
Appendix A presents the quark model calculation of the strong couplings 
among excited states transforming under the mixed symmetry representation 
of SU(4) and Appendix B gives an interpretation of our results in the
language of quark operators.

\section{Spin-flavor structure of the excited baryons}

In the large-$N_c$ limit, the s-wave baryons containing only $u,d$
quarks form $I=J$ towers of degenerate states. Both possibilities
$I=J=1/2,3/2,\cdots$ and $I=J=0,1,\cdots$ are of physical significance,
the former corresponding to the light baryons and the latter to
baryons with one heavy quark (in this case $J$ is to be interpreted
as the angular momentum of the light degrees of freedom). Baryons with
strangeness can be also incorporated in the large-$N_c$ limit as separate 
towers of states,
each labeled by a quantum number $K$ related to the number of strange
quarks as $K=\frac12 n_s$. For each $K$ tower, the spin $J$ and isospin
$I$ take values restricted by the condition $|I-J|\leq K$.

This picture is precisely the same as the one predicted by
the NRQM with SU(4) spin-flavor symmetry. In NRQM language the
s-wave baryons have orbital wavefunctions which are completely symmetrical
under permutations of two quarks. This constrains their spin-isospin
wavefunction to transform also under the completely symmetric
representation of SU(4), which contains the $I,J$ values
given above. Spin- and flavor-independence of the interquark forces
in the NRQM is responsible for the degeneracy of all these states.
Fig.1 shows the Young diagram of the totally symmetric
representation of SU(4) for $N_c=3$ and its extension to the
case of arbitrary $N_c$.

\newcommand{\drawsquare}[2]{\hbox{%
\rule{#2pt}{#1pt}\hskip-#2pt
\rule{#1pt}{#2pt}\hskip-#1pt
\rule[#1pt]{#1pt}{#2pt}}\rule[#1pt]{#2pt}{#2pt}\hskip-#2pt
\rule{#2pt}{#1pt}}
\bea
\raisebox{-5.0pt}{\drawsquare{18.0}{0.4}}\hskip-0.4pt
\raisebox{-5.0pt}{\drawsquare{18.0}{0.4}}\hskip-0.4pt
\raisebox{-5.0pt}{\drawsquare{18.0}{0.4}}\rightarrow
\overbrace{\,\raisebox{-5.0pt}{\drawsquare{18.0}{0.4}}\hskip-0.4pt
        \raisebox{-5.0pt}{\drawsquare{18.0}{0.4}}\cdots
        \raisebox{-5.0pt}{\drawsquare{18.0}{0.4}}\,}^{N_c}\nonumber
\eea
\begin{quote}
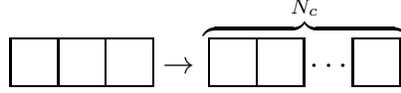
{\bf Fig.1.} Young tableaux for the SU(4) representation of the
s-wave baryons for $N_c=3$ and in the large-$N_c$ limit.
\end{quote}

The spectrum of the p-wave baryons has a more complicated structure.
The spin-flavor wavefunction
of the light baryons has mixed symmetry, transforming for $N_c=3$ as a
{\bf 70} under SU(6) and as a {\bf 20} under SU(4). 
To keep our results as general as possible and to avoid some
ambiguities connected with the identification of the large-$N_c$
states with physical states, we will not assume SU(3) symmetry.
Just as in the case of the s-wave baryons, we will divide the
p-wave states into sectors with well-defined strangeness and assume
only isospin symmetry.
We extrapolate the mixed symmetry representation from $N_c=3$ to the 
large-$N_c$ case as shown in Fig.2.

\bea
\raisebox{-24.0pt}{\drawsquare{18.0}{0.4}}\hskip-18.4pt
        \raisebox{-6pt}{\drawsquare{18.0}{0.4}}\hskip-0.4pt
\raisebox{-6pt}{\drawsquare{18.0}{0.4}}\rightarrow
\raisebox{-24.0pt}{\drawsquare{18.0}{0.4}}\hskip-18.4pt
        \raisebox{-6pt}{\drawsquare{18.0}{0.4}}\hskip-0.4pt
\raisebox{-6pt}{\drawsquare{18.0}{0.4}}\,\cdots\,
\raisebox{-6pt}{\drawsquare{18.0}{0.4}}\nonumber
\eea
\begin{quote}
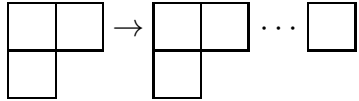
{\bf Fig.2.} SU(4) representations for light p-wave baryons, for
$N_c=3$ and in the large-$N_c$ limit.
\end{quote}

Under the isospin-spin SU(2)$\times$SU(2) group 
this representation splits into $(I,S)$ representations
which satisfy $|I-S|\leq 1$ 
(except for $I=S=N_c/2$ which is only contained in the totally symmetric 
representation). 

\bea
\overbrace{\,
\raisebox{-6pt}{\drawsquare{18.0}{0.4}}\hskip-0.4pt
\raisebox{-6pt}{\drawsquare{18.0}{0.4}}\hskip-0.4pt\,\cdots\,
\raisebox{-6pt}{\drawsquare{18.0}{0.4}}\hskip-0.4pt\,}^{N_c-1}
\,\, \otimes\,\,\,
 \raisebox{-6pt}{\drawsquare{18.0}{0.4}}\hskip-0.4pt\,\,\, =\,\,\,
\overbrace{\,
\raisebox{-6pt}{\drawsquare{18.0}{0.4}}\hskip-0.4pt
\raisebox{-6pt}{\drawsquare{18.0}{0.4}}\hskip-0.4pt\,\cdots\,
\raisebox{-6pt}{\drawsquare{18.0}{0.4}}\hskip-0.4pt\,}^{N_c}
\,\, \oplus \,\,\,
\overbrace{\,
\raisebox{-24.0pt}{\drawsquare{18.0}{0.4}}\hskip-18.4pt
        \raisebox{-6pt}{\drawsquare{18.0}{0.4}}\hskip-0.4pt
\raisebox{-6pt}{\drawsquare{18.0}{0.4}}\,\cdots\,
\raisebox{-6pt}{\drawsquare{18.0}{0.4}}\,}^{N_c-1}\nonumber
\eea
\begin{quote}
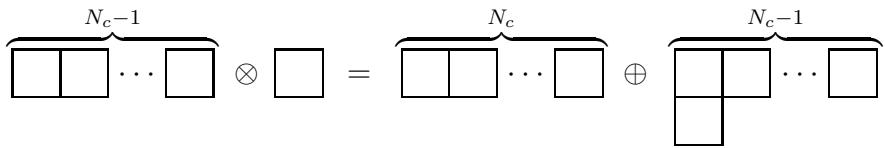
{\bf Fig.3.} Product of SU(4) representations used in the text for
the determination of the $(I,S)$ content of the mixed symmetry representation.
\end{quote}

This can be proven by considering the product of SU(4) representations
shown in Fig.3 and its decomposition into irreducible representations of
$SU(2)_{spin}\times SU(2)_{isospin}$.
For definiteness we will take $N_c$ to be odd, although the argument
is equally valid also for even values of $N_c$.
The isospin-spin $(I,S)$ content of the product of representations on
the l.h.s. can be obtained from the corresponding product
\bea
\left\{ (0,0)\,, (1,1)\,, \cdots (\frac{N_c-1}{2},\frac{N_c-1}{2})\right\}
\otimes\, (\frac12,\frac12)
\eea
and includes all representations of the form $(i\pm\frac12\,, i\pm\frac12)$
with $i=1,\cdots,(N_c-1)/2$. All the representations with $I\neq S$ occur 
with multiplicity 1. The respresentations with $(I,S)=(i+\frac12,i+\frac12)\,, 
(i-\frac12,i-\frac12)$ appear twice, except for $(I,S)=(N_c/2,N_c/2)$ which
appears only once.
On the other hand, the symmetric representation on the r.h.s. of Fig.3 
contains only the $I=S$ representations described above, but with unit multiplicity.
Subtracting them from the $(I,S)$ representations on the l.h.s. of Fig.3 
we are
left with the representation content mentioned above for the mixed symmetry
SU(4) representation.
 This can be further checked by comparing the dimensionality of
the SU(4) representation given by the Young diagram in Fig.2 for arbitrary
$N_c$ with the sum of the dimensions of the $(I,S)$ representations described
above
\bea
\mbox{dim} = \frac12(N_c-1)(N_c+1)(N_c+2) = 
\sum_{n=2}^{N_c-1}\left[ n^2+2n(n+2)\right]\,.
\eea

The total baryon spin $J$ is given by $\vec J=\vec S+\vec L$
with $L=1$.
The lowest-lying observed p-wave light baryons containing only
$u,d$ quarks are listed in Table 1 together with their quantum numbers
in the quark model \cite{PDG}. 
Note that the states $(I,S)=(3/2,3/2)$ which would be present
in the large-$N_c$ limit are forbidden in the $N_c=3$ case for 
the reason mentioned above.

\begin{center}
\begin{tabular}{|c|c|c|c|c|}
\hline
State & \quad $(I,J^P)$\quad\quad & \quad $L_{2I,2J}$\quad\quad &
\quad $(I,S)$\quad\quad & $(SU(3),SU(2))$ \\
\hline
\hline
N(1535) & $(\frac12,\frac12^-)$ & S$_{11}$ & $(\frac12,\frac12)$ & 
$({\bf 8}, {\bf 2})$ \\
N(1520) & $(\frac12,\frac32^-)$ & D$_{13}$ & &  \\
\hline
N(1650) & $(\frac12,\frac12^-)$ & S$_{11}$ & $(\frac12,\frac32)$ & 
$({\bf 8}, {\bf 4})$ \\
N(1700) & $(\frac12,\frac32^-)$ & D$_{13}$ &  &  \\
N(1675) & $(\frac12,\frac52^-)$ & D$_{15}$ &  &  \\
\hline
$\Delta(1620)$ & $(\frac32,\frac12^-)$ & S$_{31}$ & $(\frac32,\frac12)$ & 
$({\bf 10}, {\bf 2})$ \\
$\Delta(1700)$ & $(\frac32,\frac32^-)$ & D$_{33}$ &  &  \\
\hline
\end{tabular}
\end{center}
\begin{quote}
{\bf Table 1.}
The p-wave light baryons containing only $u,d$ quarks and their quantum 
numbers.
\end{quote}

It is not difficult to introduce also strangeness in this picture.
Because the strange quark is now different from the other $N_c-1$
quarks in the baryon, the Pauli principle constrains only the
symmetry properties of the wavefunction for the latter. 
In this case both SU(4) representations shown in Figs.1 and 2 are possible.
We show in Table 2 the lowest-lying observed and expected p-wave hyperons with
one strange quark together with their quark model quantum numbers.
For example, the states with $(I,S)=(1,3/2)$ in Table 2 are completely
symmetric under a permutation of the $u,d$ quarks, whereas
the states $(I,S)=(0,3/2)$ are antisymmetric under the same 
transformation (for $N_c=3$ the mixed symmetry state is in
fact antisymmetric). The symmetric representation corresponds to
{\bf 10} and the antisymmetric one to {\bf 6} of SU(4).
The other states in Table 2 are mixtures of both representations.

To construct the analogs of these states in the large-$N_c$ limit,
it is convenient to introduce two vectors $\vec K$ and $\vec P$,
which will be called $K$-spin and $P$-spin respectively. The $K$-spin 
counts the number of strange quarks as described above and takes the
value 1/2 for hyperons with one $s$ quark \cite{DJM1}. The $P$-spin labels
the type of permutational symmetry of the $u,d$ quarks' wavefunction 
in the baryon and is equal to 0 for the symmetric representation and 
to 1 for the mixed symmetric representation.

With these definitions the total quark spin $S$ of a 
p-wave baryon takes all the values compatible with
\bea
\vec S = \vec I + \vec K + \vec P\,.
\eea
In addition to this, an exclusion rule must be imposed for
$P=1$, forbidding the following symmetric states 
\bea
|\vec S-\vec K\,| = I = \frac{N_c}{2}-K\,.
\eea
This exclusion rule is operative only at the top of the
large-$N_c$ towers and therefore can be neglected in the
large-$N_c$ limit. One should keep however in mind the
fact that new unphysical states are introduced in the
large-$N_c$ limit which would be otherwise forbidden by
this rule.

The classification of the states into symmetric and mixed 
representations is even more transparent for the p-wave baryons
with one heavy quark. In the heavy mass limit the spin and parity
of the light degrees of freedom become good quantum numbers.
Furthermore, in the NRQM the total spin of the light quarks
$S_\ell$ is also conserved and can be used together with the isospin
to identify the permutational symmetry of the state.

\begin{center}
\begin{tabular}{|c|c|c|c|}
\hline
State & \quad $(I,J^P)$\quad\quad & 
\quad $(I,S)$\quad\quad & $(SU(3),SU(2))$ \\
\hline
\hline
$\Lambda(1405)$ & $(0,\frac12^-)$ & $(0,\frac12)$ & 
$({\bf 1}, {\bf 2})$ \\
$\Lambda(1520)$ & $(0,\frac32^-)$ &  &  \\
\hline
$\Lambda(1670)$ & $(0,\frac12^-)$ & $(0,\frac12)$ & 
$({\bf 1}, {\bf 2})$ \\
$\Lambda(1690)$ & $(0,\frac32^-)$ &  &  \\
$\Sigma(1620)$ & $(1,\frac12^-)$ & $(1,\frac12)$ & \\
$\Sigma(1670)$ & $(1,\frac32^-)$ & & \\
\hline
$\Lambda(1800)$ & $(0,\frac12^-)$ & $(0,\frac32)$ & 
$({\bf 8}, {\bf 4})$ \\
$\Lambda(?)$ & $(0,\frac32^-)$ &  &  \\
$\Lambda(1830)$ & $(0,\frac52^-)$ &  &  \\
$\Sigma(1750)$ & $(1,\frac12^-)$ & $(1,\frac32)$ & \\
$\Sigma(?)$ & $(1,\frac32^-)$ & & \\
$\Sigma(1775)$ & $(1,\frac52^-)$ & & \\
\hline
$\Sigma(?)$ & $(1,\frac12^-)$ & $(1,\frac12)$ & 
$({\bf 10}, {\bf 2})$ \\
$\Sigma(?)$ & $(1,\frac32^-)$ & & \\
\hline
\end{tabular}
\end{center}
\begin{quote}
{\bf Table 2.}
The p-wave hyperons containing one strange quark and their quantum 
numbers.
\end{quote}

Thus, in the large-$N_c$ limit the symmetric representation will give rise
to an $I=S_{\ell}$ tower of states, in analogy to the situation for the
light s-wave baryons (with the total spin of the light quarks $S_\ell$ taking 
the place of the total spin $J$). The mixed symmetry representation will generate 
also a tower with $|I-S_\ell|\leq 1$, as in the case of the light p-wave 
baryons. From this the states with $I=S_\ell=(N_c-1)/2$ will have to be
excluded. The total heavy baryon spin $J$ will be given in the general
case including also strangeness by
\bea
\vec J = \vec I + \vec S_\ell + \vec S_Q + \vec K + \vec P + \vec L
\eea
with $S_Q=1/2$ the heavy quark spin.

We emphasize that the use of quark model quantum numbers such as
$S,S_\ell,S_Q$, etc. does not imply any dynamical assumption on our
part and is made with the sole purpose of counting states.
All our main results below will be obtained without any assumption
of whether these quantities are conserved or not. We use the NRQM
just as a convenient language which serves to guide our intuition 
about the spin and flavor structure of the states of interest.

In the next Section we will study the strong couplings of the
excited baryons in the large-$N_c$ limit.

\section{Consistency conditions for excited baryons}

We will obtain constraints on the pion couplings of the excited
baryons by studying both elastic pion scattering on these states and 
inelastic scattering among s-wave and excited states. The results will
follow from a set of consistency conditions, derived by requiring
the total scattering amplitude to satisfy large-$N_c$ counting rules
\cite{Wi,Coleman,Lam}.
We start by reviewing the large-$N_c$ scaling properties of the
different couplings which will be needed.

Pions couple to baryons with a strength proportional to the matrix 
element of the axial current taken between the corresponding states.
In the case of the s-wave baryons this matrix element was parametrized
in \cite{DM,Jen,DJM1} as
\bea\label{Xdef}
\langle J',m',\alpha' |\bar q\gamma^i\gamma_5\frac12 \tau^aq|J,m,\alpha\rangle
= N_c g(X) \langle J',m',\alpha' |X^{ia}|J,m,\alpha\rangle
\eea
with $X^{ia}$ an irreducible tensor operator of spin and isospin 1 and
$g(X)$ a reduced matrix element of order 1 in the large-$N_c$ limit. 
$X^{ia}$ has a large-$N_c$ expansion of the form $X^{ia}=X_0^{ia} + 
X^{ia}_1/N_c + \cdots$.
The matrix element (\ref{Xdef}) grows linearly with $N_c$ because the 
axial current couples to each of the $N_c$ quarks in the baryon.

We will use a similar parametrization for the matrix element of the
axial current taken between two excited baryons
\bea\label{Zdef}
\langle J',I';m',\alpha' |\bar q\gamma^i\gamma_5\frac12 \tau^aq|J,I;m,\alpha\rangle
= N_c g(Z) \langle J',I';m',\alpha' |Z^{ia}|J,I;m,\alpha\rangle
\eea
where $Z^{ia}$ is again an irreducible tensor operator with $J=I=1$.
This matrix element grows also with $N_c$ for the same reason as 
in the preceding case.

On the other hand, the axial current matrix elements taken between
s-wave and p-wave baryons grow slower than $N_c$. We parametrize
the matrix elements of the time and space components of the
axial current as
\bea\label{Ydef}
& &\langle J',m',\alpha' |\bar q\gamma^0\gamma_5\frac12 \tau^aq|J,I;m,\alpha\rangle
= N_c^\kappa g(Y) \langle J',m',\alpha' |Y^{a}|J,I;m,\alpha\rangle\\\label{Qdef}
& &\langle J',m',\alpha' |\bar q\gamma^i\gamma_5\frac12 \tau^aq|J,I;m,\alpha\rangle
= N_c^\kappa g(Q) q^j
\langle J',m',\alpha' |Q^{ij,a}|J,I;m,\alpha\rangle
\eea
with $q^\mu$ the momentum of the current. 
In the quark model the scaling power $\kappa$ is equal to 1/2 for p-wave
baryons transforming under the completely symmetric representation of SU(4)
and 0 for baryons transforming according to the mixed symmetry
representation of SU(4). This will be proved in Section IV.
$Y^a$ is a tensor operator with
spin 0 and isospin 1 and $Q^{ij,a}$ has spin 2 and isospin 1 ($Q^{ij,a}=
Q^{ji,a}$, $Q^{ii,a}=0$).
The operators $Z,Y,Q$ have expansions in powers of $1/N_c$ of the same form
as $X$.

The pion coupling to the states appearing in (\ref{Xdef}-\ref{Qdef}) is 
obtained by dividing these matrix elements by the pion decay constant
$f_\pi=O(\sqrt{N_c})$ \cite{DM,Jen,DJM1,GervSak}.
The consistency conditions of DJM were obtained by considering pion scattering
on s-wave baryons $\pi^a(q)+B \to \pi^b(p)+B'$ \cite{DM,Jen,DJM1}. The 
leading contribution to this amplitude arises from two tree graphs
with the pions coupling in either order and is given by
\bea
{\cal T} = \frac{N_c^2 g^2(X)}{f_\pi^2}\cdot\frac{q^i p^j}{E(\vec q)}
\left( X_0^{jb\dagger}X_0^{ia} - X_0^{ia}X_0^{jb\dagger}\right)\,.
\eea
This scattering amplitude is of order $N_c$, in apparent contradiction
with the large-$N_c$ counting rules of Witten according to which it should
be of order 1. One concludes therefore that one has
\bea\label{Xcc}
\left[ X_0^{jb\dagger},\, X_0^{ia}\right] = 0\,.
\eea
This is the leading order consistency condition of DJM \cite{DM,Jen,DJM1,GervSak}.

Taking as target a p-wave baryon, the above
reasoning can be extended immediately to the couplings $Z$, for which
one obtains the analogous condition
\bea\label{Zcc}
\left[ Z_0^{jb\dagger},\, Z_0^{ia}\right] = 0\,.
\eea
The operators $Z^{ia}$ act only on the space of the p-wave states
which are degenerate among themselves and have vanishing matrix
elements between p-wave states of different mass.

We would like next to derive consistency conditions involving the
couplings $Y$ and $Q$. In order to do so we consider the scattering
amplitude for the process $\pi^a(q)+$ (p-wave) $\to \pi^b(p)+$
(s-wave). The mass splitting between s-wave and p-wave states is of
order 1 in the large-$N_c$ limit, so that the initial and final pions 
will not have the same energy. Adding together the contributions of
the diagrams with intermediate s-wave and p-wave baryons we obtain 
for this case
\bea
{\cal T} &=& \frac{N_c^{1+\kappa} g(Y)}{f_\pi^2}\left\{
-\frac{p^i E(\vec q)}{E(\vec p)}
\left(g(X) X^{ia} Y^{b\dagger} - g(Z)Y^{b\dagger} Z^{ia}\right)\right.\\
& &\qquad\qquad\qquad\left. +
\frac{q^i E(\vec p)}{E(\vec q)}
\left(g(X)X^{ib\dagger} Y^a - g(Z)Y^{a} Z^{ib\dagger} \right)\right\}\nonumber\\
&+& \frac{N_c^{1+\kappa} g(Q)}{f_\pi^2}\left\{
-\frac{p^i q^j q^k}{E(\vec p)}
\left(g(X)X^{ia} Q^{jk,b\dagger} - g(Z)Q^{jk,b\dagger} Z^{ia}\right)\right.\nonumber\\
& &\qquad\qquad\qquad\left. +
\frac{q^k p^i p^j}{E(\vec q)}
\left(g(X)X^{kb\dagger} Q^{ij,a} - g(Z)Q^{ij,a} Z^{kb\dagger} \right)\right\}
\,.\nonumber
\eea
This scattering amplitude is apparently of order $\kappa \geq 0$ which again 
violates the counting rules of Witten, according to which it should be at most
of order $N_c^{-1/2}$ \cite{Coleman}. This requires all the independent kinematical
structures to vanish to leading order
\bea\label{Ycc}
& &g(X)X^{ia}_0 Y^{b\dagger}_0 - g(Z)Y^{b\dagger}_0 Z^{ia}_0 = 0\,,\quad
g(X)X^{ib\dagger}_0 Y^{a}_0 - g(Z)Y^{a}_0 Z^{ib\dagger}_0 = 0\\
\label{Qcc}
& &g(X)X^{ia}_0 Q^{jk,b\dagger}_0 - g(Z)Q^{jk,b\dagger}_0 Z^{ia}_0 = 0\,,\quad
g(X)X^{kb\dagger}_0 Q^{ij,a}_0 - g(Z)Q^{ij,a}_0 Z^{kb\dagger}_0 = 0\,.
\eea
All of our conclusions about the pion couplings of the excited baryons in the
large-$N_c$ limit will follow from the set of consistency conditions
(\ref{Zcc},\ref{Ycc},\ref{Qcc}). In the present paper we restrict ourselves
to the leading order in the large-$N_c$ expansion. Therefore, to simplify
the notation, we will drop
the index 0 on the coupling operators throughout in the following.

\subsection{Consistency condition for $Z$}

The consistency condition for $Z^{ia}$ (\ref{Zcc}) is completely identical 
in form to the one for
$X^{ia}$ (\ref{Xcc}) which has been studied in detail in \cite{DM,DJM1}.
These authors showed that $X^{ia}$ forms, together with the generators
of the spin-isospin SU(2)$\times$SU(2) group $J^i,I^a$ a contracted
SU(4) algebra. Every possible solution for $X^{ia}$ corresponds
to a particular irreducible representation of this algebra.
The most general irreducible representation can be labeled by a
spin vector $\vec\Delta$, in terms of which the basis states of the
representation are constructed as $\vec J=\vec I + \vec \Delta$.

In principle it would be possible to take over the
results of \cite{DJM1} for $X^{ia}$ and write down directly
the matrix elements of $Z^{ia}$.
We will prefer however to construct the
solution for $Z^{ia}$ by using a NRQM-inspired ansatz.
Besides reproducing the result of \cite{DJM1}, this approach has the
advantage of suggesting a method for obtaining the solution of
the consistency conditions (\ref{Ycc},\ref{Qcc}).
In retrospect, this will furnish also a proof of the validity
of the NRQM predictions for excited baryons in the large-$N_c$ limit.

We begin by parametrizing the matrix elements of $Z^{ia}$ taken between
states belonging to $\Delta$- and $\Delta'$-towers respectively as
\bea\label{Zparam}
& &\langle J',I'; m',\alpha'|Z^{ia}|J,I;m,\alpha\rangle =\\
& &\quad (-1)^{J+I-\Delta}\sqrt{(2I+1)(2J+1)} Z(J',I';J,I)
\langle J',m'|J,1;m,i\rangle \langle I',\alpha'|I,1;\alpha,a\rangle\,.\nonumber
\eea
The notation adopted anticipates a result to be proven below,
according to which $Z$ only connects towers with $\Delta=\Delta'$.
The reduced matrix element $Z(J',I';J,I)$ depends on the common value
of $\Delta$, although for the sake of simplicity this is not made explicit.
The normalization coefficient is chosen such that the reduced
matrix element is symmetric under a permutation of the initial
and final indices $Z(J',I';J,I)=Z(J,I;J',I')$.

The consistency condition (\ref{Zcc}) can be used to obtain
constraints on the reduced matrix elements $Z(J',I';J,I)$. For this
it will be sandwiched between two general states $\langle J',I'; 
m',\alpha'|\cdots |J,I;m,\alpha\rangle$ and a complete
set of intermediate states is inserted. We obtain
\bea
& &\sum_{J_1 I_1 m_1 \alpha_1} 
\langle J',I'; m',\alpha'|Z^{jb\dagger}|J_1,I_1;m_1,\alpha_1\rangle
\langle J_1,I_1;m_1,\alpha_1 |Z^{ia}|J,I;m,\alpha\rangle\\
& &\qquad\qquad -\,(Z^{jb\dagger}\leftrightarrow Z^{ia}) = 0\nonumber\,.
\eea
This equation can be projected, as in \cite{DM}, onto the channel with
total angular momentum $H$ and isospin $K$ by multiplying it with
\bea
\langle H',h'|J',1;m',j\rangle \langle H,h|J,1;m,i\rangle
\langle K',\eta'|I',1;\alpha',b\rangle  \langle K,\eta|I,1;\alpha ,a\rangle
\eea
and summing over $m,m',i,j,\alpha,\alpha',a,b$.
The resulting consistency condition takes the form
\bea\label{Zconsistc}
& &\sum_{J_1,I_1} (2J_1+1)(2I_1+1)
\left\{ \begin{array}{ccc}
J & 1 & H \\
J' & 1 & J_1 \end{array}\right\}
\left\{ \begin{array}{ccc}
I & 1 & K \\
I' & 1 & I_1 \end{array}\right\}
Z(J',I';J_1,I_1) Z(J_1,I_1;J,I) \\
& &\qquad =\,(-)^{2(J'+I')} Z(H,K;J',I') Z(H,K;J,I)\,.\nonumber
\eea

We will try to guess the solution of this consistency condition by
using as guidance the nonrelativistic quark model. Once found,
the solution will be seen to be unique by using for example numerical
solution of the consistency condition (\ref{Zconsistc}) or the method
of the induced representations \cite{DJM1}. 

Let us consider for simplicity the case of baryons without strange quarks.
Also, let us first assume that the total baryon spin is given by 
$\vec J=\vec I + \vec L$, which is to say that the baryon will be
regarded as containing a ``core'' of $u,d$ quarks transforming
under the symmetric representation of SU(4). The ``core'' spin $S$
is therefore equal to its isospin $I$. In addition to this, the orbital
angular momentum $\vec L$ is added to make up the total spin $\vec J$.
This corresponds to the case of a heavy baryon transforming under the
symmetric representation of SU(4), provided that $J$ is interpreted as
the angular momentum of the light degrees of freedom.

The basis states can be easily constructed and are given by
\bea\label{basisL}
|I,L;J,m,\alpha\rangle = \sum_{m_S m_L}
|I,L;m_S,m_L,\alpha\rangle \langle J,m|I,L;m_S,m_L\rangle\,.
\eea
The current $Z^{ia}$ becomes in the quark model
\bea\label{QMcurrent}
Z^{ia} \to \sigma^i \otimes \tau^a
\eea
where $\sigma^i$ acts only on the spin of the $u,d$ quarks
$\vec S$ and $\tau^a$ acts only on the isospin $\vec I$.

Therefore the matrix element of $Z^{ia}$ between the states (\ref{basisL}) 
can be expressed as
\bea\label{expansion}
& &\langle I',L';J',m',\alpha' |Z^{ia} |I,L;J,m,\alpha\rangle =\\
& &\quad\sum_{m_S m_L m'_S m'_L}
\langle I',L';m'_S,m'_L,\alpha' | \sigma^i \otimes \tau^a |I,L;m_S,m_L,\alpha\rangle
\langle J',m'|I',L';m'_S,m'_L\rangle
\langle J,m|I,L;m_S,m_L\rangle\,.\nonumber
\eea
The matrix element in the basis $|I,L;m_S,m_L,\alpha\rangle$ can be
parametrized with the help of the Wigner-Eckart theorem in terms
of a new reduced matrix element $Z(I',I)$
\bea
& &\langle I',L';m'_S,m'_L,\alpha' | \sigma^i \otimes \tau^a |
I,L;m_S,m_L,\alpha\rangle=\\
& &\qquad\frac{1}{2I'+1}
Z(I',I) \langle I',m'_S|I,1;m_S,i\rangle \delta_{LL'}\delta_{m_L m_L'}
\langle I',\alpha'|I,1;\alpha,a\rangle\,.\nonumber
\eea
With this normalization the reduced matrix element is symmetric
$Z(I',I)=Z(I,I')$.

Inserting this expression in (\ref{expansion}) it is possible to
compute the matrix element of $Z^{ia}$ taken between $|I,L;J,m,\alpha\rangle$
states. Comparing with the parametrization (\ref{Zparam}) we obtain the
following connection between $Z(J'I',JI)$ and $Z(I',I)$
\bea
& &(-)^{J+I-\Delta}\sqrt{(2J+1)(2I+1)}Z(J'I',JI) =\\
& &\qquad Z(I',I)\delta_{LL'}\delta_{L\Delta}(-)^{2J'+J-I'-\Delta+1}
\sqrt{\frac{2J+1}{2I'+1}}
\left\{ \begin{array}{ccc}
I' & 1 & I \\
J & L & J' \end{array}\right\}
\,.\nonumber
\eea

We can find a consistency condition for $Z(I',I)$ by inserting this
expression into (\ref{Zconsistc}). The sum over $J_1$ can be performed
explicitly and we find
\bea
& &(2K+1) \sum_{I_1} (-)^{-2I_1}
\left\{ \begin{array}{ccc}
I & 1 & I_1 \\
I' & 1 & K \end{array}\right\}
\left\{ \begin{array}{ccc}
I & 1 & I_1 \\
\Delta & J' & I' \\
J & H & 1 \end{array}\right\}
Z(I',I_1) Z(I_1,I)\\
& &\qquad = 
(-)^{2I'-2K}
\left\{ \begin{array}{ccc}
K & 1 & I' \\
J' & \Delta & H \end{array}\right\}
\left\{ \begin{array}{ccc}
K & 1 & I \\
J & \Delta & H \end{array}\right\}
Z(K,I') Z(K,I)\,.\nonumber
\eea
It is easy to see, by making use of the relation (Eq.(6.4.8) in \cite{Edmonds})
\bea\label{Edmonds(6.4.8)}
& &\sum_{\mu} (2\mu+1)
\left\{ \begin{array}{ccc}
j_{11} & j_{12} & \mu \\
j_{23} & j_{33} & \lambda \end{array}\right\}
\left\{ \begin{array}{ccc}
j_{11} & j_{12} & \mu \\
j_{21} & j_{22} & j_{23} \\
j_{31} & j_{32} & j_{33} \end{array}\right\} = 
(-)^{2\lambda}
\left\{ \begin{array}{ccc}
j_{21} & j_{22} & j_{23} \\
j_{12} & \lambda & j_{32} \end{array}\right\}
\left\{ \begin{array}{ccc}
j_{31} & j_{32} & j_{33} \\
\lambda & j_{11} & j_{21} \end{array}\right\}\,.
\eea
that this equation is satisfied by the solution
$Z(I',I)=\sqrt{(2I+1)(2I'+1)}$ (up to a constant which
can be absorbed into $g(Z)$). 

We obtain in this way the result
\bea\label{result1}
Z(J',I';J,I) = (-)^{-I+I'+1}
\left\{ \begin{array}{ccc}
I' & 1 & I \\
J & L & J' \end{array}\right\}
\delta_{\Delta\Delta'}\delta_{L\Delta}\,.
\eea

We consider next the slightly more complicated case of the baryons 
transforming under the mixed symmetry representation of SU(4). This
is relevant for the light baryons containing only $u,d$ quarks.
In this case, the total spin of the excited baryon is given by
$\vec J=\vec I + \vec P + \vec L$. 
There is an important difference in the application of the quark
model to this situation, connected with the fact that
$\sigma^i$  in (\ref{QMcurrent}) acts on the spins of the $u,d$ 
quarks. The total spin of the $u,d$ quarks
 $\vec S=\vec I + \vec P$ is not equal to $I$ 
as before. Therefore the natural set of states for doing the
quark model calculation is $|(IP)S,L;J,m,\alpha\rangle$.

On the other hand, we would like to classify the states in
(\ref{Zparam}) according to the value of the spin vector 
$\vec \Delta$, such that $\vec I+\vec \Delta=\vec J$. This requires 
a different coupling
of the vectors $\vec I,\vec P,\vec L$: $|I,(PL)\Delta;J,m,\alpha\rangle$.
The connection between these two sets of states is a well-known
recoupling problem in the theory of angular momentum and is given by
Eq.(6.1.5) in \cite{Edmonds}
\bea\label{3Jrecoupling}
& &|I,(PL)\Delta;J,m,\alpha\rangle =\\
& &\qquad\qquad (-)^{-I-P-L-J}\sum_S
\sqrt{(2S+1)(2\Delta+1)}
\left\{ \begin{array}{ccc}
I & P & S \\
L & J & \Delta \end{array}\right\}
|(IP)S,L;J,m,\alpha\rangle\,.\nonumber
\eea

The matrix element of $Z^{ia}$ taken between the $|(IP)S,L;J,m\rangle$
states can be written as
\bea\label{(IP)S,L}
& &\langle (I'P')S',L';J',m',\alpha'|Z^{ia}|(IP)S,L;J,m,\alpha\rangle =\\
& &\quad\sum_{m_S m_L m'_S m'_L}
\langle I'S'L';m'_S,m'_L,\alpha'|\sigma^i\otimes\tau^a |ISL;m_S,m_L,\alpha\rangle
\nonumber\\
& &\qquad\qquad\times
\langle J',m'|S',L';m'_S,m'_L\rangle
\langle J,m|S,L;m_S,m_L\rangle\nonumber\,.
\eea
An application of the Wigner-Eckart theorem gives
\bea\label{WigEck}
& &\langle S' I' L';m'_S,m'_L,\alpha'|\sigma^i\otimes\tau^a |
S I L;m_S,m_L,\alpha\rangle =\\
& &\qquad \frac{1}{\sqrt{(2S'+1)(2I'+1)}}Z(S'I',SI)
\delta_{LL'}\delta_{m_L m'_L}
\langle S',m'_S|S,1;m_S,i\rangle
\langle I',\alpha'|I,1;\alpha,a\rangle\,,\nonumber
\eea
with $Z(S'I',SI)$ a new reduced matrix element.
With this choice for the normalization factor, 
it transforms under a permutation of the initial and final indices 
as
\bea\label{symm}
Z(S'I',SI)=(-)^{S+I-S'-I'}Z(SI,S'I')\,.
\eea

It is easy to compute now the matrix element of $Z^{ia}$ between
the $|I,(PL)\Delta;J,m,\alpha\rangle$ states by inserting (\ref{WigEck})
into (\ref{(IP)S,L}) and using the expansion (\ref{3Jrecoupling}).
We obtain
\bea
& &\langle I',(P'L')\Delta';J',m',\alpha'| Z^{ia} |
I, (PL)\Delta; J,m,\alpha\rangle = (-)^{-I'-P'-L'-J'-I-P-L-J}\\
& &\times\sum_{SS'}\sqrt{\frac{(2S+1)(2\Delta+1)(2\Delta'+1)}{2I'+1}}
\delta_{LL'}
\left\{ \begin{array}{ccc}
I' & P' & S' \\
L' & J' & \Delta' \end{array}\right\}
\left\{ \begin{array}{ccc}
I & P & S \\
L & J & \Delta \end{array}\right\}
Z(S'I',SI)\nonumber\\
& &\times\sum_{m_S m_L m'_S}
\langle S',m'_S|S,1;m_S,i\rangle
\langle J',m'|S',L;m'_S,m_L\rangle
\langle J,m|S,L;m_S,m_L\rangle
\langle I',\alpha'|I,1;\alpha,a\rangle\nonumber\,.
\eea
Let us pause for one moment and compare this expression with (\ref{Zparam}).
One can see that the isospin CG coefficient is the same on the r.h.s. of 
these two relations. We extract $Z(J'I',JI)$ by multiplying
both equations with $\langle J',m'|J,1;m,i\rangle$ and summing over
$(m,i)$. The resulting sum over 4 CG coefficients can be expressed
as a 6$j$ symbol. We obtain finally
\bea\label{spin-ind}
& &(-)^{J+I-\Delta}\sqrt{(2J+1)(2I+1)}Z(J',I';J,I) =\\
& &\qquad (-)^{-I'+J'-I-P'-P-L+1}
\sqrt{\frac{(2J+1)(2\Delta+1)(2\Delta'+1)}{2I'+1}}\nonumber\\
& &\quad\times\sum_{S S'} (-)^{-S'}\sqrt{(2S+1)(2S'+1)}
\left\{ \begin{array}{ccc}
I' & P' & S' \\
L & J' & \Delta' \end{array}\right\}
\left\{ \begin{array}{ccc}
I & P & S \\
L & J & \Delta \end{array}\right\}
\left\{ \begin{array}{ccc}
S' & 1 & S \\
J & L & J' \end{array}\right\}
Z(S'I',SI)\nonumber\,.
\eea

We insert the following ansatz for the reduced matrix element $Z(S'I',SI)$
(inspired by (\ref{result1}) with the identification
$(ILJ)\to (IPS)$)
\bea\label{Zansatz}
Z(S'I',SI) = (-)^{-S-I'}\sqrt{(2S+1)(2S'+1)(2I+1)(2I'+1)}
\left\{ \begin{array}{ccc}
I & P & S \\
S' & 1 & I' \end{array}\right\}
\eea
which has the required symmetry property (\ref{symm}).

If we assume that $P=P'$ we can perform the sums over $S$ and $S'$ in
(\ref{spin-ind}) with the help of the identity (Eq.(C.35e) in \cite{Messiah})
\bea\label{C.35e}
\sum_x (-)^{\phi}(2x+1)
\left\{ \begin{array}{ccc}
a & b & x \\
c & d & g \end{array}\right\}
\left\{ \begin{array}{ccc}
c & d & x \\
e & f & h \end{array}\right\}
\left\{ \begin{array}{ccc}
e & f & x \\
b & a & j \end{array}\right\} =
\left\{ \begin{array}{ccc}
g & h & j \\
e & a & d \end{array}\right\}
\left\{ \begin{array}{ccc}
g & h & j \\
f & b & c \end{array}\right\}\,,
\eea
with $\phi=a+b+c+d+e+f+g+h+x+j$.

The final result for
$Z(J'I',JI)$ is
\bea\label{result2}
Z(J',I';J,I) = (-)^{I-2J+I'+P}
\left\{ \begin{array}{ccc}
I' & 1 & I \\
J & \Delta & J' \end{array}\right\}
\delta_{\Delta\Delta'}\delta_{PP'}\,.
\eea
For $P\neq P'$ this reduced matrix element vanishes because the
operator (\ref{QMcurrent}) is totally symmetric and the initial
and final states have different permutational symmetry in the
spin-flavor of the $N_c$ quarks\footnote{In the quark model this matrix 
element receives
a nonvanishing contribution starting at order $(v/c)^2$ in the
nonrelativistic expansion \cite{TMYP}. However, a study of these
effects would take us beyond the model-independent framework of
the present work.}.

One can see that, in spite of their quite different detailed
structure, both cases considered lead to the same answer
(\ref{result1}) and (\ref{result2}), which also coincides
with the result obtained by \cite{DJM1} for the case of $X^{ia}$.
The two most important properties of this solution are now 
apparent:

\begin{itemize}
\item The excited states can be classified in towers of states 
labeled by a spin vector $\Delta$ such that $|J-I|\leq \Delta$.
To enforce the cancellation of the leading $N_c$-dependence
among the different intermediate states,
expressed by the consistency condition,
all the members of a $\Delta$-tower must
be degenerate among themselves.

\item Pions do not couple towers of excited states with different values 
of $\Delta$, as the corresponding nondiagonal matrix elements of
$Z^{ia}$ vanish.
\end{itemize}

The second property will be useful in the study of the consistency
conditions for $Y$ and $Q$ (\ref{Ycc},\ref{Qcc}), as it allows us
to consider the couplings of each $\Delta$-tower of excited baryons at a time.
We turn now to the first of them, the coupling $Y$ responsible for
S-wave pion couplings between p- and s-wave baryons.

\subsection{Consistency condition for $Y$}

We parametrize the matrix elements of the $Y^a$ operator as
\bea\label{Yparam}
& &\langle J',I';m',\alpha'| Y^a |J,I;m,\alpha \rangle =
(-)^{I+I'}\sqrt{2I+1}Y(J'I',JI)\delta_{JJ'}\delta_{m m'} 
\langle I',\alpha'|I,1;\alpha,a\rangle\,,\\
& &\quad (\mbox{s-wave})\qquad\quad (\mbox{p-wave}) \nonumber\\
\label{Ybarparam}
& & \langle J',I';m',\alpha'| Y^a |J,I;m,\alpha \rangle =
(-)^{2I}\sqrt{2I+1}\bar Y(J'I',JI)\delta_{JJ'}\delta_{m m'} 
\langle I',\alpha'|I,1;\alpha,a\rangle\,,\\
& &\quad (\mbox{p-wave})\qquad\quad (\mbox{s-wave}) \nonumber\,.
\eea
With this choice for the normalization coefficients we have
$Y(J'I',JI)=\bar Y(JI,J'I')$. The same definitions (\ref{Yparam})
and (\ref{Ybarparam}) will be used for transitions between other
orbital excitations.

We proceed next in complete analogy to the derivation of the 
consistency condition for $Z^{ia}$ (\ref{Zconsistc}). The relation
(\ref{Zcc}) is sandwiched between states
belonging to $\Delta'$- and $\Delta$-towers respectively
\bea
\langle J',I';m',\alpha'\mbox{ (s-wave)}|
rX^{ia} Y^{b\dagger} - Y^{b\dagger} Z^{ia}
|J,I;m,\alpha\mbox{ (p-wave)} \rangle = 0\,.
\eea
We denoted here $r=g(X)/g(Z)$.
Then a complete set of intermediate states is inserted between
each two operators. The necessary matrix elements of $X$ and
$Z$ are expressed with the help of the general result
(\ref{result1}).
The resulting equation is finally projected onto the particular
channel with total spin-isospin $(H,K)$. We obtain in this
way the consistency condition
\bea\label{Yconsistc}
& &r\sum_{I_1} (2I_1+1)
\left\{ \begin{array}{ccc}
I & 1 & K \\
I' & 1 & I_1 \end{array}\right\}
\left\{ \begin{array}{ccc}
I_1 & 1 & I' \\
H & \Delta' & J \end{array}\right\}
Y(JI_1,JI)\\
& &\qquad = (-)^{-I-K+\Delta'-\Delta}
\left\{ \begin{array}{ccc}
I & 1 & K \\
H & \Delta & J \end{array}\right\}
Y(HI',HK)\,.\nonumber
\eea
In addition to determining the structure of the reduced matrix
element $Y(J'I',JI)$, this relation will fix also the value of
the ratio $r$.

It is straightforward to check that the solution of the
consistency condition (\ref{Yconsistc}) is given by
\bea\label{Ysolution}
Y(JI',JI) = (-)^{I'+J+\Delta'}
\left\{ \begin{array}{ccc}
I' & 1 & I \\
\Delta & J & \Delta' \end{array}\right\}\,,
\eea
provided that $r=1$.
After substituting this solution in (\ref{Yconsistc})
the sum over $I_1$ can be done with the help of the
identity (\ref{C.35e}).

In particular, for decays into s-wave baryons containing
only $u,d$ quarks, we obtain the solution
\bea\label{Ysolutionud}
Y(J,JI) = (-)^{2J}
\left\{ \begin{array}{ccc}
J & 1 & I \\
\Delta & J & 0 \end{array}\right\} =
(-)^{1-J+I}\delta_{\Delta 1}\frac{1}{\sqrt{3(2J+1)}}\,.
\eea

Let us trace again how the same result arises in the quark model.
The quark model counterpart of the operator $Y^a$ is
\bea\label{NRQMY}
Y^a \to \sum_{i,j}\langle 0|1,1;j,i\rangle \sigma^j r^i \otimes \tau^a
= \frac{1}{\sqrt3}\sum_j (-)^{1-j} \sigma^j r^{-j} \otimes \tau^a\,,
\eea
where the light quark operator $\sigma^j$ acts only on the spins,
$r^i$ acts on the orbital degrees of freedom and $\tau^a$ acts on the
isospins.

We consider first the coupling of a p-wave state transforming under the
symmetric representation of SU(4). The matrix element (\ref{Yparam})
is written in the quark model as
\bea\label{YNRQM}
\langle J;m,\alpha'| Y^a |J,I;m,\alpha \rangle =
\sum_{m_S m_L}
\langle J;m,\alpha'| Y^a |I,L;m_S,m_L,\alpha \rangle
\langle J,m| I,L;m_S,m_L\rangle\,.
\eea
The Wigner-Eckart theorem can be used to parametrize the matrix element
of the operator on the r.h.s. of (\ref{NRQMY}) in terms of a new
reduced matrix element ${\cal T}(J,I)$
\bea
& &\langle J;m,\alpha'| \sigma^j r^i \otimes \tau^a |I,L;m_S,m_L,\alpha\rangle=\\
& &\,\frac{1}{2J+1}{\cal T}(J,I)
\langle J,m|I,1;m_S,j\rangle
\langle 0|L,1;m_L,i\rangle 
\langle J,\alpha' |I,1;\alpha,a\rangle\nonumber\,.
\eea

Inserting this expression into (\ref{YNRQM}) we obtain the result
\bea
& &\langle J;m,\alpha'| Y^a |J,I;m,\alpha \rangle \\
& &\quad=\frac{1}{3(2J+1)} {\cal T}(J,I)
\sum_{m_S m_L}
\langle J,m| I,1;m_S,m_L\rangle \langle J,m| I,L;m_S,m_L\rangle
\langle J,\alpha' |I,1;\alpha,a\rangle \delta_{L1}\nonumber\\
& &\quad=\frac{1}{3(2J+1)} {\cal T}(J,I) \langle J,\alpha' |I,1;\alpha,a\rangle
\delta_{L1}\,,
\nonumber
\eea
which can be compared with the defining matrix element of $Y^a$
(\ref{Yparam}). Taking into account the fact that for the quark model
states considered $\Delta=L$, we find the following expression for 
$Y(J,JI)$ in terms of the quark model reduced matrix element ${\cal T}(J,I)$
\bea
Y(J,JI) = (-)^{-J-I}\frac{1}{3(2J+1)\sqrt{2I+1}}\delta_{\Delta 1}
{\cal T}(J,I)\,.
\eea
It will be shown in Section IV.D by an explicit calculation in
the quark model that the reduced matrix ${\cal T}(J,I)$ is given in the large-$N_c$
limit, up to a numerical factor, by
\bea\label{Yansatz1}
{\cal T}(J,I)=(-)^{2I+1}\sqrt{3(2J+1)(2I+1)}\,.
\eea
This leads to the same expression (\ref{Ysolutionud}) for $Y(J,JI)$ as
the model-independent approach based on the consistency conditions.

It is possible to generalize this argument by keeping the
orbital angular momentum of the quark model state arbitrary $L,L'$.
They are only constrained by the requirement of parity conservation
$\pi(-)^L = \pi'(-)^{L'}$. The relevant quark model matrix 
element can be parametrized in this case as
\bea\label{NRYparam}
& &\langle I',L';m'_S,m'_L,\alpha'| 
\sigma^j r^i \otimes \tau^a |I,L;m_S,m_L,\alpha\rangle=\\
& &\,\frac{1}{(2I'+1)\sqrt{2L'+1}}{\cal T}(I'L',IL)
\langle I',m'_S|I,1;m_S,j\rangle
\langle L',m'_L|L,1;m_L,i\rangle 
\langle I',\alpha' |I,1;\alpha,a\rangle\nonumber\,,
\eea
with ${\cal T}(I'L',IL)$ another reduced matrix element. We assumed again
that the spin-flavor wavefunction of the $u,d$ quarks in the baryon 
is completely symmetric.

With the help of this relation it is possible to compute the
matrix element of $Y^a$ between eigenstates of the total spin
\bea
& &\langle J,I';m,\alpha'| Y^a |J,I;m,\alpha \rangle =
\frac{1}{(2I'+1)\sqrt{2L'+1}}{\cal T}(I'L',IL)\times\\
& &\quad \sum \langle 0|1,1;j,i\rangle
\langle J,m|I',L';m'_S,m'_L\rangle
\langle J,m|I,L;m_S,m_L\rangle\nonumber\\
& &\qquad\times\langle I',m'_S|I,1;m_S,j\rangle
\langle L',m'_L|L,1;m_L,i\rangle
\langle I',\alpha'|I,1;\alpha,a\rangle\nonumber\\
& &\quad = \frac{1}{\sqrt{3(2I'+1)}}{\cal T}(I'L',IL)
(-)^{1-I-J-L'}
\left\{ \begin{array}{ccc}
I & L & J \\
L' & I' & 1 \end{array}\right\}
\langle I',\alpha'|I,1;\alpha,a\rangle\nonumber\,.
\eea
This has the same structure as the model-independent
solution (\ref{Ysolution}), which is reproduced provided
one takes
\bea\label{ansatzY}
{\cal T}(I'L',IL) = (-)^{1+2I'}\sqrt{3(2I+1)(2I'+1)}\,.
\eea
The phase can be equivalently rewritten as $1+2I'=1+2I$ 
which gives an expression identical to (\ref{Yansatz1}).

A similar result is obtained also for the case of the excited
baryons whose spin-flavor wavefunction transforms according to 
the mixed representation of SU(4).
The relevant matrix element of $Y^a$ can be expressed, with the
help of the recoupling relation (\ref{3Jrecoupling}) in terms
of the matrix element
\bea\label{Yprelim}
& &\langle J'I'L';m',\alpha'|Y^a |(IP)S,L;J,m,\alpha \rangle = \\
& &\sum_{m_S m_L m'_S m'_L} \langle I',L';m'_S,m'_L,\alpha'|Y^a |
(IP)S,L;m_S,m_L,\alpha \rangle\langle J,m|S,L;m_S,m_L\rangle
\langle J',m'|I',L';m'_S,m'_L\rangle\,.\nonumber
\eea

The matrix element on the r.h.s. can be expressed with the
help of (\ref{NRQMY}) in terms
of a new reduced matrix element ${\cal T}(I'L',SIL)$ defined by
\bea\label{Tdef}
& &\langle I',L';m'_S,m'_L,\alpha'|\sigma^j r^i\otimes\tau^a |(IP)S,L;m_S,m_L,\alpha 
\rangle\\
& &\quad =\frac{1}{(2I'+1)\sqrt{2L'+1}}{\cal T}(I'L',SIL)
\langle I',m'_S|S,1;m_S,j\rangle \langle L',m'_L|L,1;m_L,i\rangle
\langle I',\alpha'|I,1;\alpha,a\rangle\nonumber\,.
\eea

Inserting this relation into (\ref{Yprelim}) 
we find for the matrix element of $Y^a$ in the $|(IP)S,L;J,m,\alpha\rangle$
basis
\bea\label{YIPSL}
& &\langle J'I'L';m',\alpha'|Y^a |(IP)S,L;J,m,\alpha \rangle  = 
\delta_{JJ'}\delta_{mm'}\langle I'\alpha'|I1;\alpha,a\rangle\\
& &\quad\times\frac{1}{\sqrt{3(2I'+1)}}
(-)^{1+L'-S-J}
\left\{ \begin{array}{ccc}
L & L' & 1 \\
I' & S & J \end{array}\right\}
{\cal T}(I'L',SIL)\,.\nonumber
\eea

Next we transform to the $|I,(PL)\Delta;J,m,\alpha\rangle$ basis with the
help of the recoupling relation (\ref{3Jrecoupling}).
We adopt the following ansatz for the quark model matrix element 
${\cal T}(I'L',SIL)$ 
\bea\label{Yansatz2}
{\cal T}(I'L',SIL) = (-)^{-I+I'}\sqrt{(2I+1)(2I'+1)(2S+1)}
\left\{ \begin{array}{ccc}
I & 1 & S \\
1 & I' & 1 \end{array}\right\}{\cal I}(L',L)
\eea
with ${\cal I}(L',L)$ an arbitrary function of its arguments\footnote{For
simplicity we will omit ${\cal I}(L',L)$ throughout in the following.}.
This will be derived in Section IV.D by explicit calculation in 
the quark model in the large-$N_c$ limit.
Inserting (\ref{Yansatz2}) into (\ref{3Jrecoupling}) we can perform the
sum over $S$ with the help of (\ref{C.35e}). The final result for the 
matrix element of $Y^a$ has the form, with $P=1$,
\bea\label{Yfin}
Y(J'I',JI) = \delta_{JJ'} c(LL'\Delta) (-)^{I'+J+\Delta'}
\left\{ \begin{array}{ccc}
\Delta & 1 & \Delta' \\
I' & J & I \end{array}\right\}\delta_{\Delta' L'}
\eea
with $c(LL'\Delta)$ a numerical coefficient given by
\bea
c(LL'\Delta) = \frac{1}{\sqrt3}\sqrt{2\Delta+1}(-)^{-\Delta-L'-1}
\left\{ \begin{array}{ccc}
\Delta & 1 & L' \\
1 & L & 1 \end{array}\right\}\,.
\eea
The result (\ref{Yfin}) can be seen to coincide with the general solution
of the consistency condition for $Y$ (\ref{Ysolution}).

\subsection{Consistency condition for $Q$}

The operator $Q^{ij,a}$ parametrizes pion coupling in a D-wave.
It will prove convenient to define a modified operator $Q^{ka}$  
with only one index $k=-2,-1,0,1,2$, by
\bea
Q^{ka} = \sum_{ij} \langle 2,k|1,1;i,j\rangle Q^{ij,a}\,.
\eea
It is easy to see that $Q^{ka}$ satisfies the same consistency
condition (\ref{Qcc}) as $Q^{ij,a}$.

We introduce reduced matrix elements associated with this
operator, defined by
\bea\label{Qparam}
& &\langle J',I';m',\alpha' \mbox{ (s-wave)}| Q^{ka} |
J,I;m,\alpha\mbox{ (p-wave)} \rangle =\\
& &\qquad(-)^{J+I+J'+I'}\sqrt{(2J+1)(2I+1)}Q(J'I',JI)\langle J',m'|J,2;m,k\rangle
\langle I',\alpha'|I,1;\alpha,a\rangle\,,\nonumber\\
\label{Qbarparam}
& & \langle J',I';m',\alpha' \mbox{ (p-wave)}| Q^{ka}
|J,I;m,\alpha \mbox{ (s-wave)} \rangle =\\
& &\qquad (-)^{2J+2I}\sqrt{(2J+1)(2I+1)}\bar Q(J'I',JI) \langle J',m'|J,2;m,k\rangle
\langle I',\alpha'|I,1;\alpha,a\rangle\,.\nonumber
\eea
As usual, the choice for the normalization coefficients is made
such that $Q(J'I',JI)=\bar Q(JI,J'I')$. The same definitions (\ref{Qparam})
and (\ref{Qbarparam}) apply to transitions between other orbital
excitations.

We derive a consistency condition for $Q(JI,J'I')$ by taking the
following matrix element of the relation (\ref{Qcc})
\bea
\langle J',I';m',\alpha'\mbox{ (s-wave)}|
rX^{ia} Q^{jb\dagger} - Q^{jb\dagger} Z^{ia}
|J,I;m,\alpha\mbox{ (p-wave)} \rangle = 0\,.
\eea
We insert a complete set of intermediate states between the
two operators and  project this relation onto the particular
channel with total spin-isospin $(H,K)$ by multiplication with
\bea
\langle H',h'|J',2;m',j\rangle \langle H,h|J,1;m,i\rangle
\langle K',\eta'|I',1;\alpha',b\rangle  \langle K,\eta|I,1;\alpha ,a\rangle\,.
\eea

We obtain finally the set of constraints
\bea\label{Qconsistc}
& &r\sum_{J_1 I_1}(-)^{-J'+J_1}(2J_1+1)(2I_1+1)
\left\{ \begin{array}{ccc}
I' & 1 & I_1 \\
J_1 & \Delta' & J' \end{array}\right\}
\left\{ \begin{array}{ccc}
J & 1 & H \\
J' & 2 & J_1 \end{array}\right\}
\left\{ \begin{array}{ccc}
I & 1 & K \\
I' & 1 & I_1 \end{array}\right\}Q(J_1 I_1, JI)\\
& &\qquad\qquad =(-)^{2H-I-K+\Delta'-\Delta+1}
\left\{ \begin{array}{ccc}
K & 1 & I \\
J & \Delta & H \end{array}\right\}
Q(J' I', HK)\nonumber\,.
\eea

We quote directly the solution of this consistency condition.
We will attempt later to make it plausible using a quark model
construction. The most general solution can be written as a sum of 
9$j$ symbols of the form
\bea\label{Qansatz}
Q(J'I',JI) = \sum_{y=1,2,3}c_y
\left\{ \begin{array}{ccc}
\Delta' & I' & J' \\
\Delta & I & J \\
y & 1 & 2 \end{array}\right\}
\eea
which satisfies (\ref{Qconsistc}) provided that $r=1$. In 
particular, for final s-wave states containing only $u,d$ quarks
one has $\Delta'=0$, $J'=I'$ and the 9$j$ symbols reduce to
6$j$ symbols
\bea
Q(J',JI) = c_1\frac{(-)^{J+J'}}{\sqrt{3(2J'+1)}}
\left\{ \begin{array}{ccc}
2 & J & J' \\
I & 1 & 1 \end{array}\right\}\delta_{\Delta 1} +
c_2 \frac{(-)^{J+J'+1}}{\sqrt{5(2J'+1)}}
\left\{ \begin{array}{ccc}
2 & J & J' \\
I & 1 & 2 \end{array}\right\}\delta_{\Delta 2}\,.
\eea

It is not completely straightforward to check that
(\ref{Qconsistc}) is indeed satisfied by (\ref{Qansatz}).
Therefore it might be 
useful to sketch the steps of this derivation.
First, the 9$j$ symbols on the l.h.s. are written as a sum over
3 6$j$ symbols with the help of Eq.(6.4.3) in \cite{Edmonds}
\bea\label{6.4.3}
\left\{ \begin{array}{ccc}
\Delta' & I_1 & J_1 \\
\Delta & I & J \\
y & 1 & 2 \end{array}\right\} =
\sum_x (-)^{2x} (2x+1)
\left\{ \begin{array}{ccc}
\Delta' & \Delta & y \\
1 & 2 & x \end{array}\right\} 
\left\{ \begin{array}{ccc}
I_1 & I & 1 \\
\Delta & x & J \end{array}\right\} 
\left\{ \begin{array}{ccc}
J_1 & J & 2 \\
x & \Delta' & I_1 \end{array}\right\} \,.
\eea
This allows the sum over $J_1$ to be performed with the help of
(\ref{C.35e})
\bea
& &\sum_{J_1}(-)^{-J'+J_1} (2J_1+1)
\left\{ \begin{array}{ccc}
J' & 1 & J_1 \\
I_1 & \Delta' & I' \end{array}\right\} 
\left\{ \begin{array}{ccc}
I_1 & \Delta' & J_1 \\
2 & J & x \end{array}\right\} 
\left\{ \begin{array}{ccc}
2 & J & J_1 \\
1 & J' & H \end{array}\right\} \\
& &\qquad =(-)^{\phi_1}
\left\{ \begin{array}{ccc}
I' & x & H \\
2 & J' & \Delta' \end{array}\right\} 
\left\{ \begin{array}{ccc}
I' & x & H \\
J & 1 & I_1 \end{array}\right\} \nonumber
\eea
with $\phi_1=-2J'+1+I_1+I'-\Delta'-J-x-H$. Next, the sum
over $I_1$ can be done, also with the help of (\ref{C.35e}).
As a result, the l.h.s. of (\ref{Qconsistc}) takes the form
\bea
r(-)^{\phi_2}\sum_x (-)^{2(x+\Delta')} (2x+1)
\left\{ \begin{array}{ccc}
\Delta' & \Delta & y \\
1 & 2 & x \end{array}\right\} 
\left\{ \begin{array}{ccc}
I' & K & 1 \\
\Delta & x & H \end{array}\right\} 
\left\{ \begin{array}{ccc}
J' & H & 2 \\
x & \Delta' & I' \end{array}\right\} \times
\left\{ \begin{array}{ccc}
K & H & \Delta \\
J & I & 1 \end{array}\right\} \,.
\eea
with $\phi_2 = 2H-\Delta-\Delta'-K-I+1$. 
We added a phase factor identically equal
to 1 under the summation sign, which allows the $x$-sum to be performed 
with the help of a relation analogous to (\ref{6.4.3}).
The result is a 9$j$ symbol identical to the one on the r.h.s. of 
(\ref{Qconsistc}). It is easy to check 
that also the total phase factor and the remaining 6$j$ symbol
are the same as the ones on the r.h.s. of (\ref{Qconsistc}), which
proves the validity of (\ref{Qansatz}). The
solution (\ref{Qconsistc}) satisfies the consistency condition for
$Q$ regardless of the value of $y$, which can take therefore all
values compatible with the nonvanishing of the 9$j$ symbol in which it
appears.

We will try now to make the result (\ref{Qansatz}) plausible, by 
examining the structure of this coupling in the quark model.
The operator $Q^{ka}$ is given in the quark model by
\bea\label{NRQMQ}
Q^{ka} \to \sum_{ij}\langle 2,k|1,1;j,i\rangle \sigma^j r^i \otimes \tau^a
\eea
with the same $\sigma^j, r^i, \tau^a$ as in (\ref{NRQMY}).
Let us consider first  baryons containing only $u,d$
quarks and whose flavor-spin wavefunction transforms under the symmetric
representation of SU(4). We will keep the orbital angular momenta of the
initial and final states $L,L'$ completely general, subject only to
the requirement of parity conservation $\pi(-)^L=\pi' (-)^{L'}$.

The matrix element of the quark model operators on the r.h.s. of 
(\ref{NRQMQ}) between eigenstates of $\vec S$ and $\vec L$ has been 
already parametrized in (\ref{NRYparam}) in terms of the reduced
matrix element ${\cal T}(I'L',IL)$.
The matrix element of $Q^{ka}$ between eigenstates of
$\vec J=\vec I+\vec L$ can be easily obtained as 
\bea
& &\langle J',I';m',\alpha'| Q^{ka} |J,I;m,\alpha\rangle =
\sum_{i j m_S m_L m'_S m'_L}
\langle 2,k|1,1;j,i\rangle
\langle J',m'|I',L';m'_S,m'_L\rangle\\
& &\quad\times
\langle J,m|I,L;m_S,m_L\rangle
\langle I',L';m'_S,m'_L,\alpha'| \sigma^j r^i \otimes \tau^a |
I,L;m_S,m_L,\alpha \rangle\,.\nonumber
\eea
Comparing with (\ref{Qparam}) we see that it is possible to
extract $Q(J'I',JI)$ by multiplying the r.h.s. with 
$\langle J',m'|J,2;m,k\rangle$ and summing over $m,k$.
The resulting sum over 6 CG coefficients can be written in
terms of a 9$j$ symbol by using Eq.(6.4.4) in \cite{Edmonds}. 
We obtain finally
\bea
& &(-)^{J+I+J'+I'}Q(J'I',JI)\sqrt{(2J+1)(2I+1)}\\
& &\qquad =
(-)^{-I-J+I'+J'+L+L'}\sqrt{5\frac{2J+1}{2I'+1}}
{\cal T}(I'L',IL)
\left\{ \begin{array}{ccc}
L' & I' & J' \\
L & I & J \\
1 & 1 & 2 \end{array}\right\} \,.\nonumber
\eea
For the quark model states considered one has $\Delta=L$, 
so that the 9$j$ symbol corresponds to $y=1$ in
(\ref{Qansatz}).
Requiring equality with the model-independent solution 
(\ref{Qansatz}) of the consistency condition for $Q$
gives for the quark model reduced matrix element ${\cal T}(I'L',IL)$ the
expression
\bea
{\cal T}(I'L',IL) = (-)^{L+L'}\frac{1}{\sqrt5}
\sqrt{(2I'+1)(2I+1)}\,.
\eea
This agrees, up to an unimportant overall coefficient, with the
result (\ref{ansatzY}) obtained from considering the matrix element of the
s-wave operator $Y^a$.

The 9$j$-symbol with $y=2$ in (\ref{Qansatz}) arises when considering
initial states transforming under the mixed symmetry representation
of SU(4). The calculation for this case proceeds in close analogy
to the one for the $Y^a$ operator. First we compute the matrix element
of $Q^{ka}$ in the $|(IP)S,L;J,m,\alpha\rangle$ basis with the
help of the relation
\bea\label{Qprelim}
& &\langle J'I'L';m',\alpha'|Q^{ka} |(IP)S,L;J,m,\alpha \rangle = \\
& &\sum_{m_S m_L m'_S m'_L} \langle I',L';m'_S,m'_L,\alpha'|Q^{ka} |
(IP)S,L;m_S,m_L,\alpha \rangle\langle J,m|S,L;m_S,m_L\rangle
\langle J',m'|I',L';m'_S,m'_L\rangle\nonumber
\eea
followed by the application of (\ref{NRQMQ},\ref{Tdef}). We obtain 
\bea
& &\langle J'I'L';m',\alpha'|Q^{ka} |(IP)S,L;J,m,\alpha \rangle = 
\langle J'm'|J2;mk\rangle \langle I'\alpha'|I1;\alpha a\rangle\\
& &\quad\times \sqrt{5\frac{2J+1}{2I'+1}} 
\left\{ \begin{array}{ccc}
S & 1 & I' \\
L & 1 & L' \\
J & 2 & J' \end{array}\right\}
{\cal T}(I'L',SIL)\,.\nonumber
\eea

We are eventually interested in the matrix elements of $Q^{ka}$
in the basis $|I,(PL)\Delta;J,m,\alpha\rangle$. Using the recoupling
relation (\ref{3Jrecoupling}) we get
\bea\label{3.70}
& &\langle J'I'L';m',\alpha'|Q^{ka} |I,(PL)\Delta;J,m,\alpha \rangle = 
\langle J'm'|J2;mk\rangle \langle I'\alpha'|I1;\alpha a\rangle\\
& &\qquad\times\sqrt{5(2\Delta+1)(2I+1)(2J+1)} (-)^{-L-J+2J'-I'+1}\nonumber\\
& &\qquad\times
\sum_S (-)^{-2S}(2S+1)
\left\{ \begin{array}{ccc}
I & 1 & S \\
L & J & \Delta \end{array}\right\}
\left\{ \begin{array}{ccc}
1 & 1 & 1 \\
I & I' & S \end{array}\right\}
\left\{ \begin{array}{ccc}
S & 1 & I' \\
L & 1 & L' \\
J & 2 & J' \end{array}\right\}\nonumber
\eea
where we used the ansatz (\ref{Yansatz2}) for ${\cal T}(I'L',SIL)$.
To do the sum over $S$ we first combine the two 6$j$ symbols
with the help of (\ref{C.35e}) such that $S$ appears only in one
6$j$ symbol
\bea
\left\{ \begin{array}{ccc}
I & 1 & S \\
1 & I' & 1 \end{array}\right\}
\left\{ \begin{array}{ccc}
I & 1 & S \\
L & J & \Delta \end{array}\right\} =
\sum_x (-)^\phi (2x+1)
\left\{ \begin{array}{ccc}
I' & J & x \\
\Delta & 1 & I \end{array}\right\}
\left\{ \begin{array}{ccc}
\Delta & 1 & x \\
1 & L & 1 \end{array}\right\}
\left\{ \begin{array}{ccc}
1 & L & x \\
J & I' & S \end{array}\right\}
\eea
with $\phi=I'+J+\Delta+L+I+S+1+x$.
Now the sum over $S$ can be performed using (\ref{Edmonds(6.4.8)})
\bea
\sum_S (2S+1)
\left\{ \begin{array}{ccc}
1 & I' & S \\
J & L & x \end{array}\right\}
\left\{ \begin{array}{ccc}
1 & I' & S \\
2 & J' & J \\
1 & L' & L \end{array}\right\} =
(-)^{2x}
\left\{ \begin{array}{ccc}
2 & J' & J \\
I' & x & L' \end{array}\right\}
\left\{ \begin{array}{ccc}
1 & L' & L \\
x & 1 & 2 \end{array}\right\}\,.
\eea
We obtain for the matrix element of $Q^{ka}$ in the $|I,(PL)\Delta;J,m,\alpha\rangle$
basis the following expression containing a sum over 4 6$j$ symbols
\bea\label{46js}
& &\langle J'I'L';m',\alpha'|Q^{ka} |I,(PL)\Delta;J,m,\alpha \rangle = 
\langle J'm'|J2;mk\rangle \langle I'\alpha'|I1;\alpha a\rangle\\
& &\qquad\times\sqrt{5(2\Delta+1)(2I+1)(2J+1)} (-)^{-I-J'-I'+\Delta+J+L'+L}\nonumber\\
& &\qquad\times
\sum_x (-)^{-x}(2x+1)
\left\{ \begin{array}{ccc}
I' & J & x \\
\Delta & 1 & I \end{array}\right\}
\left\{ \begin{array}{ccc}
\Delta & 1 & x \\
1 & L & 1 \end{array}\right\}
\left\{ \begin{array}{ccc}
2 & L' & x \\
I' & J & J' \end{array}\right\}
\left\{ \begin{array}{ccc}
2 & L' & x \\
L & 1 & 1 \end{array}\right\}\,.\nonumber
\eea
This can be put in a form resembling (\ref{Qansatz}) by first combining
the second and fourth 6$j$ symbols with (\ref{C.35e})
\bea
\left\{ \begin{array}{ccc}
1 & L & x \\
\Delta & 1 & 1 \end{array}\right\}
\left\{ \begin{array}{ccc}
1 & L & x \\
L' & 2 & 1 \end{array}\right\} =
\sum_{y=1,2} (-)^{\phi'}(2y+1)
\left\{ \begin{array}{ccc}
1 & 2 & y \\
1 & 1 & 1 \end{array}\right\}
\left\{ \begin{array}{ccc}
1 & 1 & y \\
\Delta & L' & L \end{array}\right\}
\left\{ \begin{array}{ccc}
\Delta & L' & y \\
2 & 1 & x \end{array}\right\}
\eea
with $\phi'=\Delta+L-L'-x+y$.
The sum over $x$ can be now done in terms of a 9$j$ symbol similar
to those in (\ref{Qansatz})
\bea
\sum_x (-)^{2x}(2x+1)
\left\{ \begin{array}{ccc}
I' & J & x \\
\Delta & 1 & I \end{array}\right\}
\left\{ \begin{array}{ccc}
\Delta & 1 & x \\
2 & L' & y \end{array}\right\}
\left\{ \begin{array}{ccc}
L' & 2 & x \\
J & I' & J' \end{array}\right\} =
\left\{ \begin{array}{ccc}
L' & I' & J' \\
\Delta & I & J \\
y & 1 & 2 \end{array}\right\}\,.
\eea
When inserted into (\ref{46js}) this gives a result
for the reduced matrix element $Q(J'I',JI)$ of the
same form as (\ref{Qansatz})
\bea
Q(J'I',JI) = \sum_{y=1,2}c_y(LL'\Delta)
\left\{ \begin{array}{ccc}
L' & I' & J' \\
\Delta & I & J \\
y & 1 & 2 \end{array}\right\}
\eea
with coefficients $c_y$ given by
\bea
c_y(LL'\Delta) = \sqrt{5(2\Delta+1)}
(-)^{-2J'+y}(2y+1)
\left\{ \begin{array}{ccc}
1 & 2 & y \\
1 & 1 & 1 \end{array}\right\}
\left\{ \begin{array}{ccc}
1 & 1 & y \\
\Delta & L' & L \end{array}\right\}\,.
\eea

Finally, the most general solution for $Q(J'I',JI)$ containing also
9$j$ symbols with $y=3$ is obtained if one considers transitions
among two states with mixed symmetry. This situation is not very
relevant from a phenomenological point of view so that its
discussion is relegated to Appendix B.

\newpage
\section{Quark model matrix elements}

\subsection{Symmetric states}

In this Section we compute the reduced matrix elements of the
operator $\sigma^i\otimes\tau^a$ on quark model states with
arbitrary number of colors. It will be seen that in the limit
$N_c\to\infty$ these reduced matrix elements coincide with those
required by the consistency conditions discussed in Section III.

We start by computing the reduced matrix element $Z(I',I)$ defined
by
\bea\label{A1}
& &\langle I',L';m'_S,m'_L,\alpha' | \sigma^i \otimes \tau^a |
I,L;m_S,m_L,\alpha \rangle=\\
& &\qquad\frac{1}{2I'+1}
Z(I',I) \langle I',m'_S|I,1;m_S,i\rangle \delta_{LL'}\delta_{m_L m_L'}
\langle I',\alpha'|I,1;\alpha,a\rangle\,.\nonumber
\eea
The states on the l.h.s. transform under the completely symmetric
representation of SU(4). For simplicity we will take them to contain
only $u$- and $d$-type quarks, although additional quark flavors can be
included in a straightforward way. In the quark model with $N_c$ colors 
they are given by
\bea\label{A2}
|I,m,\alpha\rangle &=& \sum_i \langle I,m|\frac{N_u}{2},\frac{N_d}{2};
i,m-i\rangle {\cal S}|\frac{N_u}{2},i\rangle_u 
|\frac{N_d}{2},m-i\rangle_d \\
&=& \sum_i \langle I,m|\frac{N_u}{2},\frac{N_d}{2};
i,m-i\rangle {\cal S}(u\uparrow)^{\frac{N_u}{2}+i}
(u\downarrow)^{\frac{N_u}{2}-i}
(d\uparrow)^{\frac{N_d}{2}+m-i}
(d\downarrow)^{\frac{N_d}{2}-m+i}\,,\nonumber
\eea
with
\bea\label{A3}
N_u=\frac{N_c}{2} + \alpha\,,\qquad N_d=\frac{N_c}{2} - \alpha
\eea
the numbers of $u$ and $d$ quarks, respectively, in the baryon state.
The symbol ${\cal S}$ means complete symmetrization under permutation of
all quarks. The explicit form of the wavefunction (\ref{A2}) has been 
given without proof in \cite{Soldate} and a particular case was previously 
considered in  \cite{KakToy}. 
For a simple method of computing matrix elements in the quark model with $N_c$
colors see \cite{KaPat}.
In the following, for completeness of the presentation we give a detailed
derivation of (\ref{A2}).

{\em Proof.} Any completely symmetric state of $N_c$ quarks can be constructed
as a linear combination of symmetrized products of one-quark states
\bea
{\cal S}(n_1,n_2,n_3,n_4) = \frac{1}{\sqrt{{\cal N}}}\left( 
(u\uparrow)^{n_1}
(u\downarrow)^{n_2}
(d\uparrow)^{n_3}
(d\downarrow)^{n_4} + \mbox{ permutations}\right)
\eea
with 
\bea
{\cal N} = \frac{(n_1+n_2+n_3+n_4)!}{n_1! n_2! n_3! n_4!}\,.
\eea
It is easy to see that the action of spin and isospin operators on
these states is given by
\bea
\sum_i \sigma_+^i {\cal S}(n_1,n_2,n_3,n_4) &=& 
\sqrt{n_2(n_1+1)}{\cal S}(n_1+1,n_2-1,n_3,n_4)\\
& &\qquad + \sqrt{n_4(n_3+1)}{\cal S}(n_1,n_2,n_3+1,n_4-1)\nonumber \\
\sum_i \sigma_-^i {\cal S}(n_1,n_2,n_3,n_4) &=& 
\sqrt{n_1(n_2+1)}{\cal S}(n_1-1,n_2+1,n_3,n_4)\\
& &\qquad +
\sqrt{n_3(n_4+1)}{\cal S}(n_1,n_2,n_3-1,n_4+1)\nonumber \\
\sum_i \sigma_z^i {\cal S}(n_1,n_2,n_3,n_4) &=& 
(n_1 - n_2 + n_3 - n_4) {\cal S}(n_1,n_2,n_3,n_4)\\
\sum_i \tau_+^i {\cal S}(n_1,n_2,n_3,n_4) &=& 
\sqrt{n_3(n_1+1)}{\cal S}(n_1+1,n_2,n_3-1,n_4)\\
& &\qquad +
\sqrt{n_4(n_2+1)}{\cal S}(n_1,n_2+1,n_3,n_4-1)\nonumber \\
\sum_i \tau_-^i {\cal S}(n_1,n_2,n_3,n_4) &=& 
\sqrt{n_1(n_3+1)}{\cal S}(n_1-1,n_2,n_3+1,n_4)\\
& &\qquad +
\sqrt{n_2(n_4+1)}{\cal S}(n_1,n_2-1,n_3,n_4+1)\\
\sum_i \tau_z^i {\cal S}(n_1,n_2,n_3,n_4) &=& 
(n_1 + n_2 - n_3 - n_4) {\cal S}(n_1,n_2,n_3,n_4)\,.\nonumber
\eea
It will prove more convenient to express the arguments of the symmetrized
products of one-quark states in terms of four angular momentum-like
variables defined as
\bea
n_1 &=& j_1 + m_1\\
n_2 &=& j_1 - m_1 \\
n_3 &=& j_2 + m_2 \\
n_4 &=& j_2 - m_2\,.
\eea
In terms of these variables, the action of the 
spin and isospin operators can be expressed as
\bea\label{ladder1}
\sum_i \sigma_+^i {\cal S}(j_1,j_2,m_1,m_2) &=& 
\sqrt{(j_1-m_1)(j_1+m_1+1)}{\cal S}(j_1,j_2,m_1+1,m_2)\\
& & + \sqrt{(j_2-m_2)(j_2+m_2+1)}{\cal S}(j_1,j_2,m_1,m_2+1)\nonumber \\
\sum_i \sigma_-^i {\cal S}(j_1,j_2,m_1,m_2) &=& 
\sqrt{(j_1+m_1)(j_1-m_1+1)}{\cal S}(j_1,j_2,m_1-1,m_2)\\
& & +
\sqrt{(j_2+m_2)(j_2-m_2+1)}{\cal S}(j_1,j_2,m_1,m_2-1)\nonumber \\
\sum_i \sigma_z^i {\cal S}(j_1,j_2,m_1,m_2) &=& 
2(m_1+m_2){\cal S}(j_1,j_2,m_1,m_2)\\
\sum_i \tau_+^i {\cal S}(j_1,j_2,m_1,m_2) &=& 
\sqrt{(j_2+m_2)(j_1+m_1+1)}{\cal S}(j_1+\frac12,j_2-\frac12,m_1+\frac12,m_2-\frac12)\\
& & + \sqrt{(j_2-m_2)(j_1-m_1+1)}
{\cal S}(j_1+\frac12,j_2-\frac12,m_1-\frac12,m_2+\frac12)\nonumber \\
\sum_i \tau_-^i {\cal S}(j_1,j_2,m_1,m_2) &=& 
\sqrt{(j_1+m_1)(j_2+m_2+1)}{\cal S}(j_1-\frac12,j_2+\frac12,m_1-\frac12,m_2+\frac12)\\
& & + \sqrt{(j_1-m_1)(j_2-m_2+1)}
{\cal S}(j_1-\frac12,j_2+\frac12,m_1+\frac12,m_2-\frac12)\nonumber\\
\sum_i \tau_z^i {\cal S}(j_1,j_2,m_1,m_2) &=& 
2(j_1-j_2){\cal S}(j_1,j_2,m_1,m_2)\,.\label{ladder2}
\eea

A state of well-defined spin is constructed by taking appropriate linear 
combinations of symmetrized products of one-particle states
\bea\label{cdef}
|I,m,\alpha\rangle = \sum_{m_1,m_2}c(m_1,m_2,m,j_1,j_2) {\cal S}(j_1,j_2,m_1,m_2)
\eea
with $m_1+m_2=m$.
The quantum numbers of the state fix $j_1$ and $j_2$ through the conditions
\bea
\alpha &=& j_1 - j_2\\
N_c &=& 2(j_1 + j_2)
\eea
which give (\ref{A3}) with $j_1=N_u/2$ and $j_2=N_d/2$.

The coefficients $c$ in (\ref{cdef}) can be determined by requiring the
states $|I,m,\alpha\rangle$ to satisfy the relations
\bea
J_\pm |I,m,\alpha\rangle &=& \sqrt{(I\mp m)(I\pm m+1)}|I,m\pm 1,\alpha\rangle\,.
\eea
Inserting the expansion (\ref{cdef}) one finds, with the help of 
(\ref{ladder1}-\ref{ladder2}), the following recursion relations among
the coefficients $c$
\bea
& &\sqrt{(I-m)(I+m+1)}c(m_1,m_2,m+1,j_1,j_2)  \\
& &\qquad=\sqrt{(j_1+m_1)(j_1-m_1+1)} c(m_1-1,m_2,m,j_1,j_2)\nonumber\\
& &\qquad+ \sqrt{(j_2+m_2)(j_2-m_2+1)} c(m_1,m_2-1,m,j_1,j_2)\nonumber\\
& &\sqrt{(I+m)(I-m+1)}c(m_1,m_2,m-1,j_1,j_2) \\
& &\qquad=\sqrt{(j_1-m_1)(j_1+m_1+1)} c(m_1+1,m_2,m,j_1,j_2)\nonumber\\
& &\qquad+ \sqrt{(j_2-m_2)(j_2+m_2+1)} c(m_1,m_2+1,m,j_1,j_2)\nonumber\,.
\eea
These relations can be seen to coincide with the familiar recursion
relations for the Clebsch-Gordan coefficients, with the identification
\bea
c(m_1,m_2,m,j_1,j_2) = \langle I,m|j_1,j_2; m_1, m_2\rangle\,.
\eea
It is known that these recursion relations fix uniquely the CG coefficients
up to an overall phase. To complete our proof of (\ref{A2}) we still have
to show that this state is also an eigenstate of $\vec I\,^2$, with the same 
eigenvalue as $\vec J\,^2$. This can be done by comparing the action 
of $\vec J\,^2$ on the state (\ref{A2}) with that of $\vec I\,^2$. We obtain
\bea
\vec J\,^2|I,m,\alpha\rangle &=& 
\left(\frac12 J_+ J_- + \frac12 J_- J_+ + J_z^2\right)|I,m,\alpha\rangle\\
&=& \sum_{m_1, m_2}c(m_1,m_2,m,j_1,j_2)
\left\{ \left(j_1(j_1+1)+j_2(j_2+1)+2m_1 m_2\right){\cal S}(j_1,j_2,m_1,m_2)\right.
\nonumber\\
&+&\left. \sqrt{(j_1+m_1)(j_1-m_1+1)(j_2-m_2)(j_2+m_2+1)}
{\cal S}(j_1,j_2,m_1-1,m_2+1)\right.\nonumber\\
&+& \left. \sqrt{(j_1-m_1)(j_1+m_1+1)(j_2+m_2)(j_2-m_2+1)}
{\cal S}(j_1,j_2,m_1+1,m_2-1)\right\}\nonumber
\eea
which also coincides with the result of applying $\vec I\,^2$ on the same state.

The knowledge of the states (\ref{A2}) can be used to calculate the matrix 
element (\ref{A1}). We will choose for this calculation the spherical 
component $(i,a)=(0,0)$ of the current in (\ref{A1}). The corresponding 
quark model operator can be written as a sum over $N_c$ one-quark operators
\bea
\sigma^0 \otimes \tau^0 = \sum_{i=1}^{N_c} \sigma^i_3\tau^i_3\,.
\eea

Because of the symmetry property of $Z(I',I)$, there are only two independent
quantities to calculate: $Z(I,I)$ and $Z(I,I-1)$. We obtain for them
the results
\bea\label{Zres1}
Z(I,I) &=& (2I+1)(N_c + 2)\\\label{Zres2}
Z(I,I-1) &=& \sqrt{(2I-1)(2I+1)}
\sqrt{(N_c+2+2I)(N_c+2-2I)}\,.
\eea

In order to obtain $Z(I,I)$ we consider the following matrix element of 
the type (\ref{A1})
\bea\label{A5}
\langle I,I,I |\sum_{i=1}^{N_c} \sigma^i_3\tau^i_3  |
I,I,I \rangle=\frac{I}{(I+1)(2I+1)} Z(I,I)\,.
\eea
The quark model matrix element on the l.h.s. can be computed with the help of
the wavefunction (\ref{A2}) with the result
\bea\label{A6}
\langle I,I,I |\sum_{i=1}^{N_c} \sigma^i_3\tau^i_3  |I,I,I \rangle &=&
\sum_m |\langle I,I|\frac{N_u}{2},\frac{N_d}{2};m,I-m\rangle |^2(4m-2I)\\
&=& \frac{N_u(N_u+2)-N_d(N_d+2)}{2(I+1)}\nonumber\,.
\eea
$N_u,N_d$ are given by (\ref{A3}). Inserting this expression into (\ref{A5})
one obtains the result (\ref{Zres1}) for $Z(I,I)$.

For $Z(I,I-1)$ we consider the matrix element
\bea\label{A7}
\langle I,I-1,I-1 |\sum_{i=1}^{N_c} \sigma^i_3\tau^i_3  |
I-1,I-1,I-1 \rangle=\frac{1}{I(2I+1)} Z(I,I-1)\,.
\eea
The quark model matrix element can be computed with the result
\bea
& &\langle I,I-1,I-1 |\sum_{i=1}^{N_c} \sigma^i_3\tau^i_3  |
I-1,I-1,I-1 \rangle \\
& &\qquad =4\sum_m m\langle I,I-1|\frac{N_u}{2},\frac{N_d}{2};m,I-1-m\rangle
\langle I-1,I-1|\frac{N_u}{2},\frac{N_d}{2};m,I-1-m\rangle\nonumber\\
& &\qquad = \frac{2}{I}\sqrt{\frac{2I-1}{2I+1}\left(\frac{N_c}{2}+1+I\right)
\left(\frac{N_c}{2}+1-I\right)}\nonumber\,.
\eea
Comparing with (\ref{A7}) gives immediately the result (\ref{Zres2}).

The results (\ref{Zres1}), (\ref{Zres2}) can be put into a common form
\bea\label{common}
Z(I',I) &=& \sqrt{(2I'+1)(2I+1)}\sqrt{(N_c+2)^2-(I'-I)^2(I'+I+1)^2}\\
&=& (N_c+2)\sqrt{(2I'+1)(2I+1)} + {\cal O}(1/N_c)\,.\nonumber
\eea
We have made here apparent the fact that the corrections to the lowest-order
result come only at sub-subleading order in $1/N_c$. This is an illustration, 
on the example of the
quark model, of a model-independent result obtained by Dashen and Manohar \cite{DM}
using the counting rules for pion-baryon scattering.

\subsection{Mixed symmetry states}

In this Section we construct quark model states whose spin-flavor
wavefunctions transform under the mixed symmetry representation of
SU(4) shown in Fig.2. They can be built using the procedure described
in Sect.II, by adding one extra quark to a symmetric state of $N_c-1$
quarks. We write the state obtained by adding the $j^{th}$ quark
to a symmetric state of $N_c-1$ quarks with spin and isospin $i$, as
\bea
|SI,m,\alpha\rangle_j = \sum_{m_1 m_2 \alpha_1 \alpha_2}
\langle S,m|i,\frac12; m_1,m_2\rangle 
\langle I,\alpha|i,\frac12; \alpha_1,\alpha_2\rangle 
|i,m_1,\alpha_1\rangle \otimes |\frac12,m_2,\alpha_2\rangle_j\,.
\eea

The states of mixed symmetry under SU(4) must be antisymmetric under
permutations of the two quarks corresponding to the first column of
the Young diagram.
There are $N_c(N_c-1)/2$ ways to choose such a pair of quarks,
but not all the states obtained in this way will be linearly independent.
In fact there are only $N_c-1$ independent states with mixed symmetry,
and we will choose them such that they are antisymmetric under a
permutation of the first quark with any of the remaining $N_c-1$
quarks in the baryon. The corresponding spin-flavor wavefunction
will be denoted as
\bea\label{4.39}
|SI,m,\alpha\rangle_{[j,1]} = \frac{1}{\sqrt2}\left(
|SI,m,\alpha\rangle_j - |SI,m,\alpha\rangle_1\right)\,,\qquad
j=2,3,\dots,N_c\,.
\eea

The space part of the wavefunction must transform also under the
mixed symmetry representation of the permutation group, corresponding
to the same Young diagram as in Fig.2. There are again $N_c-1$
linearly independent wavefunctions, which can be chosen to be
antisymmetric under a permutation of the $j^{th}$ and $1^{st}$ quarks.
Their generic form is
\bea\label{mixedsp}
|L,m_L\rangle_{[j,1]} = \frac{1}{\sqrt2}\left(\psi( r_j) 
Y_{Lm_L}(\hat r_j)\phi( r_1) - \psi( r_1) 
Y_{Lm_L}(\hat r_1)\phi( r_j)\right)\phi_S( r_2,\cdots, r_{j-1}, r_{j+1},\cdots
, r_{N_c})
\eea
with $\phi_S(r_2, \cdots ,r_{j-1}, r_{j+1}, \cdots, r_{N_c})$ a symmetric function 
of its arguments. In (\ref{mixedsp}), we have assumed that the orbital angular
momentum is carried by a single quark. This is strictly true only for the lowest
orbital excitations.

It is easy to combine now the spatial and the spin-flavor parts
into a completely symmetric wavefunction of well-defined
spin and isospin. Our final result for such a quark model state is
\bea
|JI,m,\alpha\rangle = \sum_{m_S,m_L}\langle J,m|S,L;m_S,m_L\rangle
|SIL,m_S,m_L,\alpha\rangle
\eea
with
\bea\label{SImixed}
|SIL,m_S,m_L,\alpha\rangle = (-)^{\psi(SIi)}
\frac{1}{\sqrt{N_c-1}}\sum_{j=2}^{N_c}
|SI,m_S,\alpha\rangle_{[j,1]}\otimes |L,m_L\rangle_{[j,1]}\,.
\eea
The phase of these states $\psi(SIi)$ will be chosen later for
convenience.

These states have a peculiar normalization, due to the fact that
the spatial wavefunctions (\ref{mixedsp}) with $j\neq j'$ are
not orthogonal. They satisfy instead
\bea\label{4.43}
_{[j',1]} \langle L,m'_L|L,m_L\rangle_{[j,1]} = \frac12(\delta_{jj'}+1)
\delta_{m_Lm'_L}{\cal I}
\eea
with ${\cal I}$ an overlap integral.
Using this expression we obtain the following exact result for the norm 
of the states (\ref{SImixed})
\bea\label{4.44}
& &\langle S'I'1,m'_S,m'_L,\alpha'|SI1,m_S,m_L,\alpha\rangle =
\delta_{SS'}\delta_{m_S m'_S}\delta_{m_L m'_L}
\delta_{II'}\delta_{\alpha\alpha'}\frac{N_c+2}{4}
{\cal I}\\
& &\qquad\times
\left\{ 3(2i+1)
\left\{ \begin{array}{ccc}
S & I & 1 \\
\frac12 & \frac12 & i \end{array}\right\}^2 -
\frac{2(2i+1)}{N_c-1}
\left\{ \begin{array}{ccc}
S & I & 0 \\
\frac12 & \frac12 & i \end{array}\right\}^2\right.\nonumber\\
& &\qquad\qquad\qquad\left. +
\frac{1}{2(N_c-1)}
\left[\frac52 + 2i(i+1) - S(S+1) - I(I+1)\right]\right\}\,.\nonumber
\eea

The derivation of this relation will be presented in some detail, as it
illustrates a few techniques useful in dealing with the mixed symmetry
states. We start by computing the scalar product of two direct
product states
\bea\label{dirproduct}
_{j'}\langle S'I',m',\alpha'|SI,m,\alpha\rangle_j &=&
\sum \langle S',m'|i',\frac12; m'_1,m'_2\rangle 
\langle I',\alpha'|i',\frac12; \alpha'_1,\alpha'_2\rangle 
\langle S,m|i,\frac12; m_1,m_2\rangle \\
& & \hspace{-2cm}\times\langle I,\alpha|i,\frac12; \alpha_1,\alpha_2\rangle 
_{j'}\langle \frac12,m'_2,\alpha'_2|\otimes \langle i',m'_1,\alpha'_1
|i,m_1,\alpha_1\rangle \otimes |\frac12,m_2,\alpha_2\rangle_j\,.\nonumber
\eea
The matrix element on the r.h.s. can be written as
\bea
& &_{j'}\langle \frac12,m'_2,\alpha'_2|\otimes \langle i',m'_1,\alpha'_1
|i,m_1,\alpha_1\rangle \otimes |\frac12,m_2,\alpha_2\rangle_j\\
& &\qquad =\, _{j}\langle \frac12,m'_2,\alpha'_2|\otimes \langle i',m'_1,\alpha'_1|
P_{jj'}|i,m_1,\alpha_1\rangle \otimes |\frac12,m_2,\alpha_2\rangle_j\nonumber
\eea
where 
\bea\label{4.47}
P_{jj'} = \frac14\left(1 + \vec\sigma_j\cdot\vec\sigma_{j'}\right)
\left(1 + \vec\tau_j\cdot\vec\tau_{j'}\right)
\eea
is an operator which exchanges the spins and isospins of the $j,j'$
quarks. We obtain in this way
\bea\label{4.48}
& &_{j'}\langle \frac12,m'_2,\alpha'_2|\otimes \langle i',m'_1,\alpha'_1
|i,m_1,\alpha_1\rangle \otimes |\frac12,m_2,\alpha_2\rangle_j\\
& &\qquad = \frac14\, _{j}\langle\frac12,m'_2,\alpha'_2|\frac12,m_2,\alpha_2\rangle_j
\langle i',m'_1,\alpha'_1|i,m_1,\alpha_1\rangle \nonumber\\
& &\qquad + \frac14 \sum_k (-)^k\,
_{j}\langle\frac12,m'_2,\alpha'_2|\sigma_j^{-k}|\frac12,m_2,\alpha_2\rangle_j
\langle i',m'_1,\alpha'_1|\sigma_{j'}^k|i,m_1,\alpha_1\rangle\nonumber\\
& &\qquad + \frac14 \sum_b (-)^b\,
_{j}\langle\frac12,m'_2,\alpha'_2|\tau_j^{-b}|\frac12,m_2,\alpha_2\rangle_j
\langle i',m'_1,\alpha'_1|\tau_{j'}^b|i,m_1,\alpha_1\rangle\nonumber\\
& &\qquad + \frac14 \sum_{k,b} (-)^{k+b}\,
_{j}\langle\frac12,m'_2,\alpha'_2|\sigma_j^{-k}\tau_j^{-b}|\frac12,m_2,\alpha_2\rangle_j
\langle i',m'_1,\alpha'_1|\sigma_{j'}^k\tau_{j'}^b|i,m_1,\alpha_1\rangle\nonumber\,.
\eea
The matrix elements on the one-quark states are computed easily with the 
results
\bea
& &_{j}\langle\frac12,m'_2,\alpha'_2|\frac12,m_2,\alpha_2\rangle_j =
\delta_{m_2 m'_2}\delta_{\alpha_2 \alpha'_2}\\
& &_{j}\langle\frac12,m'_2,\alpha'_2|\sigma_j^{k}|\frac12,m_2,\alpha_2\rangle_j
= \sqrt3 \langle \frac12, m'_2|\frac12, 1; m_2,k\rangle \delta_{\alpha_2 \alpha'_2}\\
& &_{j}\langle\frac12,m'_2,\alpha'_2|\tau_j^{b}|\frac12,m_2,\alpha_2\rangle_j
= \sqrt3 \langle \frac12, \alpha'_2|\frac12, 1; \alpha_2,b\rangle 
\delta_{m_2 m'_2}\\
& &_{j}\langle\frac12,m'_2,\alpha'_2|\sigma_j^k\tau_j^b|\frac12,m_2,\alpha_2\rangle_j
= 3\langle \frac12, m'_2|\frac12, 1; m_2,k\rangle
\langle \frac12, \alpha'_2|\frac12, 1; \alpha_2,b\rangle \,.
\eea

The matrix elements of the one-quark operators taken on symmetric states
containing $N_c-1$ quarks can be obtained with the help of the wavefunction
(\ref{A2}) of these states.
For example, the matrix element of $\sigma_j^k$ is parametrized as
\bea\label{sndef}
\langle i',m',\alpha'|\sigma_j^k|i,m,\alpha\rangle
= F(i)\delta_{ii'}\delta_{\alpha\alpha'}\langle i,m'|i,1;m,k\rangle\,.
\eea
The state $|i,m,\alpha\rangle$ has the explicit form
\bea
|i,m,\alpha\rangle = \sum_k \langle i,m|j_1,j_2;k,m-k\rangle
{\cal S}_{N_c-1}(j_1+k,j_1-k,j_2+m-k,j_2-m+k)
\eea
with $j_1=N_u/2$, $j_2=N_d/2$ and $N_{u,d}=(N_c-1)/2\pm\alpha$.
Next, we single out the quark $j$ by using the relation
\bea\label{nsingled}
& &{\cal S}_{N_c}(n_1,n_2,n_3,n_4) = \\
& &\qquad \sqrt{\frac{n_1}{N_c}}(u\uparrow)_j {\cal S}_{N_c-1}(n_1-1,n_2,n_3,n_4) +
\sqrt{\frac{n_2}{N_c}}(u\downarrow)_j {\cal S}_{N_c-1}(n_1,n_2-1,n_3,n_4)
\nonumber\\
& &\qquad+ \sqrt{\frac{n_3}{N_c}}(d\uparrow)_j {\cal S}_{N_c-1}(n_1,n_2,n_3-1,n_4) +
\sqrt{\frac{n_4}{N_c}}(d\downarrow)_j {\cal S}_{N_c-1}(n_1,n_2,n_3,n_4-1)\,.
\nonumber
\eea
The reduced matrix element $F(i)$ can be computed by taking the spherical
component $k=0$ in (\ref{sndef}).
The matrix element on the l.h.s. of this relation can be written with the help 
of (\ref{nsingled}) as
\bea
& &\langle i,m,\alpha|\sigma_j^0|i,m,\alpha\rangle =
\sum_k |\langle i,m|j_1,j_2;k,m-k\rangle |^2\\
& &\qquad \times\left\{\frac{j_1+k}{N_c-1} - \frac{j_1-k}{N_c-1} +
\frac{j_2+m-k}{N_c-1} - \frac{j_2-m+k}{N_c-1}\right\} =
\frac{2m}{N_c-1}\nonumber\,.
\eea
Comparing with (\ref{sndef}) we obtain
\bea
F(i) = \frac{2}{N_c-1}\sqrt{i(i+1)}\,.
\eea
In a completely analogous way we write the other needed matrix elements as
\bea
& &\langle i',m',\alpha'|\tau_j^a|i,m,\alpha\rangle
= G(i)\delta_{ii'}\delta_{mm'}\langle i,\alpha'|i,1;\alpha,a\rangle\\
& &\langle i',m',\alpha'|\sigma_j^k\tau_j^a|i,m,\alpha\rangle
= \sqrt{\frac{2i+1}{2i'+1}}
H(i',i)\langle i',m'|i,1;m,k\rangle\langle i',\alpha'|i,1;\alpha,a\rangle\,.
\eea
The corresponding reduced matrix elements can be computed with the results
\bea
& &G(i) = \frac{2}{N_c-1}\sqrt{i(i+1)}\\
& &H(i,i) = \frac{N_c+1}{N_c-1}\\
& &H(i,i-1) = \frac{2}{N_c-1}
\sqrt{\left(\frac{N_c-1}{2}+i+1\right)\left(\frac{N_c-1}{2}-i+1\right)}\\
& &H(i',i) = 1 + \frac{2}{N_c} + {\cal O}(1/N_c^2)\,.
\eea

We note from these results that only the unit operator 1 and 
$\sigma_j^k \tau_j^a$ give leading contributions to
(\ref{dirproduct}) in the large-$N_c$ limit. Inserting the
individual expressions for the matrix elements into (\ref{dirproduct})
we obtain
\bea\label{dirprod1}
& & _{j'}\langle S'I'i,m'\alpha'|SIi,m\alpha\rangle_j =
\delta_{SS'}\delta_{mm'}\delta_{II'}\delta_{\alpha\alpha'}\\
& &\times\frac14\left[ 1 + 
\sqrt{6(2i+1)} F(i)(-)^{i+\frac12+S}
\left\{ \begin{array}{ccc}
\frac12 & i & S \\
i & \frac12 & 1 \end{array}\right\} + 
\sqrt{6(2i+1)} G(i)(-)^{i+\frac12+I}
\left\{ \begin{array}{ccc}
\frac12 & i & I \\
i & \frac12 & 1 \end{array}\right\}\right.\nonumber\\
& &\qquad\left. +
6(-)^{2i+1+S+I}(2i+1)H(i,i)
\left\{ \begin{array}{ccc}
S & i & \frac12 \\
1 & \frac12 & i \end{array}\right\}
\left\{ \begin{array}{ccc}
I & i & \frac12 \\
1 & \frac12 & i \end{array}\right\}\right]\,.\nonumber
\eea
The product of two 6$j$-symbols can be transformed with the
help of the identity (\ref{C.35e}) into the form
\bea
\left\{ \begin{array}{ccc}
S & i & \frac12 \\
1 & \frac12 & i \end{array}\right\}
\left\{ \begin{array}{ccc}
I & i & \frac12 \\
1 & \frac12 & i \end{array}\right\} =
\frac12 (-)^{S+I+1+2i}
\left[
\left\{ \begin{array}{ccc}
S & I & 0 \\
\frac12 & \frac12 & i \end{array}\right\}^2 -
\left\{ \begin{array}{ccc}
S & I & 1 \\
\frac12 & \frac12 & i \end{array}\right\}^2\right]\,.
\eea
Furthermore, the second 6$j$-symbol on the r.h.s. can be eliminated
by using the relation
\bea\label{4.65}
\sum_{x=0,1}(2x+1)
\left\{ \begin{array}{ccc}
I & S & x \\
\frac12 & \frac12 & i \end{array}\right\}^2 = \frac{1}{2i+1}\,.
\eea
We obtain finally for the scalar product of tensor product states
(\ref{dirprod1}) the simple result
\bea\label{I2}
& &_{j'}\langle S'I',m',\alpha'|SI,m,\alpha\rangle_j =
\delta_{SS'}\delta_{mm'}\delta_{II'}\delta_{\alpha\alpha'}\\
& &\times\left\{ (2i+1)\frac{N_c+1}{N_c-1}
\left\{ \begin{array}{ccc}
S & I & 0 \\
\frac12 & \frac12 & i \end{array}\right\}^2 -
\frac{1}{2(N_c-1)}\left[\frac52 + 2i(i+1) - S(S+1) - I(I+1)\right]
\right\} \,.\nonumber
\eea
This result only holds if the two external quarks are different 
$j\neq j'$. If they are identical, only the first term in (\ref{4.48})
contributes (without the factor 1/4). This gives
\bea\label{I1}
_{j}\langle S'I',m',\alpha'|SI,m,\alpha\rangle_j =
\delta_{SS'}\delta_{mm'}\delta_{II'}\delta_{\alpha\alpha'}\,.
\eea

We can use (\ref{I1}) and (\ref{I2}) to compute the norm of the states
$|SIL,m_S,m_L,\alpha\rangle$. With the help of the definition (\ref{SImixed}),
it can be written as
\bea
& &\langle S'I'1,m'_S,m'_L,\alpha'|SI1,m_S,m_L,\alpha\rangle =\\
& &\qquad = \frac{1}{N_c-1}\sum_{j,j'=2}^{N_c}\,
_{[j',1]}\langle S'I',m'_S,\alpha'|SI,m_S,\alpha\rangle_{[j,1]}\,
_{[j',1]}\langle 1,m'_L|1,m_L\rangle_{[j,1]} \nonumber\\
& & \qquad = \frac{N_c+2}{4} \delta_{m_L m'_L} {\cal I} \left(
_{j}\langle S'I',m',\alpha'|SI,m,\alpha\rangle_j -\,
_{j'}\langle S'I',m',\alpha'|SI,m,\alpha\rangle_j\right)\nonumber\,,
\eea
where we used (\ref{4.39}) and (\ref{4.43}). To bring this into
the final form (\ref{4.44}) we only need to insert the expressions 
(\ref{I1}) and (\ref{I2}) for the scalar products on the r.h.s. 
and simplify the resulting expression with the help of (\ref{4.65}).

\subsection{Matrix elements of $Z^{ka}$ on mixed symmetry states}

In this Section we will compute the matrix element (\ref{WigEck})
of $Z^{ka}$ taken between quark model states with mixed symmetry.
It will be shown that the ansatz for $Z(S'I',SI)$ introduced in 
Sect.III.A can in fact be obtained by an explicit calculation in the
quark model.

We parametrize the matrix element of $Z^{ka}$ between the quark
model states (\ref{SImixed}) as
\bea\label{4.69}
& &\langle S' I' 1;m'_S,m'_L,\alpha'|\sum_{n=1}^{N_c}\sigma_n^k\tau_n^a |
S I 1;m_S,m_L,\alpha\rangle =\\
& &\qquad \frac{1}{\sqrt{(2S'+1)(2I'+1)}}Z(S'I',SI)
\delta_{m_L m'_L} \langle S',m'_S|S,1;m_S,k\rangle
\langle I',\alpha'|I,1;\alpha,a\rangle\,.\nonumber
\eea
We obtain for the reduced matrix element the following result
\bea\label{4.70}
Z(S'I',SI) &=& \frac34N_c(N_c+2)\sqrt{(2i+1)(2i'+1)}
\sqrt{(2S+1)(2S'+1)(2I+1)(2I'+1)}\\
&\times& (-)^{i'-i+S'+I'+\psi(SIi)+\psi(S'I'i')} 
\left\{ \begin{array}{ccc}
1 & S' & S \\
1 & I & I' \end{array}\right\}
\left\{ \begin{array}{ccc}
i & S & \frac12 \\
1 & \frac12 & I \end{array}\right\}
\left\{ \begin{array}{ccc}
i' & S' & \frac12 \\
1 & \frac12 & I' \end{array}\right\}{\cal I}\,.\nonumber
\eea
This has to be divided with the square roots of the norms of the initial 
and final states (\ref{4.44}). To leading order in $N_c$ the result takes 
exactly the form
(\ref{Zansatz}) provided the phase $\psi(SIi)$ of the quark model states
(\ref{SImixed}) is chosen as
\bea
\psi(SIi) = i+I+\frac12\,.
\eea

The derivation of (\ref{4.70}) proceeds in close analogy to the
computation of the norm of the mixed symmetry states. First, we
express the matrix element (\ref{4.69}) of $Z^{ka}$ in terms of  
matrix elements on direct product states as
\bea\label{4.72}
& &\langle S' I' 1;m'_S,m'_L,\alpha'|\sum_{n=1}^{N_c}\sigma_n^k\tau_n^a |
S I 1;m_S,m_L,\alpha\rangle \\
& & = (-)^{\psi(SIi)+\psi(S'I'i')}\frac{1}{N_c-1}\sum_{j,j'=2}^{N_c}\,
_{[j',1]}\langle S'I',m'_S,\alpha'|\sum_{n=1}^{N_c}\sigma_n^j\tau_n^a|
SI,m_S,\alpha\rangle_{[j,1]}\,\,
_{[j',1]}\langle 1,m'_L|1,m_L\rangle_{[j,1]} \nonumber\\
& & = (-)^{\psi(SIi)+\psi(S'I'i')}\frac{N_c+2}{4} \delta_{m_L m'_L} 
{\cal I}\left(Z_1 - Z_2\right)\nonumber\,,
\eea
where we denoted the diagonal and nondiagonal matrix elements of $Z^{ia}$ on
direct product states by
\bea
Z_1 &=& _{j}\langle S'I',m',\alpha'|\sum_{n=1}^{N_c}\sigma_n^k\tau_n^a|
SI,m,\alpha\rangle_j\\
Z_2 &=& _{j'}\langle
S'I',m',\alpha'|\sum_{n=1}^{N_c}\sigma_n^k\tau_n^a|SI,m,\alpha\rangle_j\,.
\eea

The nondiagonal matrix element on direct product states $(j'\neq j)$ can
be transformed into a diagonal one with the help of the exchange operator
(\ref{4.47})
\bea
& &_{j'}\langle \frac12,m'_2,\alpha'_2|\otimes \langle i',m'_1,\alpha'_1|
\sum_{n=1}^{N_c}\sigma_n^k\tau_n^a
|i,m_1,\alpha_1\rangle \otimes |\frac12,m_2,\alpha_2\rangle_j\\
& &\qquad = _{j}\langle \frac12,m'_2,\alpha'_2|\otimes \langle i',m'_1,\alpha'_1|P_{jj'}
\sum_{n=1}^{N_c}\sigma_n^k\tau_n^a
|i,m_1,\alpha_1\rangle \otimes |\frac12,m_2,\alpha_2\rangle_j\nonumber
\eea
This expression can be computed by expanding the $P_{jj'}$ operator and 
inserting a complete set of intermediate states.
\bea\label{4.74}
& &_{j'}\langle \frac12,m'_2,\alpha'_2|\otimes \langle i',m'_1,\alpha'_1|
\sum_{n=1}^{N_c}\sigma_n^k\tau_n^a
|i,m_1,\alpha_1\rangle \otimes |\frac12,m_2,\alpha_2\rangle_j =\\
& & \frac14\, _{j}\langle\frac12,m'_2,\alpha'_2|\frac12,m_2,\alpha_2\rangle_j
\langle i',m'_1,\alpha'_1|\sum_{n=1}^{N_c}\sigma_n^k\tau_n^a|
i,m_1,\alpha_1\rangle \nonumber\\
&+& \frac14 \sum_l (-)^l\,
_{j}\langle\frac12,m'_2,\alpha'_2|\sigma_j^{-l}|\frac12,m_2,\alpha_2\rangle_j\nonumber\\
& &\qquad\qquad\times
\sum_{m''_1}\langle i',m'_1,\alpha'_1|\sigma_{j'}^l|i',m''_1,\alpha'_1\rangle
\langle i',m''_1,\alpha'_1|
\sum_{n=1}^{N_c}\sigma_n^k\tau_n^a
|i,m_1,\alpha_1\rangle\nonumber\\
&+&  \frac14 \sum_b (-)^b\,
_{j}\langle\frac12,m'_2,\alpha'_2|\tau_j^{-b}|\frac12,m_2,\alpha_2\rangle_j\nonumber\\
& &\qquad\qquad\times
\sum_{\alpha''_1}\langle i',m'_1,\alpha'_1|\tau_{j'}^b|i',m'_1,\alpha''_1\rangle
\langle i',m'_1,\alpha''_1|
\sum_{n=1}^{N_c}\sigma_n^k\tau_n^a
|i,m_1,\alpha_1\rangle\nonumber\\
&+& \frac14 \sum_{l,b} (-)^{l+b}\,
_{j}\langle\frac12,m'_2,\alpha'_2|\sigma_j^{-l}\tau_j^{-b}|\frac12,m_2,\alpha_2\rangle_j
\nonumber\\
& &\qquad\qquad\times
\sum_{i'',m''_1,\alpha''_1}\langle i',m'_1,\alpha'_1|\sigma_{j'}^l\tau_{j'}^b|
i'',m''_1,\alpha''_1\rangle
\langle i'',m''_1,\alpha''_1|\sum_{n=1}^{N_c}\sigma_n^k\tau_n^a
|i,m_1,\alpha_1\rangle\nonumber\,.
\eea
Only completely symmetric states of $N_c-1$ quarks contribute to the
sum over intermediate states since both the operator and the initial state
in the last matrix elements of each term are symmetric under permutations of
any quarks. 
One notes that keeping only the first and the last term in this relation
is sufficient to obtain the large-$N_c$ limit of this matrix 
element\footnote{The exact result for arbitrary $N_c$ is presented in the
Appendix A.}.
Furthermore, in the sum over quarks in $Z^{ka}$ one can omit the
term acting on the $j'^{th}$ quark, as this will only change the result
by an amount nonleading in $N_c$. This allows us to compute these
matrix elements by using the results of Sec.IV.A. Putting all pieces together
one obtains for the matrix element of $Z^{ka}$ between quark-model states 
with well-defined spin and isospin $(S,I)$ the following result
\bea\label{4.75}
& &_{j'}\langle S'I',m'_S,\alpha'|Z^{ka}|SI,m_S,\alpha\rangle_j =\\
& &\qquad N_c\sqrt{(2i+1)(2i'+1)}\sqrt{(2S+1)(2I+1)}
\langle S',m'_S|S,1;m_S,k\rangle \langle I',\alpha'|I,1;\alpha,a\rangle\nonumber\\
& &\qquad\times\left[ \frac14 (-)^{-2i'+I+S-1}
\left\{ \begin{array}{ccc}
1 & S & S' \\
\frac12 & i' & i \end{array}\right\}
\left\{ \begin{array}{ccc}
1 & I & I' \\
\frac12 & i' & i \end{array}\right\}\right.\nonumber\\
& &\left.\qquad + \frac32 (-)^{I+I'+S+S'}
\sum_{i''}(2i''+1)
\left\{ \begin{array}{ccc}
i'' & \frac12 & S' \\
\frac12 & i' & 1 \end{array}\right\}
\left\{ \begin{array}{ccc}
S & 1 & S' \\
i'' & \frac12 & i \end{array}\right\}
\left\{ \begin{array}{ccc}
i'' & \frac12 & I' \\
\frac12 & i' & 1 \end{array}\right\}
\left\{ \begin{array}{ccc}
I & 1 & I' \\
i'' & \frac12 & i \end{array}\right\}\right]\nonumber\,.
\eea
Each of the last two lines corresponds to the contributions of the
first and fourth terms in (\ref{4.74}), respectively.
They can be transformed into the following form by a repeated application
of (\ref{C.35e})
\bea\label{4.76}
& &\left\{ \begin{array}{ccc}
1 & S & S' \\
\frac12 & i' & i \end{array}\right\}
\left\{ \begin{array}{ccc}
1 & I & I' \\
\frac12 & i' & i \end{array}\right\} =
(-)^{i+i'+I+S+I'+S'}\\
& &\qquad\times\left[
\left\{ \begin{array}{ccc}
S' & S & 1 \\
I & I' & 0 \end{array}\right\}
\left\{ \begin{array}{ccc}
S & i & \frac12 \\
\frac12 & 0 & I \end{array}\right\}
\left\{ \begin{array}{ccc}
S' & \frac12 & i' \\
\frac12 & I' & 0 \end{array}\right\} - 3
\left\{ \begin{array}{ccc}
S' & S & 1 \\
I & I' & 1 \end{array}\right\}
\left\{ \begin{array}{ccc}
S & i & \frac12 \\
\frac12 & 1 & I \end{array}\right\}
\left\{ \begin{array}{ccc}
S' & \frac12 & i' \\
\frac12 & I' & 1 \end{array}\right\}\right]\nonumber\\
\label{4.77}
& &\sum_{i''}(2i''+1)\left\{ \begin{array}{ccc}
i'' & \frac12 & S' \\
\frac12 & i' & 1 \end{array}\right\}
\left\{ \begin{array}{ccc}
S & 1 & S' \\
i'' & \frac12 & i \end{array}\right\}
\left\{ \begin{array}{ccc}
i'' & \frac12 & I' \\
\frac12 & i' & 1 \end{array}\right\}
\left\{ \begin{array}{ccc}
I & 1 & I' \\
i'' & \frac12 & i \end{array}\right\} = (-)^{1+i'-i+S+I}\\
& &\qquad\times\frac12
\left[
\left\{ \begin{array}{ccc}
S' & S & 1 \\
I & I' & 0 \end{array}\right\}
\left\{ \begin{array}{ccc}
S & i & \frac12 \\
\frac12 & 0 & I \end{array}\right\}
\left\{ \begin{array}{ccc}
S' & \frac12 & i' \\
\frac12 & I' & 0 \end{array}\right\} +
\left\{ \begin{array}{ccc}
S' & S & 1 \\
I & I' & 1 \end{array}\right\}
\left\{ \begin{array}{ccc}
S & i & \frac12 \\
\frac12 & 1 & I \end{array}\right\}
\left\{ \begin{array}{ccc}
S' & \frac12 & i' \\
\frac12 & I' & 1 \end{array}\right\}\right]\nonumber\,.
\eea
Inserting these expressions into (\ref{4.75}) we obtain the
following result for the nondiagonal matrix element of $Z^{ka}$ 
between direct product states
\bea\label{4.78}
Z_2 &=& N_c\sqrt{(2i+1)(2i'+1)}\sqrt{(2S+1)(2I+1)}
\langle S',m'_S|S,1;m_S,k\rangle \langle I',\alpha'|I,1;\alpha,a\rangle\\
& &\qquad\times(-)^{1+i'-i+S'+I'}
\left\{ \begin{array}{ccc}
S' & S & 1 \\
I & I' & 0 \end{array}\right\}
\left\{ \begin{array}{ccc}
S & i & \frac12 \\
\frac12 & 0 & I \end{array}\right\}
\left\{ \begin{array}{ccc}
S' & \frac12 & i' \\
\frac12 & I' & 0 \end{array}\right\}\,.\nonumber
\eea
For the diagonal case, only the first term in (\ref{4.74}) survives (without the
factor of 1/4). Using (\ref{4.76}) we can write for this case
\bea\label{4.79}
Z_1 &=& (-)^{1-i'+i+S'+I'}\\
& &\qquad\times N_c\sqrt{(2i+1)(2i'+1)}\sqrt{(2S+1)(2I+1)}
\langle S',m'_S|S,1;m_S,k\rangle 
\langle I',\alpha'|I,1;\alpha,a\rangle\nonumber\\
& &\times
\left[
\left\{ \begin{array}{ccc}
S' & S & 1 \\
I & I' & 0 \end{array}\right\}
\left\{ \begin{array}{ccc}
S & i & \frac12 \\
\frac12 & 0 & I \end{array}\right\}
\left\{ \begin{array}{ccc}
S' & \frac12 & i' \\
\frac12 & I' & 0 \end{array}\right\} - 3
\left\{ \begin{array}{ccc}
S' & S & 1 \\
I & I' & 1 \end{array}\right\}
\left\{ \begin{array}{ccc}
S & i & \frac12 \\
\frac12 & 1 & I \end{array}\right\}
\left\{ \begin{array}{ccc}
S' & \frac12 & i' \\
\frac12 & I' & 1 \end{array}\right\}\right]\,.\nonumber
\eea

Inserting (\ref{4.78}) and (\ref{4.79}) into (\ref{4.72}) and using the 
definition of $Z(S'I',SI)$ (\ref{4.69}) gives the final result for the 
matrix element of $Z$ (\ref{4.70}).

We will present in the following an alternative method of calculating
the matrix element of a current between states with mixed symmetry.
Besides reproducing the result (\ref{4.70}), this method has the
advantage of simplifying very much the computation of transition
matrix elements between excited and ground state baryons, to be
discussed in the next Section. We start by writing the matrix element
of the current $Z^{ka}$ taken between two states (\ref{SImixed}) as
\bea\label{4.81}
\langle S'I'| Z^{ka}|SI\rangle &=& (-)^{\psi(SIi)+\psi'(S'I'i')}
\frac{1}{2(N_c-1)}\\
& &\times\left\{
\sum_{jj'=2}^{N_c}\, _{[j',1]}\langle S'I'|Z^{ka}|SI\rangle_{[j,1]} +
\sum_{j=2}^{N_c}\,
_{[j,1]}\langle S'I'|Z^{ka}|SI\rangle_{[j,1]} \right\}{\cal I}\,.\nonumber
\eea
We consider the two terms of this relation in turn. The first sum can be written as
\bea\label{4.82}
\sum_{jj'=2}^{N_c}\, _{[j',1]}\langle S'I'|Z^{ka}|SI\rangle_{[j,1]} =\,
 _1\langle S'I'|\sum_{j'=2}^{N_c}A[j',1]Z^{ka}\sum_{j=2}^{N_c}A[j,1]|SI\rangle_1\,,
\eea
where
\bea
A[j,1] = \frac{1}{\sqrt2}\left(1 - P_{j1}\right)
\eea
is the antisymmetrization operator for quarks $[j,1]$ and $P_{j1}$ has been
defined in (\ref{4.47}). The spin states defined in (\ref{4.39}) can
be written in terms of it as $|SI\rangle_{[i,1]}=-A[i,1]|SI\rangle_1$.

An important relation we will use extensively in the following expresses 
the result of symmetrizing a direct product state $|SI\rangle_1$ under
a permutation of any two quarks
\bea\label{4.86}
\Pi |SI,m\alpha\rangle_1 = \left(1 + P_{12} + \cdots + P_{1N_c}\right)
|SI,m\alpha\rangle_1
= \delta_{SI} B(Ii) |I,m,\alpha\rangle
\eea
with $|I,m,\alpha\rangle$ the completely symmetric state constructed in
Section IV.A. The normalization constant $B(Ii)$ can be computed by
taking the norm of the both sides of this relation. We obtain
\bea
\delta_{SI}B^2(Ii) &=& \sum_{nn'=1}^{N_c}\,
 _{n'}\langle SI,m\alpha|SI,m\alpha\rangle_n \\
&=& N_c + N_c(N_c-1) 
 _{n'}\langle SI,m\alpha|SI,m\alpha\rangle_n\,.\qquad (n\neq n')\nonumber
\eea
The nondiagonal matrix element appearing on the r.h.s. has been calculated
previously and is given by (\ref{I2}). We obtain finally
\bea\label{Asquared}
B^2(Ii) &=& N_c(N_c+1)(2i+1)
\left\{ \begin{array}{ccc}
\frac12 & \frac12 & 0 \\
I & S & i \end{array}\right\}^2\\
&-& \frac{N_c}{2}
\left[\frac12 + 2i(i+1) - 2I(I+1)\right]\,.\nonumber
\eea
It will be shown below that the phase of $B(Ii)$ can be chosen such that the
leading term in $N_c$ is positive.

The sums over the antisymmetrization
operators in (\ref{4.82}) can be written in terms of the complete
symmetrization operator $\Pi$ as
\bea\label{4.89}
\sum_{j=2}^{N_c} A[j,1] = \frac{1}{\sqrt2}(N_c - \Pi)\,.
\eea
We will need also the following matrix element
\bea\label{4.88}
N_c\, _1\langle S'I'|\Pi Z^{ka}|SI\rangle_1 &=&\,
_1\langle S'I'|\Pi Z^{ka}[1+P_{12}^2+\cdots +P_{1N_c}^2]|SI\rangle_1\\
&=& _1\langle S'I'|\Pi Z^{ka}\Pi|SI\rangle_1\,.\nonumber
\eea
We used in the first line the property of the permutation operator $P_{ij}^2=1$.
The second equality is obtained by writing $\Pi Z^{ka} P_{1j}^2 = 
\Pi P_{1j}Z^{ka} P_{1j}$. 
When acting to the
left on $_1\langle S'I'|$, this gives 
\bea
_1\langle S'I'|\Pi P_{1j} =\,
_1\langle S'I'|\Pi
\eea
since $_1\langle S'I'|\Pi$ is completely symmetric under any permutation of 
the $N_c$ quarks. This completes the proof of (\ref{4.88}).

The relations (\ref{4.89}) and (\ref{4.88}) allow us to express the
sum of matrix elements (\ref{4.82}) as
\bea\label{4.90}
\sum_{jj'=2}^{N_c}\, _{[j',1]}\langle S'I'|Z^{ka}|SI\rangle_{[j,1]} &=&\,
\frac12 N_c^2\, _1\langle S'I'|Z^{ka}|SI\rangle_1 -
\frac12 \, _1\langle S'I'|\Pi Z^{ka}\Pi |SI\rangle_1\\
& &\hspace{-2cm}= \frac12 N_c^2\, _1\langle S'I'|Z^{ka}|SI\rangle_1 -
\frac{1}{2}\delta_{SI}\delta_{S'I'}B(Ii) B(I'i')
\langle S'=I'|Z^{ka}|S=I\rangle\,.\nonumber
\eea

The second term in (\ref{4.81}) can be computed in an analogous way.
We note for this the following useful properties of the 
antisymmetrization operator $A[j,1]$.

\begin{enumerate}
\item $A[j,1]$ commutes with $Z^{ka}$
\bea
[A[j,1]\,, Z^{ka}] = 0\,.
\eea
This follows from the fact that $Z^{ka}$ is completely symmetric under
a permutation of two quarks and commutes therefore with the $P$ operator
(\ref{4.47}).

\item The square of $A[j,1]$ is given by
\bea
A[j,1]^2 =  \sqrt2 A[j,1]\,.
\eea
\end{enumerate}

With the help of these relations and (\ref{4.89}) we can write
\bea
\sum_{j=2}^{N_c}\, _{[j,1]}\langle S'I'|Z^{ka}|SI\rangle_{[j,1]} &=&
\sum_{j=2}^{N_c}\, _1\langle S'I'|A[j,1]Z^{ka} A[j,1]|SI\rangle_1\\
&=&\, _1\langle S'I'|Z^{ka} \sum_{j=2}^{N_c} A[j,1]^2|SI\rangle_1\nonumber\\
&=&\, _1\langle S'I'|Z^{ka} \left(N_c-\Pi\right) |SI\rangle_1\,.\nonumber
\eea
Using a relation similar to (\ref{4.88}) for the second term, this equation
can be put into the form
\bea\label{4.94}
& &\sum_{j=2}^{N_c}\, _{[j,1]}\langle S'I'|Z^{ka}|SI\rangle_{[j,1]} =\\
& &\qquad N_c\, _1\langle S'I'|Z^{ka}|SI\rangle_1 -
\frac{1}{N_c}\delta_{SI}\delta_{S'I'}B(Ii) B(I'i')
\langle S'=I'|Z^{ka}|S=I\rangle\,.\nonumber
\eea

Combining the two results (\ref{4.90}) and (\ref{4.94}) gives the following
general expression for the matrix element of the current $Z^{ka}$ taken
between two mixed symmetry states
\bea\label{4.97}
& &\langle S'I'|Z^{ka}|SI\rangle = (-)^{\psi(SIi)+\psi'(S'I'i')}\\
& &\qquad\times\frac{N_c(N_c+2)}{4(N_c-1)}{\cal I}
\left\{ \,
_1\langle S'I'|Z^{ka}|SI\rangle_1 -
\frac{1}{N_c^2}\delta_{SI}\delta_{S'I'}B(Ii) B(I'i')
\langle S'=I'|Z^{ka}|S=I\rangle\right\}\nonumber\,.
\eea

We are now in a position to compute the phase of the normalization constant
$B(Ii)$. This can be done by comparing the two expressions (\ref{4.72}) and
(\ref{4.97}) for the matrix element $\langle Z^{ia}\rangle$. We obtain in this way
the following exact relation
\bea
\frac{1}{N_c^2}\delta_{SI}\delta_{S'I'}B(Ii) B(I'i')
\langle S'=I'|Z^{ka}|S=I\rangle = Z_2 + \frac{1}{N_c}(Z_1 - Z_2)\,.
\eea
Using (\ref{4.78}) for $Z_2$ one finds to leading order in $N_c$
\bea\label{4.99}
& &(N_c+2)\sqrt{\frac{2I+1}{2I'+1}}B(Ii) B(I'i') =
N_c^3\sqrt{(2i+1)(2i'+1)}\sqrt{(2S+1)(2I+1)}\\
& &\quad\times(-)^{1+i'-i+S'+I'}
\left\{ \begin{array}{ccc}
S' & S & 1 \\
I & I' & 0 \end{array}\right\}
\left\{ \begin{array}{ccc}
S & i & \frac12 \\
\frac12 & 0 & I \end{array}\right\}
\left\{ \begin{array}{ccc}
S' & \frac12 & i' \\
\frac12 & I' & 0 \end{array}\right\} = 
N_c^3 \frac{\sqrt{(2i+1)(2i'+1)}}{2(2I'+1)}\delta_{IS}
\delta_{I'S'}\,.\nonumber
\eea
From this follows that $B(Ii)$ can be chosen to be positive for all values
of its arguments.

It is easy to see now with the help of (\ref{4.79}) and (\ref{4.99}) that 
(\ref{4.97}) gives, to leading order in $N_c$,
the same result for $\langle S'I'|Z^{ka}|SI\rangle$ as (\ref{4.70}).

\subsection{Matrix elements of $Y^{a}$ and $Q^{ka}$ in the quark model}

As already mentioned in Section III, the matrix elements of the operators
$Y^{a}$ and $Q^{ka}$ in the quark model can be reduced to
those of the operator $\sum_{n=1}^{N_c}r^i_n \sigma^j_n \tau^a_n$. Here $r_n,
\sigma_n,\tau_n$ are vector operators acting on the orbital, spin and 
isospin degrees of freedom of the $n^{th}$ quark respectively.
In this Section we prove that the quark model reproduces, in the large-$N_c$
limit, the results (\ref{Yansatz1}) and (\ref{Yansatz2})
expected from the model-independent treatment of Sections
III.B and III.C.

We consider first the transitions from an excited baryon state transforming
under the symmetric representation of SU(4) to another symmetric
baryon state. For generality we leave the orbital momenta of the initial
and final states completely arbitrary $L,L'$.
The dependence on the spin-isospin quantum numbers is contained in the 
reduced matrix element ${\cal T}(I',I)$ defined by
\bea\label{4.100}
\langle I'L',m'_S m'_L\alpha'|\sum_{n=1}^{N_c}r^i_n \sigma_n^j\tau_n^a|
IL,m_S m_L\alpha\rangle &=&
\frac{1}{(2I'+1)\sqrt{2L'+1}}{\cal T}(I',I){\cal I}(L',L)\\
& &\hspace{-3cm}\times\langle I',m'_S|I,1;m_S,j\rangle 
\langle I',\alpha'|I,1;\alpha,a\rangle
\langle L',m'_L|L,1;m_L,i\rangle\,.\nonumber
\eea
We will restrict our considerations to baryon states for which all the
orbital angular momentum is carried by one quark at a time. This is strictly true
only for the lowest orbital excitations. In Hartree
language the spatial part of the wavefunction for these states has the
form
\bea\label{4.101}
|L,m_L\rangle = \frac{1}{\sqrt{N_c}}\sum_{p=1}^{N_c}
\phi( r_1\,)\phi( r_2\,)\cdots \psi_{L,m_L}(\vec r_p\,)\cdots
\phi( r_{N_c}\,)
\eea
where $\phi( r\,)$ is a s-wave one-particle wavefunction and
$\psi_{L,m_L}(\vec r\,)$ carries angular momentum $(L,m_L)$.
The spatial part of the matrix element (\ref{4.100}) can be written
in terms of the matrix element
\bea
\langle L',m'_L|r_n^i|L,m_L\rangle = \frac{1}{N_c}\frac{1}{\sqrt{2L'+1}}
{\cal I}(L',L) \langle L',m'_L|L,1;m_L,i\rangle\,,
\eea
with ${\cal I}(L',L)$ an overlap integral of order $N_c^0$. The case
$L'=0$ of a s-wave baryon in the final state is special, as the scaling
law with $N_c$ is different
\bea
\langle 0|r_n^i|L,m_L\rangle = \frac{1}{\sqrt{N_c}}{\cal I}
\delta_{L1} \langle 0|L,1;m_L,i\rangle\,,
\eea
For both these cases the matrix element of $r_n^i$ is independent of $n$
due to the symmetry of the wavefunction under any permutation of two quarks.
Therefore the spin-isospin part of the matrix element (\ref{4.100})
decouples completely from the spatial part and is given exactly by
the formula (\ref{common}) for the ground state baryons.
We obtain in this way for the reduced matrix element
${\cal T}(I',I)$
\bea
{\cal T}(I',I) = 
\left\{ \begin{array}{c}
\frac{N_c+2}{N_c}\sqrt{(2I+1)(2I'+1)}\,,\qquad L'\neq 0\\
\frac{N_c+2}{\sqrt{N_c}}\sqrt{(2I+1)(2I'+1)}\,,\qquad L'= 0\,.\end{array}\right.
\eea
which can be seen to coincide, up to an unimportant phase and numerical 
factor, with the result (\ref{Yansatz1}) anticipated in Section III.

We consider next the case of an excited baryon transforming under the
mixed symmetry representation of SU(4) in the initial state.
The final state corresponds to the completely symmetric representation.
We write the matrix element relevant for this case as
\bea\label{4.105}
\langle I'L',m'm'_L\alpha'|\sum_{n=1}^{N_c}r^i_n \sigma_n^j\tau_n^a|
SIL,m_Sm_L\alpha\rangle &=& \frac{1}{(2I'+1)\sqrt{2L'+1}}{\cal T}(I',SI){\cal I}(L',L)\\
& &\hspace{-3cm}\times\langle I',m'|S,1;m_S,j\rangle 
\langle I',\alpha'|I,1;\alpha,a\rangle
\langle L',m'_L|L,1;m_L,i\rangle\,.\nonumber
\eea
The scaling law with $N_c$ of the spatial part of this matrix element 
is again different, depending on whether
$L'\neq 0$ or $L'=0$. Both cases can be considered together by writing
it as
\bea
& &\langle L',m'_L|r^i_n |SIL,m_Sm_L\alpha\rangle =
\frac{N_c^\kappa (-)^{\psi(SIi)}}{\sqrt{2N_c(N_c-1)}}\sum_{k,k'=1}^{N_c}
\left(\delta_{k'n}\delta_{nk} - \delta_{k'n}\delta_{n1}\right)\\
& &\qquad\times\frac{1}{\sqrt{2L'+1}}{\cal I}(L',L)
\langle L',m'_L|L1,m_L,i\rangle |SI,m_S\alpha\rangle_{[k,1]}
\nonumber\\
& & = \frac{N_c^\kappa (-)^{\psi(SIi)}}{\sqrt{2N_c(N_c-1)}} 
\frac{1}{\sqrt{2L'+1}}{\cal I}(L',L)
\langle L',m'_L|L1,m_L,i\rangle
\left\{ |SI,m_S\alpha\rangle_{[n,1]} -
\delta_{n1}\sum_{k=2}^{N_c}|SI,m_S\alpha\rangle_{[k,1]}\right\}\,.\nonumber
\eea
Here $\kappa=1/2$ for $L'=0$ and $\kappa=0$ for $L'\neq 0$.
Adding the spin-isospin part of the operator and summing over the
$N_c$ quarks in the baryon gives
\bea\label{4.107}
& &\langle I'L',m'm'_L\alpha'|\sum_{n=1}^{N_c}r^i_n \sigma_n^j\tau_n^a|
SIL,m_Sm_L\alpha\rangle =\\
& &\qquad
\frac{N_c^\kappa (-)^{\psi(SIi)}}{\sqrt{2N_c(N_c-1)}} 
\frac{1}{\sqrt{2L'+1}}{\cal I}(L',L)
\langle L',m'_L|L1,m_L,i\rangle\nonumber\\
& &\qquad\times
\left\{ \sum_{n=1}^{N_c} \langle I',m'\alpha'|\sigma_n^j\tau_n^a
|SI,m_S\alpha\rangle_{[n,1]} -
\langle I',m'\alpha'|\sigma_1^j\tau_1^a
\sum_{k=2}^{N_c}|SI,m_S\alpha\rangle_{[k,1]}\right\}\,.\nonumber
\eea
The first term in the braces can be written as
\bea\label{4.108}
& &\sum_{n=1}^{N_c} \langle I',m'\alpha'|\sigma_n^j\tau_n^a
|SI,m_S\alpha\rangle_{[n,1]} =\\
& &\qquad \frac{1}{\sqrt2}
\left( \sum_{n=1}^{N_c} \langle I',m'\alpha'|\sigma_n^j\tau_n^a
|SI,m_S\alpha\rangle_n - 
\langle I',m'\alpha'|\sum_{n=1}^{N_c} \sigma_n^j\tau_n^a
|SI,m_S\alpha\rangle_1\right) =\nonumber\\
& &\qquad \frac{1}{\sqrt2}
\left( N_c \langle I',m'\alpha'|\sigma_1^j\tau_1^a|SI,m_S\alpha\rangle_1
- \frac{1}{N_c}\delta_{SI}B(Ii)
\langle I',m'\alpha'|\sum_{n=1}^{N_c} \sigma_n^j\tau_n^a
|I,m_S\alpha\rangle\right)\,.\nonumber
\eea
In the second line we used the identity
\bea\label{4.109}
& &N_c \langle I',m'\alpha'|\sum_{n=1}^{N_c} \sigma_n^j\tau_n^a
|SI,m_S\alpha\rangle_1 =
\langle I',m'\alpha'|\sum_{n=1}^{N_c} \sigma_n^j\tau_n^a
(1 + P_{12}^2 + \cdots + P_{1N_c}^2)|SI,m_S\alpha\rangle_1 \\
& &\quad = \sum_{k=1}^{N_c}\langle I',m'\alpha'|P_{1k}\sum_{n=1}^{N_c} \sigma_n^j\tau_n^a
P_{1k}|SI,m_S\alpha\rangle_1 = \langle I',m'\alpha'|
\sum_{n=1}^{N_c} \sigma_n^j\tau_n^a\Pi |SI,m_S\alpha\rangle_1\nonumber
\eea
followed by the application of the relation (\ref{4.86}). In (\ref{4.109})
we have defined $P_{11}=1$.

The second term in (\ref{4.107}) can be put into the following form through an
application of (\ref{4.89}) and (\ref{4.86})
\bea\label{4.110}
& &\langle I',m'\alpha'|\sigma_1^j\tau_1^a
\sum_{k=2}^{N_c}|SI,m_S\alpha\rangle_{[k,1]} =\\
& &\qquad
\frac{1}{\sqrt2}\left(\delta_{SI}B(Ii) 
\langle I',m'\alpha'|\sigma_1^j\tau_1^a|I,m_S,\alpha\rangle -
 N_c \langle I',m'\alpha'|\sigma_1^j\tau_1^a
|SI,m_S\alpha\rangle_1 \right)\,.\nonumber
\eea

Combining (\ref{4.108}) and (\ref{4.110}) together we obtain the 
following general formula for the matrix element (\ref{4.105})
\bea\label{4.111}
& &\langle I'L',m'm'_L\alpha'|\sum_{n=1}^{N_c}r^i_n \sigma_n^j\tau_n^a|
SIL,m_Sm_L\alpha\rangle =\\
& &\qquad
\frac{N_c^\kappa (-)^{\psi(SIi)}}
{\sqrt{N_c(N_c-1)}} \frac{1}{\sqrt{2L'+1}}{\cal I}(L',L)
\langle L',m'_L|L1,m_L,i\rangle\nonumber\\
& &\qquad\times
\left\{ N_c \langle I',m'\alpha'|\sigma_1^j\tau_1^a
|SI,m_S\alpha\rangle_1 - \frac{1}{N_c}\delta_{SI} B(Ii)
\langle I',m'\alpha'|\sum_{n=1}^{N_c}\sigma_n^j\tau_n^a
|I,m_S,\alpha\rangle\right\}\,.\nonumber
\eea
The second term in (\ref{4.111}) is already known
from our analysis of the symmetric states in Section IV.A. The first
matrix element is new. In the following we present the details of
its calculation.

Using (\ref{4.86}) one can write
\bea\label{4.112}
\langle I',m'\alpha'|\sigma_1^j\tau_1^a|SI,m\alpha\rangle_1 =
\frac{1}{B(I'i')}\,
_1\langle I'I',m'\alpha'|\sum_{k=1}^{N_c}P_{1k}\sigma_1^j \tau_1^a
|SI,m\alpha\rangle_1\,.
\eea
A typical term of the sum over $k$  has the form
\bea\label{4.113}
& &_1\langle I'I',m'\alpha'|P_{1k}\sigma_1^j \tau_1^a|SI,m\alpha\rangle_1 =\\
& &\quad \frac14\,
_1\langle I'I',m'\alpha'|\sigma_1^j\tau_1^a|SI,m\alpha\rangle_1 +
\frac14\,
_1\langle I'I',m'\alpha'|(\vec\sigma_1\cdot\vec\sigma_k\,)
(\vec\tau_1\cdot\vec\tau_k\,)\sigma_1^j \tau_1^a|SI,m\alpha\rangle_1 +
{\cal O}(1/N_c)
\nonumber
\eea
where we used the definition of the $P$ operator (\ref{4.47}) and the
fact that $F(i)$ and $G(i)$ computed in Section IV.B are nonleading
in $1/N_c$. The first matrix element is easily calculated with the result
\bea\label{4.114}
& &_1\langle I'I',m'\alpha'|\sigma_1^j\tau_1^a|SI,m\alpha\rangle_1 =
\langle I',m'|S,1;m,j\rangle \langle I',\alpha'|I,1;\alpha,a\rangle\\
& &\qquad\times 6\delta_{ii'}\sqrt{(2S+1)(2I+1)}
\left\{ \begin{array}{ccc}
1 & S & I' \\
i & \frac12 & \frac12 \end{array}\right\}
\left\{ \begin{array}{ccc}
1 & I & I' \\
i & \frac12 & \frac12 \end{array}\right\}\,.\nonumber
\eea
The second matrix element can be reduced to quantities already known
by simplifying the products of two spin and isospin Pauli matrices
with the help of the identity
\bea
\sigma_a \sigma_b = -\sqrt3 \langle 0|11;ab\rangle\, 1 -
\sqrt2 \sum_c \langle 1c|11;ab\rangle \sigma_c\,.
\eea
We obtain in this way
\bea\label{4.116}
& &_1\langle I'I',m'\alpha'|(\vec\sigma_1\cdot\vec\sigma_k\,)
(\vec\tau_1\cdot\vec\tau_k\,)\sigma_1^j \tau_1^a|SI,m\alpha\rangle_1 =\\
& &\quad\langle I',m'|S,1;m,j\rangle \langle I',\alpha'|I,1;\alpha,a\rangle
 \sqrt{(2i+1)(2i'+1)(2S+1)(2I+1)}\nonumber\\
& &\qquad\times\left[ (-)^{-i+\frac12+S}
\left\{ \begin{array}{ccc}
S & 1 & I' \\
i' & \frac12 & i \end{array}\right\}
+ 6(-)^{-S-I'}
\left\{ \begin{array}{ccc}
\frac12 & 1 & \frac12 \\
i & 1 & i' \\
S & 1 & I' \end{array}\right\}\right]\nonumber\\
& &\qquad\times\left[ (-)^{-i+\frac12+I}
\left\{ \begin{array}{ccc}
I & 1 & I' \\
i' & \frac12 & i \end{array}\right\}
+ 6(-)^{-I-I'}
\left\{ \begin{array}{ccc}
\frac12 & 1 & \frac12 \\
i & 1 & i' \\
I & 1 & I' \end{array}\right\}\right]\nonumber\\
& &= 36\langle I',m'|S,1;m,j\rangle \langle I',\alpha'|I,1;\alpha,a\rangle
 \sqrt{(2i+1)(2i'+1)(2S+1)(2I+1)}\nonumber\\
& &\times
\left\{ \begin{array}{ccc}
i & i' & 1 \\
\frac12 & \frac12 & I' \end{array}\right\}^2
\left\{ \begin{array}{ccc}
S & I' & 1 \\
\frac12 & \frac12 & i \end{array}\right\}
\left\{ \begin{array}{ccc}
I & I' & 1 \\
\frac12 & \frac12 & i \end{array}\right\}\,.\nonumber
\eea
We used in the second equality the identity
\bea
\left\{ \begin{array}{ccc}
\frac12 & \frac12 & 1 \\
i & i' & 1 \\
S & I' & 1 \end{array}\right\} =
\left\{ \begin{array}{ccc}
i & i' & 1 \\
\frac12 & \frac12 & I' \end{array}\right\} 
\left\{ \begin{array}{ccc}
S & I' & 1 \\
\frac12 & \frac12 & i \end{array}\right\}  +
\frac16 (-)^{i'+S+\frac32}
\left\{ \begin{array}{ccc}
i & i' & 1 \\
I' & S & \frac12 \end{array}\right\}
\eea
and a similar one with $S\to I$,
which can be obtained from (\ref{Edmonds(6.4.8)}) by
taking $\lambda=1/2$ and the $j$'s are the same as in
the 9$j$ symbol on the l.h.s.

Combining (\ref{4.114}) and (\ref{4.116}) we find the following
result for the matrix element (\ref{4.113})
\bea\label{4.118}
& & _1\langle I'I',m'\alpha'|P_{1k}\sigma_1^j \tau_1^a|SI,m\alpha\rangle_1 =\\
& &\qquad\sqrt{(2i+1)(2i'+1)(2I+1)(2S+1)}
\langle I',m'|S,1;m,j\rangle \langle I',\alpha'|I,1;\alpha,a\rangle\nonumber\\
& &\qquad\times 3\frac{1}{2I'+1}
\left\{ \begin{array}{ccc}
S & I' & 1 \\
\frac12 & \frac12 & i \end{array}\right\}
\left\{ \begin{array}{ccc}
I & I' & 1 \\
\frac12 & \frac12 & i \end{array}\right\}\,,\nonumber
\eea
where we have rewritten the $\delta_{ii'}$ symbol in (\ref{4.114}) as
\bea
\delta_{ii'} = 2\sqrt{(2i+1)(2i'+1)}
\left\{ \begin{array}{ccc}
i & i' & 0 \\
\frac12 & \frac12 & I' \end{array}\right\}^2
\eea
and added the two terms with the help of the identity
\bea
\sum_{x=0,1}(2x+1)
\left\{ \begin{array}{ccc}
i & i' & x \\
\frac12 & \frac12 & I' \end{array}\right\}^2 =
\frac{1}{2I'+1}\,.
\eea

Next we use (\ref{C.35e}) to write the product of 2 6$j$-symbols in
(\ref{4.118}) as
\bea\label{prod6j}
& &\left\{ \begin{array}{ccc}
\frac12 & I' & i \\
S & \frac12 & 1 \end{array}\right\}
\left\{ \begin{array}{ccc}
\frac12 & I' & i \\
I & \frac12 & 1 \end{array}\right\} =
\frac{\delta_{SI}}{6\sqrt{(2S+1)(2I+1)}} \\
& &\qquad + (-)^{-\frac12+S+I+I'+i}
\left\{ \begin{array}{ccc}
1 & 1 & 1 \\
S & I & I' \end{array}\right\}
\left\{ \begin{array}{ccc}
S & I & 1 \\
\frac12 & \frac12 & i \end{array}\right\}\nonumber\,.
\eea
The sum over $k$ in (\ref{4.112}) is dominated by the terms with $k\neq 1$, 
of which there are $N_c-1$. Neglecting the contribution of the $k=1$ term,
we obtain for the matrix element (\ref{4.112}) to leading order in $N_c$
\bea
& &\langle I',m'\alpha'|\sigma_1^j \tau_1^a |SI,m\alpha\rangle_1 =
3\sqrt{2(2i+1)} \sqrt{\frac{(2S+1)(2I+1)}{2I'+1}}\\
& &\qquad\times\langle I',m'|S,1;m,j\rangle 
\langle I',\alpha'|I,1;\alpha,a\rangle\nonumber\\
& &\qquad \times \left\{
\frac{\delta_{SI}}{6\sqrt{(2S+1)(2I+1)}} + (-)^{-\frac12+S+I+I'+i}
\left\{ \begin{array}{ccc}
1 & 1 & 1 \\
S & I & I' \end{array}\right\}
\left\{ \begin{array}{ccc}
S & I & 1 \\
\frac12 & \frac12 & i \end{array}\right\}\right\}\nonumber\,.
\eea
When inserted into (\ref{4.111}), the term proportional to $\delta_{SI}$
will be cancelled exactly by the second term in (\ref{4.111}).
As a result, we obtain for the reduced matrix element ${\cal T}(I',SI)$ 
taken between the unnormalized quark model states
\bea
{\cal T}(I',SI) &=& N_c^\kappa 3\sqrt{2(2i+1)}
\sqrt{(2S+1)(2I+1)(2I'+1)}(-)^{-\frac12+S+I+I'+i}
(-)^{\psi(SIi)}\\
& &\qquad\times\left\{ \begin{array}{ccc}
1 & 1 & 1 \\
S & I & I' \end{array}\right\}
\left\{ \begin{array}{ccc}
S & I & 1 \\
\frac12 & \frac12 & i \end{array}\right\}\,.\nonumber
\eea
The physical value of this reduced matrix element is obtained
by dividing with the square root of the norm of the initial state
(\ref{4.44}). This gives
\bea\label{calTfinal}
[{\cal T}(I',SI)]_{norm} &=& N_c^{\kappa-1/2}2\sqrt6
\sqrt{(2S+1)(2I+1)(2I'+1)}(-)^{-\frac12+S+I+I'+i}(-)^{\phi(SIi)+\psi(SIi)}\\
& &\qquad\times\left\{ \begin{array}{ccc}
1 & 1 & 1 \\
S & I & I' \end{array}\right\}\,.\nonumber
\eea
Finally we insert here $\phi(SIi)=1+I+S$ the phase of the 6$j$-symbol appearing
in the formula for the norm (\ref{4.44}) and $\psi(SIi)=i+I+\frac12$
the phase of the mixed symmetry state, which gives for the total phase 
$(-)^{-I+I'}$. Thus (\ref{calTfinal}) can be seen to coincide exactly, up to
a numerical factor, with the expression (\ref{Yansatz2}) expected from
the model-independent treatment of Section III.B.

An important by-product of this calculation is the large-$N_c$ scaling law
of the transition matrix elements into final states in an s-wave.
We obtain that 
the matrix elements of $Y^a$ and $Q^{ka}$ from an initial state with mixed 
symmetry scales 
like $N_c^0$. On the other hand, the same matrix elements with a symmetric
excited state in the initial state scale as $N_c^{\frac12}$.
This dependence of the scaling law on the symmetry type of the
excited state is a new feature, unnoticed previously.
As discussed in Section III, for both cases the scaling law for the total
scattering amplitude is sufficiently restrictive to allow the derivation of
useful consistency conditions.
In spite of their different $N_c$ scaling, the solutions for these matrix elements
have the same dependence on spin and flavor quantum numbers.

The quark model computations of Appendix A illustrate another important asymmetry
between the symmetric and the mixed symmetry states. The $1/N_c$ corrections
to the large-$N_c$ results for coupling ratios vanish for the former \cite{DM} 
but not for the latter. Such dependence on the symmetry properties of these 
states raises the question of how to distinguish states with different permutational 
symmetry beyond the framework of the quark model. 

The exact large-$N_c$ scaling law for matrix elements of $Y$ and $Q$ following 
from the calculations of this Section is strictly correct only
for the case of the baryons made of heavy quarks, for which
the constituent quark picture is known to be exactly valid. 
Our results following from the consistency conditions discussed in
Sect.III rest on the assumption that no important changes occur
as the quarks become light and that the modified scaling law corresponding to
this situation still allows the derivation of consistency conditions.
While this assumption seems plausible and is similar to smoothness arguments
commonly used in other large-$N_c$ studies \cite{Wi,CarGeOso}, it is important 
to keep it in mind as one of the vulnerable points of an analysis of this type.

\section{Quark model matrix elements for arbitrary $N_c$}

In this Section we compute the full expressions for
reduced matrix elements in the quark model with arbitrary number of
colors $N_c$. The results presented in the preceding Section are obtained
from these expressions by keeping only the leading terms in $N_c$.
We take advantage of our ability to derive exact relations for the
quark model matrix elements to study the $1/N_c$ corrections to the
large-$N_c$ predictions.
By examining a few simple particular cases we conclude that the
results obtained in Section III in the large-$N_c$ limit will
receive, in general, $1/N_c$ corrections.

\subsection{$Z(S'I',SI)$}

We begin by giving the result for the matrix element $Z(S'I',SI)$
defined by
\bea
& &\langle S' I'L' ;m'_S,m'_L,\alpha'|\sum_{n=1}^{N_c}\sigma_n^k\tau_n^a |
S IL ;m_S,m_L,\alpha\rangle =\\
& &\qquad \frac{1}{\sqrt{(2S'+1)(2I'+1)}}Z(S'I',SI)\delta_{LL'}
\delta_{m_L m'_L}
\langle S',m'_S|S,1;m_S,k\rangle
\langle I',\alpha'|I,1;\alpha,a\rangle\,.\nonumber
\eea
According to (\ref{4.72}) this matrix element is completely determined in
terms of the diagonal and the nondiagonal matrix elements of the
current on direct product states. These will be characterized by two
quantities $z_1,z_2$ defined by
\bea
Z_1 &=& _{j}\langle S'I',m',\alpha'|\sum_{n=1}^{N_c}\sigma_n^k\tau_n^a|
SI,m,\alpha\rangle_j\\
&=& \sqrt{(2S+1)(2I+1)}z_1
\langle S',m'|S,1;m,k\rangle
\langle I',\alpha'|I,1;\alpha,a\rangle\nonumber\\
Z_2 &=& _{j'}\langle
S'I',m',\alpha'|\sum_{n=1}^{N_c}\sigma_n^k\tau_n^a|SI,m,\alpha\rangle_j\\
&=& \sqrt{(2S+1)(2I+1)}z_2
\langle S',m'|S,1;m,k\rangle
\langle I',\alpha'|I,1;\alpha,a\rangle\nonumber\,.
\eea
The reduced matrix element $Z(S'I',SI)$ (taken between unnormalized quark model
states) is expressed in terms of $z_1$ and
$z_2$ as
\bea
Z(S'I',SI) &=& (-)^{\psi(S'I'i')+\psi(SIi)}\sqrt{(2S+1)(2I+1)(2S'+1)(2I'+1)}
\frac{N_c+2}{4}(z_1-z_2)\,.
\eea

We obtain for the diagonal matrix element the simple result
\bea
z_1 &=&\sqrt{(2i+1)(2i'+1)}z(i',i) (-)^{-1+2i'+S+I}
\left\{ \begin{array}{ccc}
1 & S & S' \\
\frac12 & i' & i \end{array}\right\}
\left\{ \begin{array}{ccc}
1 & I & I' \\
\frac12 & i' & i \end{array}\right\}\\
&+& 6\delta_{ii'}(-)^{1-2i-S'-I'}
\left\{ \begin{array}{ccc}
1 & S & S' \\
i & \frac12 & \frac12 \end{array}\right\}
\left\{ \begin{array}{ccc}
1 & I & I' \\
i & \frac12 & \frac12 \end{array}\right\}\nonumber\,,
\eea
with
\bea
z(i',i) = \sqrt{(N_c+1)^2-(i'-i)^2(i'+i+1)^2} = N_c+1 + {\cal O}(1/N_c)\,.
\eea

The nondiagonal matrix element $z_2$ can be written as a sum over the
four terms into which it can be decomposed with the help of (\ref{4.74})
\bea
z_2 = \frac14(T_1 + T_2 + T_3 + T_4)\,.
\eea
We find
\bea
T_1 &=& z_1\\
T_2 &=& \sqrt{6(2i+1)(2i'+1)^2}F(i')z(i',i)
(-)^{-\frac12+2i+i'+I+S+S'}\\
& &\times\left\{ \begin{array}{ccc}
i' & \frac12 & S' \\
\frac12 & i' & 1 \end{array}\right\}
\left\{ \begin{array}{ccc}
1 & S & S' \\
\frac12 & i' & i \end{array}\right\}
\left\{ \begin{array}{ccc}
1 & I & I' \\
\frac12 & i' & i \end{array}\right\}\nonumber\\
&-& 6\sqrt6 F(i)\delta_{ii'}\sqrt{2i'+1} (-)^{\frac12-i-I'}
\left\{ \begin{array}{ccc}
i & i' & 1 \\
\frac12 & \frac12 & S' \end{array}\right\}
\left\{ \begin{array}{ccc}
1 & S & S' \\
i & \frac12 & \frac12 \end{array}\right\}
\left\{ \begin{array}{ccc}
1 & I & I' \\
i & \frac12 & \frac12 \end{array}\right\}\nonumber\\
T_3 &=& \sqrt{6(2i+1)(2i'+1)^2}F(i')z(i',i)
(-)^{-\frac12+2i+i'+I+S+I'}\\
& &\times\left\{ \begin{array}{ccc}
i' & \frac12 & I' \\
\frac12 & i' & 1 \end{array}\right\}
\left\{ \begin{array}{ccc}
1 & S & S' \\
\frac12 & i' & i \end{array}\right\}
\left\{ \begin{array}{ccc}
1 & I & I' \\
\frac12 & i' & i \end{array}\right\}\nonumber\\
&-& 6\sqrt6 F(i)\delta_{ii'}\sqrt{2i'+1} (-)^{\frac12-i-S'}
\left\{ \begin{array}{ccc}
i & i' & 1 \\
\frac12 & \frac12 & I' \end{array}\right\}
\left\{ \begin{array}{ccc}
1 & S & S' \\
i & \frac12 & \frac12 \end{array}\right\}
\left\{ \begin{array}{ccc}
1 & I & I' \\
i & \frac12 & \frac12 \end{array}\right\}\nonumber\\
T_4 &=& 6\sqrt{(2i+1)(2i'+1)} (-)^{S+I+S'+I'}\\
& &\times\sum_{i''}\frac{z(i',i'')z(i'',i)}{N_c-1}(2i''+1)
\left\{ \begin{array}{ccc}
i'' & \frac12 & S' \\
\frac12 & i' & 1 \end{array}\right\}
\left\{ \begin{array}{ccc}
S & 1 & S' \\
i'' & \frac12 & i \end{array}\right\}
\left\{ \begin{array}{ccc}
i'' & \frac12 & I' \\
\frac12 & i' & 1 \end{array}\right\}
\left\{ \begin{array}{ccc}
I & 1 & I' \\
i'' & \frac12 & i \end{array}\right\}\nonumber\\
&+& 36\frac{z(i',i)}{N_c-1}\sqrt{(2i+1)(2i'+1)}
\left\{ \begin{array}{ccc}
i & i' & 1 \\
\frac12 & \frac12 & S' \end{array}\right\}
\left\{ \begin{array}{ccc}
i & i' & 1 \\
\frac12 & \frac12 & I' \end{array}\right\}
\left\{ \begin{array}{ccc}
S & S' & 1 \\
\frac12 & \frac12 & i \end{array}\right\}
\left\{ \begin{array}{ccc}
I & I' & 1 \\
\frac12 & \frac12 & i \end{array}\right\}\nonumber\,.
\eea

The expression for $z_2$ greatly simplifies if only terms of order 1 
are kept, in addition to the leading ones of order $N_c$, due to the fact
that in this approximation the $z(i',i)$ factors are constants. This allows 
the sum over $i''$ to be performed with the help of (\ref{4.77}). We obtain
\bea
& &\frac{z_1}{\sqrt{(2i+1)(2i'+1)}} = (N_c+1)
 (-)^{\phi_0} [\{6j_0\}^3 - 3\{6j_1\}^3]\\
& &\qquad +12
\left\{ \begin{array}{ccc}
i & i' & 0 \\
\frac12 & \frac12 & S' \end{array}\right\}
\left\{ \begin{array}{ccc}
i & i' & 0 \\
\frac12 & \frac12 & I' \end{array}\right\}
\left\{ \begin{array}{ccc}
1 & S & S' \\
i & \frac12 & \frac12 \end{array}\right\}
\left\{ \begin{array}{ccc}
1 & I & I' \\
i & \frac12 & \frac12 \end{array}\right\}+{\cal O}(N_c^{-1})\nonumber\\
& &\frac{z_2}{\sqrt{(2i+1)(2i'+1)}} = (N_c+2)
(-)^{\phi_0} \{6j_0\}^3 +
3(-)^{\phi_0} \{6j_1\}^3\\
& &\qquad + 3  \frac{\delta_{S'I'}}{2I'+1}
\left\{ \begin{array}{ccc}
1 & S & S' \\
i & \frac12 & \frac12 \end{array}\right\}
\left\{ \begin{array}{ccc}
1 & I & I' \\
i & \frac12 & \frac12 \end{array}\right\}\nonumber\\
& &\qquad - \frac12  (-)^{\phi_0}
[\frac12 + 2i'(i'+1) - S'(S'+1) - I'(I'+1)]
[\{6j_0\}^3 - 3\{6j_1\}^3] + {\cal O}(N_c^{-1})\nonumber
\eea
with
$\phi_0=1+i-i'+I'+S'$.
We denoted here the products of 6$j$ symbols encountered in Section III
\bea
\{6j_{0(1)}\}^3 &=& 
\left\{ \begin{array}{ccc}
S' & S & 1 \\
I & I' & 0(1) \end{array}\right\}
\left\{ \begin{array}{ccc}
S & i & \frac12 \\
\frac12 & 0(1) & I \end{array}\right\}
\left\{ \begin{array}{ccc}
S' & \frac12 & i' \\
\frac12 & I' & 0(1) \end{array}\right\}\,.
\eea

The difference of $z_1$ and $z_2$ can be finally written as
\bea\label{zdiff}
& &\frac{z_1-z_2}{\sqrt{(2i+1)(2i'+1)}} =  -3(N_c+2) (-)^{\phi_0}
\{6j_1\}^3 - (-)^{\phi_0}\{6j_0\}^3\\
& &\qquad + \frac12 (-)^{\phi_0}
[\frac12 + 2i'(i'+1) - S'(S'+1) - I'(I'+1)]
[\{6j_0\}^3 - 3\{6j_1\}^3]\nonumber\\
& & \qquad - 18(-)^{i+i'+I'+S'}
\left\{ \begin{array}{ccc}
S' & I' & 1 \\
\frac12 & \frac12 & i \end{array}\right\}
\left\{ \begin{array}{ccc}
S' & I' & 1 \\
\frac12 & \frac12 & i' \end{array}\right\}
\left\{ \begin{array}{ccc}
1 & S & S' \\
i & \frac12 & \frac12 \end{array}\right\}
\left\{ \begin{array}{ccc}
1 & I & I' \\
i & \frac12 & \frac12 \end{array}\right\}+{\cal O}(N_c^{-1})
\,.\nonumber
\eea

The alternative method presented in Section IV.C can be also used to
give an exact expression for the reduced matrix element of $Z^{ia}$.
We find from (\ref{4.97}) the following result
\bea\label{A16}
Z(S'I',SI) &=& (-)^{\psi(S'I'i')+\psi(SIi)}\sqrt{(2S+1)(2I+1)(2S'+1)(2I'+1)}
\frac{N_c(N_c+2)}{4(N_c-1)}\\
& &\qquad\times\left(z_1 - \frac{1}{N_c^2}B(Ii)B(I'i')
\frac{Z(I',I)}{(2I+1)(2I'+1)}\delta_{SI}\delta_{S'I'}
\right)\,,\nonumber
\eea
where $Z(I',I)$ has been defined in (\ref{common}) and $B(Ii)$ is given by
(\ref{Asquared}). We have checked explicitly that both methods lead to
the same answer for $Z(S'I',SI)$ up to the next-to-leading order in $1/N_c$.
We notice that (\ref{A16}) does not involve any summation over intermediate
state quantum numbers.

The leading order term in (\ref{zdiff}) is written as proportional to $N_c+2$,
which was seen to give the correct result to two orders in the
$1/N_c$ expansion for the case of the symmetric baryons. 
It is natural to ask whether a similar result holds also for the
reduced matrix element $Z(S'I',SI)$. In the following we will
argue that no result of comparable simplicity can be obtained for 
the $1/N_c$ corrections to this quantity. Strictly speaking
this still does not prove that there are nonvanishing $1/N_c$ corrections to
$Z(J'I',JI)$ (which is the true physical coupling with a meaning 
beyond the quark model) which is related to $Z(S'I',SI)$ by (\ref{spin-ind}).
It is still conceivable that the $1/N_c$ corrections to $Z(S'I',SI)$ 
add up to zero when inserted into (\ref{spin-ind}), although we
have not been able to prove it.

We will consider for simplicity the case when the quantum numbers of
the initial and final states satisfy
\bea\label{SneqI}
S\neq I\,, S'\neq I'
\eea
and examine the structure of the $1/N_c$ corrections in the following
two particular cases: a) $S=S'\,,I=I'$ and b) $S=I'\,,I=S'$.
This constrains $i,i'$ to be equal: $i=i'$.

The norm of a state satisfying
(\ref{SneqI}) can be obtained from (\ref{4.44}) and is given exactly by
\bea\label{normSneqI}
\langle SI|SI\rangle = \frac{N_c+2}{4}\frac{N_c}{N_c-1}\,.
\eea
This will have to be divided out from the quantity on the r.h.s.
of (\ref{zdiff}). We obtain 
\bea\label{ZSneqI}
& &\frac{1}{\sqrt{\langle SIi|SIi\rangle \langle S'I'i'|S'I'i'\rangle}}
\frac{N_c+2}{4}(z_1-z_2) = \sqrt{(2i+1)(2i'+1)}
\left\{-3(N_c+2) (-)^{\phi_0} \{6j_1\}^3\right.\\
& &\qquad\left. - 18(-)^{i+i'+I'+S'}
\left\{ \begin{array}{ccc}
S' & I' & 1 \\
\frac12 & \frac12 & i \end{array}\right\}
\left\{ \begin{array}{ccc}
S' & I' & 1 \\
\frac12 & \frac12 & i' \end{array}\right\}
\left\{ \begin{array}{ccc}
1 & S & S' \\
i & \frac12 & \frac12 \end{array}\right\}
\left\{ \begin{array}{ccc}
1 & I & I' \\
i & \frac12 & \frac12 \end{array}\right\}
+ 3 (-)^{\phi_0} \{6j_1\}^3 \right\}\nonumber\,.
\eea
Next, we note that in the limit (\ref{SneqI}) the product
of 4 $6j$ symbols on the r.h.s. can be written as
\bea
& &(-)^{i+i'+I'+S'}
\left\{ \begin{array}{ccc}
S' & I' & 1 \\
\frac12 & \frac12 & i \end{array}\right\}
\left\{ \begin{array}{ccc}
S' & I' & 1 \\
\frac12 & \frac12 & i' \end{array}\right\}
\left\{ \begin{array}{ccc}
1 & S & S' \\
i & \frac12 & \frac12 \end{array}\right\}
\left\{ \begin{array}{ccc}
1 & I & I' \\
i & \frac12 & \frac12 \end{array}\right\} \\
& &\qquad\qquad\qquad =
\left\{ \begin{array}{cc}
\frac{1}{36(2i+1)^2}\sqrt{\frac{(2i-1)(2i+3)}{i(i+1)}} &
\mbox{case a)}\\
\frac{1}{9(2i+1)^2} &
\mbox{case b)} \end{array}\right.\nonumber
\eea

On the other hand, the leading order term is proportional to
$\{6j_1\}^3$.
From (\ref{4.76}) we obtain
in the limit (\ref{SneqI})
\bea
\{6j_1\}^3 = (-)^{1+2i}\frac13
\left\{ \begin{array}{ccc}
1 & S & S' \\
\frac12 & i' & i \end{array}\right\}
\left\{ \begin{array}{ccc}
1 & I & I' \\
\frac12 & i' & i \end{array}\right\}
= \left\{ \begin{array}{cc}
(-)^{2i}\frac{1}{3(2i+1)^2}\sqrt{\frac{(2i-1)(2i+3)}{2i(2i+2)}} &
\mbox{case a)}\\
(-)^{1+2i}\frac{1}{6(2i+1)^2i(i+1)} &
\mbox{case b)} \end{array}\right.
\eea
One can see that for case b)
the terms of order 1 in (\ref{ZSneqI}) do 
not have the same structure as the leading term of order $N_c$ 
and therefore cannot be generally absorbed into a rescaling of the latter.

\subsection{${\cal T}(I',SI)$}

Next we present the exact calculation of the reduced matrix element
${\cal T}(I',SI)$ defined by
\bea
\langle I'L',m'm'_L\alpha'|\sum_{n=1}^{N_c}r^i_n \sigma_n^j\tau_n^a|
SIL,m_Sm_L\alpha\rangle &=& \frac{1}{(2I'+1)\sqrt{2L'+1}}{\cal T}(I',SI){\cal I}(L',L)\\
& &\hspace{-3cm}\times\langle I',m'|S,1;m_S,j\rangle 
\langle I',\alpha'|I,1;\alpha,a\rangle
\langle L',m'_L|L,1;m_L,i\rangle\,,\nonumber
\eea
relevant for the transitions from orbital excitations with mixed symmetry
to symmetric states.
We calculate this matrix element starting from the general formula
(\ref{4.111}) and proceeding along the same steps as in Section IV.D.
The first term in (\ref{4.111}) can be expressed with the help of
(\ref{4.112}) in terms of the two matrix elements ($k\neq 1$)
\bea
& &_1\langle I'I',m'\alpha'|\sigma_1^j \tau_1^a|SI,m\alpha\rangle_1 =
\sqrt{(2S+1)(2I+1)}t_1
\langle I',m'|S1;m,j\rangle \langle I',\alpha'|I1;\alpha,a\rangle\\
& &_1\langle I'I',m'\alpha'|P_{1k}\sigma_1^j \tau_1^a|SI,m\alpha\rangle_1 =
\sqrt{(2S+1)(2I+1)}t_2
\langle I',m'|S1;m,j\rangle \langle I',\alpha'|I1;\alpha,a\rangle\,.
\eea

Expanding the permutation operator $P_{1k}$ and evaluating the resulting
matrix elements with the help of the results of Section IV.B we obtain the 
following exact expressions for the coefficients $t_1,t_2$
\bea
t_1 &=& 6\delta_{ii'}
\left\{ \begin{array}{ccc}
1 & S & I' \\
i & \frac12 & \frac12 \end{array}\right\}
\left\{ \begin{array}{ccc}
1 & I & I' \\
i & \frac12 & \frac12 \end{array}\right\}\\
t_2 &=& \frac32 \delta_{ii'}
\left\{ \begin{array}{ccc}
1 & S & I' \\
i & \frac12 & \frac12 \end{array}\right\}
\left\{ \begin{array}{ccc}
1 & I & I' \\
i & \frac12 & \frac12 \end{array}\right\}\\
&-& 3\sqrt{\frac32}F(i)\delta_{ii'}\sqrt{2i'+1}(-)^{I'-\frac12+i}
\left\{ \begin{array}{ccc}
i & i' & 1 \\
\frac12 & \frac12 & I' \end{array}\right\}
\left\{ \begin{array}{ccc}
S & I' & 1 \\
\frac12 & \frac12 & i \end{array}\right\}
\left\{ \begin{array}{ccc}
1 & I & I' \\
i & \frac12 & \frac12 \end{array}\right\}\nonumber\\
&-& 3\sqrt{\frac32}G(i)\delta_{ii'}\sqrt{2i'+1}(-)^{I'-\frac12+i}
\left\{ \begin{array}{ccc}
i & i' & 1 \\
\frac12 & \frac12 & I' \end{array}\right\}
\left\{ \begin{array}{ccc}
S & I' & 1 \\
\frac12 & \frac12 & i \end{array}\right\}
\left\{ \begin{array}{ccc}
1 & I & I' \\
i & \frac12 & \frac12 \end{array}\right\}\nonumber\\
&+& 9H(i',i)\sqrt{(2i+1)(2i'+1)}
\left\{ \begin{array}{ccc}
i & i' & 1 \\
\frac12 & \frac12 & I' \end{array}\right\}^2
\left\{ \begin{array}{ccc}
S & I' & 1 \\
\frac12 & \frac12 & i \end{array}\right\}
\left\{ \begin{array}{ccc}
I & I' & 1 \\
\frac12 & \frac12 & i \end{array}\right\}\nonumber\,.
\eea
The result for $t_2$ simplifies considerably when only the terms of
order 1 and $1/N_c$ are kept
\bea
& &\frac{t_2}{\sqrt{(2i+1)(2i'+1)}} =\\
& &\qquad \left\{ \frac{3}{2I'+1}
 + \frac{1}{N_c}\frac{6}{2I'+1} 
 - \frac{6}{N_c}\left[\frac52 + 2i'(i'+1) -
2I'(I'+1)\right] \left\{ \begin{array}{ccc}
i & i' & 0 \\
\frac12 & \frac12 & I' \end{array}\right\}^2\right\}\nonumber\\
& &\qquad\times
\left\{ \begin{array}{ccc}
S & I' & 1 \\
\frac12 & \frac12 & i \end{array}\right\}
\left\{ \begin{array}{ccc}
I & I' & 1 \\
\frac12 & \frac12 & i \end{array}\right\} + {\cal O}(N_c^{-2})\nonumber\,.
\eea
The matrix element on the r.h.s. of (\ref{4.112}) is proportional to
the combination
\bea
& & t_1+(N_c-1) t_2 =\sqrt{(2i+1)(2i'+1)}\\
& &\quad\times\left[\frac{3}{2I'+1}N_c + \frac{3}{2I'+1} -
6\left[ \frac12 + 2i'(i'+1) -2I'(I'+1)\right]
\left\{ \begin{array}{ccc}
i & i' & 0 \\
\frac12 & \frac12 & I' \end{array}\right\}^2\right]\nonumber\\
& &\qquad\times 
\left\{ \begin{array}{ccc}
S & I' & 1 \\
\frac12 & \frac12 & i \end{array}\right\}
\left\{ \begin{array}{ccc}
I & I' & 1 \\
\frac12 & \frac12 & i \end{array}\right\} + {\cal O}(N_c^{-1})\nonumber\,.
\eea
After dividing this expression with $B(I'i')$ we find for the first 
matrix element in (\ref{4.111}) the following result
\bea\label{A28}
& &\langle I',m'\alpha'|\sigma_1^j\tau_1^a|SI,m_S\alpha\rangle_1 =\\
& &\qquad\sqrt{(2S+1)(2I+1)}
\langle I',m'|S1;m,j\rangle \langle I',\alpha'|I1;\alpha,a\rangle\,\,
3\sqrt2 \sqrt{\frac{2i+1}{2I'+1}}\nonumber\\
& &\quad\times
\left\{ 1 + \frac{1}{2N_c} + \frac{2I'+1}{3N_c}
\left[ \frac12 + 2i'(i'+1) -2I'(I'+1)\right]
\left(\frac{3}{2(2i'+1)} - 6
\left\{ \begin{array}{ccc}
i & i' & 0 \\
\frac12 & \frac12 & I' \end{array}\right\}^2\right)\right\}\nonumber\\
& &\qquad\times
\left\{ \begin{array}{ccc}
S & I' & 1 \\
\frac12 & \frac12 & i \end{array}\right\}
\left\{ \begin{array}{ccc}
I & I' & 1 \\
\frac12 & \frac12 & i \end{array}\right\} + {\cal O}(N_c^{-1})\nonumber\,.
\eea
The second matrix element in (\ref{4.111}) is given by
\bea\label{A29}
& &\delta_{SI} B(Ii)
\langle I',m'\alpha'|\sum_{n=1}^{N_c}\sigma_n^j\tau_n^a
|I,m_S,\alpha\rangle =\\
& &\quad  \delta_{SI}
N_c^2 \sqrt{(2S+1)(2I+1)}
\langle I',m'|S1;m,j\rangle \langle I',\alpha'|I1;\alpha,a\rangle\,\,
3\sqrt2 \sqrt{\frac{2i+1}{2I'+1}}\nonumber\\
& &\quad\times
\frac{1}{6(2I+1)}\left\{ 1 + \frac{5}{2N_c} - \frac{2I+1}{2N_c(2i+1)}
\left[ \frac12 + 2i(i+1) -2I(I+1)\right]\right\}\nonumber\,.
\eea

The result for the reduced matrix element ${\cal T}(I',SI)$ valid to 
next-to-leading order in $1/N_c$ is obtained by
inserting (\ref{A28}) and (\ref{A29}) into (\ref{4.111}) and
making use of (\ref{prod6j}) for the product of 6$j$ symbols in (\ref{A28}).

Let us examine closer the structure of the $1/N_c$ corrections to the
leading order result for ${\cal T}(I',SI)$ on the simple particular case when 
$S\neq I$. After
dividing with the norm of the initial state (\ref{normSneqI})
we find for this case
\bea
{\cal T}(I',SI) &=& N_c^{\kappa-\frac12} 2\sqrt6 \sqrt{(2S+1)(2I+1)(2I'+1)}
(-)^{-I+I'}
\left\{ \begin{array}{ccc}
1 & 1 & 1 \\
S & I & I' \end{array}\right\}\\
& &\hspace{-3cm}\times
\left\{ 1 - \frac{1}{2N_c} + \frac{2I'+1}{3N_c}
\left[ \frac12 + 2i'(i'+1) -2I'(I'+1)\right]
\left(\frac{3}{2(2i'+1)} - 6
\left\{ \begin{array}{ccc}
i & i' & 0 \\
\frac12 & \frac12 & I' \end{array}\right\}^2\right)\right\}\nonumber\,.
\eea
The last term in the braces has the explicit expression
\bea
& &\frac{2I'+1}{3N_c}
\left[ \frac12 + 2i'(i'+1) -2I'(I'+1)\right]
\left(\frac{3}{2(2i'+1)} - 6
\left\{ \begin{array}{ccc}
i & i' & 0 \\
\frac12 & \frac12 & I' \end{array}\right\}^2\right)  \\
& & \qquad\qquad =\frac{2I'+1}{2N_c}(-)^{\frac12-i'+I'}
(1 - 2\delta_{ii'})\nonumber
\eea
which shows that it cannot be absorbed into a rescaling of the leading order term.
This quark model calculation suggests therefore that the ratios of the 
$Y$ and $Q$ couplings of the mixed symmetry states
predicted by the consistency conditions in the large-$N_c$ limit
will receive nontrivial $1/N_c$ corrections.

\section{Conclusions and Outlook}

We have studied the strong couplings of the excited baryons in the
large-$N_c$ limit with the help of consistency conditions on pion-baryon
scattering amplitudes. This method is similar to the one used by
Dashen, Jenkins and Manohar \cite{DM,Jen,DJM1,DJM2} in their analysis
of the strong couplings of the s-wave baryons. In extending their
analysis to the excited baryons' sector one has to deal with additional
complications, related to the more complex structure of the spectrum of 
these states. 

The consistency conditions are very effective in constraining the large-$N_c$
spin-isospin dependence of the strong vertices of these states, especially for 
the $S$-wave pion coupling, which is completely determined in terms of
just one unknown constant. The allowed form of the strong vertices turns out to
be exactly the same as the one following from the constituent quark model.
In addition to constraining the structure of the strong vertex, the
consistency conditions predict also the equality of the pion couplings to
excited and to s-wave baryons respectively. This is again what is expected
from the constituent quark model. 

Our findings extend therefore the results obtained in \cite{DM,Jen,DJM1,DJM2}
for the strong couplings of the $s$-wave baryons and give a natural explanation
for the successes of the quark model when applied to strong decays of the
excited baryons \cite{Close,FaiPla,Hey,CGKM} in terms of the large-$N_c$ expansion.
For example, this lends additional support to some predictions made recently for 
strong decays of excited heavy baryons \cite{TMYP,FKT} with the help of the
quark model.
However, as discussed in Appendix A, the quark model predictions for 
ratios of strong couplings for these states cannot be expected to hold to
the same accuracy as in the s-wave sector, as these ratios are not in general
protected against $1/N_c$ corrections. 
The exact results in Appendix A provide a specific framework to study quantitatively
how good the large-$N_c$ approximation is by examining their complete $N_c$
dependence as $N_c$ varies from the physical value $N_c=3$ to infinity.

The results of the present paper can be expanded in a number of directions.
We recall that our analysis has only assumed isospin symmetry. Thus, one can
attempt to incorporate SU(3) with some amount of symmetry breaking, by 
studying consistency conditions following from large-$N_c$ counting rules for
kaon-baryon scattering amplitudes
\cite{DJM1,DJM2}. In this way one should be able to relate the strong couplings
of different towers of states with different strangeness quantum numbers,
which in our present analysis are left completely unrelated.
Second, we have only discussed excited states transforming under the
symmetric and mixed symmetric representations of SU(4). It is known
that excited states exist which transform also under the antisymmetric
representation. Extending our analysis to this case should be completely
straightforward. Finally, a similar analysis could be performed for
the electromagnetic couplings of the excited baryons, with the help of
consistency conditions for photon-baryon scattering amplitudes.
For the s-wave baryons such constraints on the magnetic moments have been 
worked out in \cite{JenMan}.
We plan to return to some of these problems in a future publication.

\acknowledgements

The research of 
D.P. was supported by the Ministry of Science and the Arts of Israel.
The work of T.M.Y. was supported in part by the National Science 
Foundation.

\newpage
\appendix
\section{Transition matrix elements between states with mixed symmetry}

We present in this Appendix the computation in the quark model of the matrix 
elements of $Y^a$ and $Q^{ka}$ between excited baryon states with mixed symmetry.
This quantity is phenomenologically relevant for strong decays of positive-parity
excited baryons into negative-parity states in the {\bf 70}. We include this 
calculation here merely for the sake of completeness and because the result provides an 
explicit realization for the most general solution of the consistency
condition for $Q(J'I',JI)$ (\ref{Qansatz}).

We start by computing the quark model reduced matrix element ${\cal T}(S'I',SI)$
defined by
\bea
& &\langle S'I'L',m'_S m'_L\alpha'| \sum_{n=1}^{N_c} r_n^i \sigma_n^j \tau_n^a|
SIL,m_S m_L\alpha\rangle =
\frac{1}{\sqrt{(2S'+1)(2I'+1)(2L'+1)}}\\ 
& &\qquad\times{\cal T}(S'I',SI) {\cal I}(L',L)
\langle S',m'_S|S1;m_S, j\rangle
\langle L',m'_L|L1;m_L, i\rangle
\langle I',\alpha'|I1;\alpha, a\rangle\,.\nonumber
\eea
We proceed in close analogy to the calculation of ${\cal T}(I',SI)$ in Section
IV.D. First we take the matrix element of the spatial part of the operator, which
is parametrized by the overlap integral ${\cal I}(L',L)$
\bea
& &\langle S'I'L',m'_S m'_L\alpha'| r_n^i \sigma_n^j \tau_n^a|
SIL,m_S m_L\alpha\rangle\\
& &\qquad = \frac{(-)^{\psi(S'I'i')+\psi(SIi)}}{2(N_c-1)}\sum_{k,k'=2}^{N_c}\,
_{[k',1]}\langle S'I',m'_S \alpha'|\sigma_n^j \tau_n^a|SI,m_S\alpha\rangle_{[k,1]}
\nonumber \\
& &\qquad\times
\frac{1}{\sqrt{2L'+1}}
\langle L',m'_L|L1;m_L, i\rangle {\cal I}(L',L)
[\delta_{kn}\delta_{kk'} + \delta_{n1}]\,.\nonumber
\eea
After summing over the contributions of the $N_c$ quarks to the transition
operator we obtain the following general expression for the reduced matrix
element ${\cal T}(S'I',SI)$
\bea
& &\frac{1}{\sqrt{(2S'+1)(2I'+1)}}{\cal T}(S'I',SI)
\langle S',m'_S|S1;m_S, j\rangle
\langle I',\alpha'|I1;\alpha, a\rangle =
\frac{(-)^{\psi(S'I'i')+\psi(SIi)}}{2(N_c-1)}\\
& &\quad\times
\left\{ \sum_{n=1}^{N_c}\, 
_{[n,1]}\langle S'I',m'_S \alpha'|\sigma_n^j \tau_n^a|SI,m_S\alpha\rangle_{[n,1]}
+ \sum_{k'=2}^{N_c}\, _{[k',1]}\langle S'I',m'_S \alpha'|\sigma_1^j \tau_1^a
\sum_{k=2}^{N_c}|SI,m_S\alpha\rangle_{[k,1]}\right\}\,.\nonumber
\eea
The first term in the braces is of order $N_c$ and is therefore suppressed relative to the
second one, which is of order $N_c^2$. 
In this section we work only to leading order in $N_c$ so we keep only the contribution
of the second term. It can be computed by expressing the sums over $k,k'$ with the help
of (\ref{4.89})
\bea
& &\sum_{k'=2}^{N_c}\, _{[k',1]}\langle S'I',m'_S \alpha'|\sigma_1^j \tau_1^a
\sum_{k=2}^{N_c}|SI,m_S\alpha\rangle_{[k,1]} =\\
& &\frac12 \left\{
\delta_{SI}\delta_{S'I'}B(Ii) B(I'i') 
\langle I',m'_S, \alpha'|\sigma_1^j \tau_1^a|I,m_S,\alpha\rangle -
N_c^2\, _1\langle S'I';m'_S, \alpha'|P_{1k}\sigma_1^j \tau_1^a|SI;m_S,\alpha\rangle_1\right.
\nonumber\\
& &\left. \quad -
N_c^2\, _1\langle S'I';m'_S, \alpha'|\sigma_1^j \tau_1^a P_{1k}|SI;m_S,\alpha\rangle_1
+ N_c^2 \, _1\langle S'I';m'_S, \alpha'|\sigma_1^j \tau_1^a|SI;m_S,\alpha\rangle_1\right\}
+ {\cal O}(N_c)\,.\nonumber
\eea
Each of the terms on the r.h.s. can be evaluated using the methods of Sec.IV.D. We
obtain for the reduced matrix element ${\cal T}(S'I',SI)$
\bea
& &\frac{1}{\sqrt{(2S'+1)(2I'+1)}}{\cal T}(S'I',SI) = \\
& &\qquad(-)^{\psi(S'I'i')+\psi(SIi)}\frac{N_c^2}{4(N_c-1)}\sqrt{(2i+1)(2i'+1)(2S+1)(2I+1)}
\nonumber\\
& &\times\left\{ \frac{\delta_{SI}\delta_{S'I'}}{2(2I+1)(2I'+1)} -
3\frac{\delta_{S'I'}}{2I'+1}
\left\{ \begin{array}{ccc}
S & S' & 1 \\
\frac12 & \frac12 & i \end{array}\right\}
\left\{ \begin{array}{ccc}
I & I' & 1 \\
\frac12 & \frac12 & i \end{array}\right\}\right.\nonumber\\
& &\left.\qquad \qquad
- 3(-)^{-I-S-I'-S'}
\frac{\delta_{SI}}{2I+1}
\left\{ \begin{array}{ccc}
S & S' & 1 \\
\frac12 & \frac12 & i' \end{array}\right\}
\left\{ \begin{array}{ccc}
I & I' & 1 \\
\frac12 & \frac12 & i' \end{array}\right\}\right.\nonumber\\
& &\left.\qquad
+ 12 (-)^{-I-S-I'-S'}
\left\{ \begin{array}{ccc}
i & i' & 0 \\
\frac12 & \frac12 & I \end{array}\right\}
\left\{ \begin{array}{ccc}
i & i' & 0 \\
\frac12 & \frac12 & S \end{array}\right\}
\left\{ \begin{array}{ccc}
S & S' & 1 \\
\frac12 & \frac12 & i' \end{array}\right\}
\left\{ \begin{array}{ccc}
I & I' & 1 \\
\frac12 & \frac12 & i' \end{array}\right\}\right\}\nonumber\,.
\eea
The first two terms can be combined together by using (\ref{prod6j}) for
the product of two 6$j$-symbols in the second term. The last two terms
can be also written together such that we obtain for the total sum 
of the four terms in the braces
\bea\label{B6}
& &\{\cdots\} = 
\left\{ \begin{array}{ccc}
S & I & 1 \\
\frac12 & \frac12 & i \end{array}\right\}
\left[
-3(-)^{-\frac12+S+I+I'+i}
\frac{\delta_{S'I'}}{2I'+1}
\left\{ \begin{array}{ccc}
1 & 1 & 1 \\
S & I & I' \end{array}\right\}\right.\\
& &\left.\qquad\qquad +
18(-)^{1+i+i'-I'-S'}
\left\{ \begin{array}{ccc}
S & I & 1 \\
\frac12 & \frac12 & i' \end{array}\right\}
\left\{ \begin{array}{ccc}
S & S' & 1 \\
\frac12 & \frac12 & i' \end{array}\right\}
\left\{ \begin{array}{ccc}
I & I' & 1 \\
\frac12 & \frac12 & i' \end{array}\right\}\right]\,.\nonumber
\eea
The product of three 6$j$-symbols on the r.h.s. can be transformed into
the following form by repeated application of (\ref{prod6j})
\bea
& &\left\{ \begin{array}{ccc}
S & I & 1 \\
\frac12 & \frac12 & i' \end{array}\right\}
\left\{ \begin{array}{ccc}
S & S' & 1 \\
\frac12 & \frac12 & i' \end{array}\right\}
\left\{ \begin{array}{ccc}
I & I' & 1 \\
\frac12 & \frac12 & i' \end{array}\right\} =\\
& &\qquad\left\{ \begin{array}{ccc}
S' & I' & 1 \\
\frac12 & \frac12 & i' \end{array}\right\}
\left[
\frac{\delta_{SI'}}{6(2I'+1)} +
(-)^{2S+S'+I}
\left\{ \begin{array}{ccc}
1 & 1 & 1 \\
S & I' & I \end{array}\right\}
\left\{ \begin{array}{ccc}
1 & 1 & 1 \\
I' & S' & S \end{array}\right\}\right]\nonumber\\
& &\qquad +
\frac{\delta_{S'I'}}{6(2I'+1)} 
(-)^{-\frac12+S+S'+I+i'}
\left\{ \begin{array}{ccc}
1 & 1 & 1 \\
S & I' & I \end{array}\right\}\,.\nonumber
\eea
When inserted into (\ref{B6}), the last term in this relation exactly cancels the 
first term in (\ref{B6}). We obtain in this way the following expression for the
reduced matrix element ${\cal T}(S'I',SI)$ taken between unnormalized mixed
symmetry states
\bea
& & {\cal T}(S'I',SI) = 
\frac92 N_c (-)^{\psi(S'I'i')+\psi(SIi)}\sqrt{(2i+1)(2i'+1)(2S+1)(2I+1)(2S'+1)(2I'+1)}\\
& &\qquad\times
(-)^{1+i+i'-I'-S'}
\left\{ \begin{array}{ccc}
S & I & 1 \\
\frac12 & \frac12 & i \end{array}\right\}
\left\{ \begin{array}{ccc}
S' & I' & 1 \\
\frac12 & \frac12 & i' \end{array}\right\}\nonumber\\
& &\qquad\times\left[
\frac{\delta_{SI'}}{6(2I'+1)} +
(-)^{2S+S'+I}
\left\{ \begin{array}{ccc}
1 & 1 & 1 \\
S & I' & I \end{array}\right\}
\left\{ \begin{array}{ccc}
1 & 1 & 1 \\
I' & S' & S \end{array}\right\}\right]\,.\nonumber
\eea

The physical value of ${\cal T}(S'I',SI)$ is obtained after dividing this
expression with the squared roots of the norms for the initial and final states
(\ref{4.44}). Inserting the appropriate phases of the mixed symmetry states
$\psi(SIi)=\frac12+I+i$ we obtain our final result
\bea\label{B9}
& & [{\cal T}(S'I',SI)]_{norm} = 
6(-)^{S+2I+I'}\sqrt{(2S+1)(2I+1)(2S'+1)(2I'+1)}\\
& &\qquad\times
\left[
\frac{\delta_{SI'}}{6(2I'+1)} +
(-)^{2S+S'+I}
\left\{ \begin{array}{ccc}
1 & 1 & 1 \\
S & I' & I \end{array}\right\}
\left\{ \begin{array}{ccc}
1 & 1 & 1 \\
I' & S' & S \end{array}\right\}\right]\,.\nonumber
\eea
This will be used in the following to compute the matrix elements of
$Y^a$ and $Q^{ka}$ between states with mixed symmetry.

\subsection{Matrix elements of $Y^a$}

The matrix element of $Y^a$ takes its simplest form in the 
$|(IP)S,L;J,m,\alpha\rangle$ basis, where it is directly proportional to 
${\cal T}(S'I',SI)$ 
\bea\label{Yquarkmodel}
& &\langle (I'P)S',L';J',m',\alpha'|Y^a|(IP)S,L;J,m,\alpha\rangle =\\
& &\qquad \sqrt{\frac{2J+1}{2I'+1}}{\cal T}(S'I',SI) \delta_{JJ'}\delta_{mm'}
\left\{ \begin{array}{ccc}
S & 1 & S' \\
L & 1 & L' \\
J & 0 & J' \end{array}\right\}
\langle I',\alpha |I1;\alpha,a\rangle\,.\nonumber
\eea
The 9$j$-symbol with one value of 0 can be reduced to a 6$j$-symbol. We decided
to write it in this form, as it allows us to read off 
the results from the corresponding expressions for $Q^{ka}$ given below 
by making the replacement $2\to 0$ in the Wigner symbols.

We are interested finally in the matrix elements of $Y^a$ in the
$|I,(PL)\Delta;J,m,\alpha\rangle$ basis, which is reached through the
recoupling relation (\ref{3Jrecoupling}). With $P=P'=1$, we have
\bea
& &\langle I',(P'L')\Delta';J',m',\alpha'|Y^a|I,(PL)\Delta;J,m,\alpha\rangle =\\
& &\qquad (-)^{-I-L-J-I'-L'-J'}\sqrt{\frac{2J+1}{2I'+1}} \delta_{JJ'}\delta_{mm'}
\langle I',\alpha |I1;\alpha,a\rangle\nonumber\\
&\times&
\sum_{S S'}\sqrt{(2S+1)(2S'+1)(2\Delta+1)(2\Delta'+1)}
\left\{ \begin{array}{ccc}
I & 1 & S \\
L & J & \Delta \end{array}\right\}
\left\{ \begin{array}{ccc}
I' & 1 & S' \\
L' & J' & \Delta' \end{array}\right\}
\left\{ \begin{array}{ccc}
S & 1 & S' \\
L & 1 & L' \\
J & 0 & J' \end{array}\right\}
{\cal T}(S'I',SI)\,.\nonumber
\eea
All what is left to do is insert here the result of the quark model calculation
of ${\cal T}(S'I',SI)$ (\ref{B9}) and perform the summations over $S,S'$.

We write the total result for the reduced matrix element $Y(JI',JI)$ as
\bea
Y(JI',JI) = [Y(JI',JI)]_1 + [Y(JI',JI)]_2
\eea
where $[Y(JI',JI)]_{1,2}$ stand for the contributions of the two terms in 
${\cal T}(S'I',SI)$  (\ref{B9}). We find
\bea
[Y(JI',JI)]_1 &=& (-)^{1+2J}\sqrt{2\Delta+1}\delta_{\Delta' L}
\left\{ \begin{array}{ccc}
L & \Delta' & 0 \\
1 & 1 & L' \end{array}\right\}\times
(-)^{J+I'+\Delta'}
\left\{ \begin{array}{ccc}
I & 1 & I' \\
\Delta' & J & \Delta \end{array}\right\}
\eea
\bea
[Y(JI',JI)]_2 &=& 2\sqrt3 (-)^{2J+\Delta+\Delta'}
\sqrt{(2\Delta+1)(2\Delta'+1)}
\left\{ \begin{array}{ccc}
1 & 1 & 1 \\
L & \Delta' & L' \end{array}\right\}
\left\{ \begin{array}{ccc}
1 & 1 & 1 \\
\Delta' & \Delta & L \end{array}\right\}\\
&\times&
(-)^{J+I'+\Delta'}
\left\{ \begin{array}{ccc}
I & 1 & I' \\
\Delta' & J & \Delta \end{array}\right\}\,.\nonumber
\eea
Their sum can be seen to have the same form as the model-independent
solution of the corresponding consistency condition (\ref{Ysolution}).

\subsection{Matrix elements of $Q^{ka}$}

The matrix element of $Q^{ka}$ is given, in the $|(IP)S,L;J,m,\alpha\rangle$ basis,
by an expression similar to (\ref{Yquarkmodel})
\bea\label{Qquarkmodel}
& &\langle (I'P)S',L';J',m',\alpha'|Q^{ka}|(IP)S,L;J,m,\alpha\rangle =\\
& &\qquad \sqrt{5\frac{2J+1}{2I'+1}}{\cal T}(S'I',SI) 
\left\{ \begin{array}{ccc}
S & 1 & S' \\
L & 1 & L' \\
J & 2 & J' \end{array}\right\}
\langle J',m' |J2;m,k\rangle
\langle I',\alpha |I1;\alpha,a\rangle\,.\nonumber
\eea
This can be transformed to the $|I,(PL)\Delta;J,m,\alpha\rangle$ basis
with the help of the recoupling relation (\ref{3Jrecoupling}).
Again with $P=P'=1$, we have
\bea\label{B16}
& &\langle I',(P'L')\Delta';J',m',\alpha'|Q^{ka}|I,(PL)\Delta;J,m,\alpha\rangle =\\
& &\qquad (-)^{-I-L-J-I'-L'-J'}\sqrt{5\frac{2J+1}{2I'+1}} \langle J',m' |J2;m,k\rangle
\langle I',\alpha |I1;\alpha,a\rangle\nonumber\\
&\times&
\sum_{S S'}\sqrt{(2S+1)(2S'+1)(2\Delta+1)(2\Delta'+1)}
\left\{ \begin{array}{ccc}
I & 1 & S \\
L & J & \Delta \end{array}\right\}
\left\{ \begin{array}{ccc}
I' & 1 & S' \\
L' & J' & \Delta' \end{array}\right\}
\left\{ \begin{array}{ccc}
S & 1 & S' \\
L & 1 & L' \\
J & 2 & J' \end{array}\right\}
{\cal T}(S'I',SI)\,.\nonumber
\eea

In the following we consider the contributions of the two terms in ${\cal T}(S'I',SI)$
(\ref{B9}) to this relation in turn. For the first term the summation over $S$ is trivial 
and amounts to the substitution $S\to I'$. The remaining sum over $S'$ can be readily 
done by using (\ref{Edmonds(6.4.8)}), which gives for the contribution of this term to 
$Q(J'I',JI)$
\bea
[Q(J'I',JI)]_1 &=& (-)^{1+2\Delta'}
\sqrt{5(2\Delta+1)(2\Delta'+1)}
\left\{ \begin{array}{ccc}
I & 1 & I' \\
L & J & \Delta \end{array}\right\}
\left\{ \begin{array}{ccc}
L & \Delta' & 2 \\
1 & 1 & L' \end{array}\right\}
\left\{ \begin{array}{ccc}
J & J' & 2 \\
\Delta' & L & I' \end{array}\right\}\,.\nonumber\\
\eea
This can be put into a form similar to (\ref{Qansatz}) by expressing the
product of the first and last 6$j$-symbols with the help of (\ref{Edmonds(6.4.8)})
\bea
\left\{ \begin{array}{ccc}
I & 1 & I' \\
L & J & \Delta \end{array}\right\}
\left\{ \begin{array}{ccc}
J & J' & 2 \\
\Delta' & L & I' \end{array}\right\} =
\sum_{y=1,2,3} (2y+1)
\left\{ \begin{array}{ccc}
1 & 2 & y \\
\Delta' & \Delta & L \end{array}\right\}
\left\{ \begin{array}{ccc}
\Delta' & I' & J' \\
\Delta & I & J \\
y & 1 & 2 \end{array}\right\}\,.
\eea

The contribution of the second term is proportional to the double sum over
$S,S'$
\bea
I_{SS'} &=& \sum_{SS'}(-)^{S+S'+I+I'}(2S+1)(2S'+1)
\left\{ \begin{array}{ccc}
I & 1 & S \\
L & J & \Delta \end{array}\right\}
\left\{ \begin{array}{ccc}
I' & 1 & S' \\
L' & J' & \Delta' \end{array}\right\}\\
& &\qquad \times
\left\{ \begin{array}{ccc}
1 & 1 & 1 \\
S & I' & I \end{array}\right\}
\left\{ \begin{array}{ccc}
1 & 1 & 1 \\
I' & S' & S \end{array}\right\}
\left\{ \begin{array}{ccc}
S & 1 & S' \\
L & 1 & L' \\
J & 2 & J' \end{array}\right\}\,.\nonumber
\eea

The summations over $S$ and $S'$ are analogous to the sum over $S$ encountered
in Sec.III.C in Eq.(\ref{3.70}) and can be performed along similar lines.
A slight generalization of the sum over $S$ in (\ref{3.70}) gives the identity
\bea\label{B19}
& &\sum_S (-)^{2S}(2S+1)
\left\{ \begin{array}{ccc}
I & 1 & S \\
L & J & \Delta \end{array}\right\}
\left\{ \begin{array}{ccc}
1 & 1 & 1 \\
I & I' & S \end{array}\right\}
\left\{ \begin{array}{ccc}
S & 1 & I' \\
L & y & L' \\
J & 2 & J' \end{array}\right\} =\\
& &\qquad
(-)^{I+J'+L+1} \sum_{z=1,2,3}(-)^z (2z+1)
\left\{ \begin{array}{ccc}
y & 1 & z \\
1 & 2 & 1 \end{array}\right\}
\left\{ \begin{array}{ccc}
y & 1 & z \\
\Delta & L' & L \end{array}\right\}
\left\{ \begin{array}{ccc}
L' & I' & J' \\
\Delta & I & J \\
z & 1 & 2 \end{array}\right\}\nonumber
\eea
where $y$ can take the values $y=1,2,3$. Applying (\ref{B19}) twice we obtain
\bea
I_{SS'} &=& \sum_{z=1,2,3}(-)^z (2z+1)
\sum_{y=1,2}(-)^y (2y+1)
\left\{ \begin{array}{ccc}
1 & 1 & y \\
1 & 2 & 1 \end{array}\right\}
\left\{ \begin{array}{ccc}
1 & 1 & y \\
\Delta' & L & L' \end{array}\right\}\\
& &\qquad\times
\left\{ \begin{array}{ccc}
y & 1 & z \\
1 & 2 & 1 \end{array}\right\}
\left\{ \begin{array}{ccc}
y & 1 & z \\
\Delta & \Delta' & L \end{array}\right\}
\left\{ \begin{array}{ccc}
\Delta' & I' & J' \\
\Delta & I & J  \\
z & 1 & 2 \end{array}\right\}\,.\nonumber
\eea
The contribution of the second term in (\ref{B9}) to the reduced matrix
element $Q(J'I',JI)$ is given, in terms of $I_{SS'}$, by
\bea
[Q(J'I',JI)]_2 = 6(-)^{L+L'}\sqrt{5(2\Delta+1)(2\Delta'+1)}I_{SS'}\,.
\eea
The total expression for $Q(J'I',JI)$ is given by 
\bea
Q(J'I',JI) = [Q(J'I',JI)]_1 + [Q(J'I',JI)]_2
\eea
which can be seen to have the form of the general solution (\ref{Qansatz}).

\newpage
\section{Quark operators for pion couplings of excited states}

There exists an alternative description of the baryon states and of their 
couplings in the large-$N_c$ expansion, based on the use of quark operators 
\cite{DJM1,MR1,CarGeOso,CGKM}.
Compared to the method of the consistency conditions used in the main
text, this approach has the advantage of making a direct connection with
the quark structure of the baryons. This connection is obvious for the
case of baryons containing heavy quarks, but the validity of the method
is not restricted to this case and extends also to baryons
made up of light quarks.

In this Appendix we give a partial proof of the equivalence of the method of the
consistency conditions as used in the main text with the method
of the quark operators. More precisely, we show that the two-body
operators introduced in \cite{CGKM} to parametrize the pion couplings of the
$L=1$ excited baryons to the ground state baryons, give the same contribution
as the one-body operators in the large-$N_c$ limit. This clarifies the
relation of our results to those of \cite{CGKM}. This proof can probably be
made complete along the lines of \cite{DJM1} to include the contributions of
all $n$-body quark operators.

We begin by briefly describing the basic idea of the $1/N_c$ expansion expressed
in the language of quark operators. Any QCD operator ${\cal O}$, such as the 
axial current or the pion coupling to baryons, can be expanded as \cite{DJM1}
\bea\label{qop}
{\cal O} = \sum_{n,k} c_k^{(n)}\frac{1}{N_c^{n-1}}{\cal O}_k^{(n)}\,.
\eea
Here ${\cal O}_k^{(n)}$ are all possible $n$-body operators with the
same quantum numbers as the QCD operator ${\cal O}$. The contribution of an 
$n$-body operator to the matrix element of ${\cal O}$
involves, in the language of the Feynman diagrams, at least 
$n-1$ gluon lines connecting different quarks in the baryon. This supplies a
factor of $\alpha_s^{n-1}$ which translates, in the large-$N_c$ limit, into 
the suppression factor $1/N_c^{n-1}$ in (\ref{qop}). 

Counting powers of $1/N_c$ with the help of (\ref{qop}) is obscured by the
fact that the matrix elements of ${\cal O}_k^{(n)}$ can be proportional to
powers of $N_c$. This can happen if the contributions of the $N_c$ quarks 
in the baryon add up coherently into the matrix element of ${\cal O}_k^{(n)}$.
For the case of excited baryons in the initial state transforming under the
mixed symmetry representation, it has been pointed out in \cite{CGKM} that
there are infinitely many operators contributing to leading order in $1/N_c$.
This means that some $n$-body operators ${\cal O}_k^{(n)}$ will have matrix 
elements of order $N_c^{n-1}$, which will compensate the suppression factor in
(\ref{qop}).

We will consider in the following all quark operators which contribute 
to leading order in $1/N_c$ to the $S$-wave and $D$-wave pion couplings up 
to and including 2-body operators. These will be denoted as in \cite{CGKM}
\bea
Y^a &=& aA^a + \frac{1}{N_c}bB^a + \cdots\\
Q^{ka} &=& dD^{ka} + \frac{1}{N_c}\left( eE^{ka} + fF^{ka}\right) + \cdots\,.
\eea
with $a,b,d,e,f$ unknown coefficients of the order of unity.
The 1-body operators $A^a$ and $D^{ka}$ are identical to the ones introduced
already in Sec.III
\bea
A^a  &=& \langle 0|11;ji\rangle \sum_{n=1}^{N_c} r_n^i \sigma_n^j \tau_n^a\\
D^{ka} &=& \langle 2k|11;ji\rangle \sum_{n=1}^{N_c} r_n^i \sigma_n^j \tau_n^a\,.
\eea
The 2-body operators are defined as \cite{CGKM}
\bea\label{2bodyB}
B^a &=& \langle 0|11;dc\rangle \sum_{n\neq n'=1}^{N_c}
\langle 1c|11;ij\rangle r_n^i \sigma_n^j \sigma_{n'}^d \tau_{n'}^a\\
\label{2bodyE}
E^{ka} &=& 
\langle 2k|11;ij\rangle \sum_{n\neq n'=1}^{N_c}
r_n^i \langle 1j|11;pq\rangle \sigma_n^p \sigma_{n'}^q \tau_{n'}^a\\
\label{2bodyF}
F^{ka} &=&
\langle 2k|11;qj\rangle \sum_{n\neq n'=1}^{N_c}
\langle 1q|11;ip\rangle r_n^i \sigma_n^p \sigma_{n'}^j \tau_{n'}^a\,.
\eea
In addition to these operators, the authors of \cite{CGKM} include also two other 
2-body quark operators $C^a$ and $G^{ka}$. However, when considering only SU(2)
pion couplings as in the present paper, their matrix elements are not enhanced
by a factor of $N_c$ so they will not be included.

We are interested in the matrix elements of $Y^a$ and $Q^{ka}$ taken between
mixed symmetry excited states and symmetric states.
The matrix elements of the 1-body quark operators $A^a$ and $D^{ka}$ have been 
computed already in Sec.IV to
leading order in $1/N_c$ and in Sec.V to all orders in $1/N_c$. In the following
we describe the computation of the matrix elements of the 2-body operators
$B^a$, $E^{ka}$ and $F^{ka}$.

The matrix elements of the 2-body operators can be expressed in terms of the
quantity ${\cal T}_{s}(I',SI)$ defined by
\bea\label{calTs}
& &\langle I'L';m'm'_L\alpha'|\sum_{n\neq n'=1}^{N_c} r_n^i
\langle s j|11;kl\rangle \sigma_n^k \sigma_{n'}^l \tau_{n'}^a|
SI;m_S m_L \alpha\rangle = 
\frac{1}{(2I'+1)\sqrt{2L'+1}}{\cal T}_{s}(I',SI)\\
& &\qquad\qquad
\langle I'm'|S s;m_S j\rangle \langle L'm'_L|L1;m_L i\rangle
\langle I'\alpha'|I1;\alpha a\rangle\,.\nonumber
\eea
The reduced matrix element ${\cal T}_{s}(I',SI)$ can be computed
using the methods applied in Sec.IV.D for the computation of
${\cal T}(I',SI)$. We obtain in this way to leading order in $1/N_c$
\bea
{\cal T}_{s}(I',SI)
= 2N_c^{\kappa+\frac12}\sqrt{2s+1}\sqrt{(2S+1)(2I+1)(2I'+1)}(-)^{-I+I'}
\left\{ \begin{array}{ccc}
s & 1 & 1 \\
I & S & I' \end{array}\right\}\,.
\eea
Note the additional factor of $N_c$ compared with the corresponding result
for the 1-body operator (\ref{calTfinal}), which can overcome the suppression 
inherent to the 2-body operators.

We will put now the 2-body operators (\ref{2bodyB},\ref{2bodyE},\ref{2bodyF}) into
a form involving the operator in (\ref{calTs}). For $B^a$ this can be done
by writing the successive couplings of the vectors entering the definition of the 
operator as
\bea\label{recoup}
B^a &=& ``|(r_n\sigma_n)1,\sigma_{n'};0\rangle '' =
-\sum_s \sqrt{3(2s+1)}
\left\{ \begin{array}{ccc}
1 & 1 & 1 \\
1 & 0 & s \end{array}\right\}
``|r_n,(\sigma_n \sigma_{n'})s;0\rangle''\\
&=& ``|r_n,(\sigma_n \sigma_{n'})1;0\rangle'' =
\langle 0|11;ij\rangle \sum_{n\neq n'=1}^{N_c}r_n^i
\langle 1j|11;kl\rangle  \sigma_n^k \sigma_{n'}^l \tau_{n'}^a\,.\nonumber
\eea
We used here the recoupling relation for 3 angular momenta \cite{Edmonds}.
This result is the analog for spherical coordinates of the well-known vector
identity $(\vec r_n\times \vec\sigma_n)\cdot \vec\sigma_{n'} = \vec r_n\cdot 
(\vec\sigma_n\times \vec\sigma_{n'})$.
Writing $B^a$ in this form one can see that its contribution to the $S$-wave amplitude 
is related to ${\cal T}_{s=1}(I',SI)$ in the same way the matrix element of $Y^a$
is related to ${\cal T}(I',SI)$ (\ref{calTfinal}). Furthermore, the two reduced matrix
elements ${\cal T}(I',SI)$ (\ref{calTfinal}) and ${\cal T}_{s=1}(I',SI)$ (\ref{calTs})
are identical, up to a trivial numerical factor, so that their contributions to $Y^a$
will be also identical. This completes the equivalence proof for the quark 
operators mediating $S$-wave transitions.

This argument can be extended to the 2-body quark operators mediating $D$-wave
couplings. For $E^{ka}$ the proof is immediate, because its matrix elements are
related to ${\cal T}_{s=1}(I',SI)$ (\ref{calTs}) in the same way the matrix elements
of $D^{ka}$ are related to ${\cal T}(I',SI)$ (\ref{calTfinal}). Since the ${\cal T}$
matrix elements are proportional, so will be their contributions to $Q^{ka}$ too.
The corresponding proof for $F^{ka}$ is slightly more complicated, and involves
first casting this operator in a different form with the help of the recoupling 
relation (\ref{recoup}).
\bea
F^{ka} &=& ``|(r_n\sigma_n)1,\sigma_{n'};2k\rangle '' =
-\sum_s \sqrt{3(2s+1)}
\left\{ \begin{array}{ccc}
1 & 1 & 1 \\
1 & 2 & s \end{array}\right\}
``|r_n,(\sigma_n \sigma_{n'})s;2k\rangle''\\
&=& -\sum_s \sqrt{3(2s+1)}
\left\{ \begin{array}{ccc}
1 & 1 & 1 \\
1 & 2 & s \end{array}\right\}
\langle 2k|1s;ij\rangle \sum_{n\neq n'=1}^{N_c}r_n^i
\langle sj|11;kl\rangle  \sigma_n^k \sigma_{n'}^l \tau_{n'}^a\,.\nonumber
\eea

The matrix element of $F^{ka}$ can be now written in the $|(IP)S,L;Jm\alpha\rangle$
basis in terms of ${\cal T}_{s}(I',SI)$ (\ref{calTs}) as
\bea
& &\langle J'I',m'\alpha'|F^{ka}|(IP)S,L;Jm\alpha\rangle = \sqrt{5\frac{2J+1}{2I'+1}}
\langle J'm'|J2;mk\rangle \langle I'\alpha'|I1;\alpha a\rangle\\
& &\qquad\times\sum_s (-)^s \sqrt{3(2s+1)}
\left\{ \begin{array}{ccc}
1 & 1 & 1 \\
1 & 2 & s \end{array}\right\}
{\cal T}_s(I',SI)
\left\{ \begin{array}{ccc}
S & s & I' \\
L & 1 & L' \\
J & 2 & J' \end{array}\right\}\,.\nonumber
\eea
This can be transformed to the basis $|I,(PL)\Delta;Jm\alpha\rangle$ with the
help of the recoupling relation (\ref{3Jrecoupling}). The resulting sum
over $S$ can be performed with the result
\bea
\Sigma_S &=& \sum_S (2S+1)
\left\{ \begin{array}{ccc}
s & 1 & 1 \\
I & S & I' \end{array}\right\}
\left\{ \begin{array}{ccc}
I & 1 & S \\
L & J & \Delta \end{array}\right\}
\left\{ \begin{array}{ccc}
S & s & I' \\
L & 1 & L' \\
J & 2 & J' \end{array}\right\}\\
&=& (-)^{J'-I-s-L}
\sum_y (-)^y (2y+1)
\left\{ \begin{array}{ccc}
1 & 1 & y \\
L' & \Delta & L \end{array}\right\}
\left\{ \begin{array}{ccc}
1 & 1 & y \\
2 & 1 & s \end{array}\right\}
\left\{ \begin{array}{ccc}
L' & I' & J' \\
\Delta & I & J \\
y & 1 & 2 \end{array}\right\}\,.\nonumber
\eea
The remaining sum over $s$ can be done with the help of the orthogonality
relation for 6$j$-symbols. We obtain finally the following result for the
matrix element of $F^{ka}$
\bea
& &\langle J'I',m'\alpha'|F^{ka}|I,(PL)\Delta;Jm\alpha\rangle = 
\langle J'm'|J2;mk\rangle \langle I'\alpha'|I1;\alpha a\rangle\\
& &\qquad\times 2N_c^{\kappa+\frac12} \sqrt{15(2J+1)(2I+1)(2\Delta+1)}
(-)^{I-J+I'+J'}
\left\{ \begin{array}{ccc}
1 & 1 & 1 \\
L' & \Delta & L \end{array}\right\}
\left\{ \begin{array}{ccc}
L' & I' & J' \\
\Delta & I & J \\
1 & 1 & 2 \end{array}\right\}\nonumber
\eea
which can be seen to have again the same form as the general solution of
the consistency condition for $Q^{ka}$ (\ref{Qansatz}).

\end{document}